%% file: ReiSch08.tex
\documentclass[seceqn,dvips]{elsart}
\usepackage{ifpdf}
\usepackage{graphicx,amssymb,lineno}

\makeatletter
\def\elsartstyle{%
    \def\normalsize{\@setfontsize\normalsize\@xiipt{14.5}}
    \def\small{\@setfontsize\small\@xipt{13.6}}
    \let\footnotesize=\small
    \def\large{\@setfontsize\large\@xivpt{18}}
    \def\Large{\@setfontsize\Large\@xviipt{22}}
    \skip\@mpfootins = 18\p@ \@plus 2\p@
    \normalsize
}
\@ifundefined{square}{}{}

\usepackage{epsfig} 
\usepackage{bbm} 
\usepackage{float,a4wide} 
\usepackage{latexsym}
\usepackage{theorem}
\usepackage{wrapfig}
\usepackage[latin1]{inputenc}
\usepackage[T1]{fontenc}
\usepackage{graphics}
\usepackage[intlimits]{amsmath}
\usepackage{xspace} 
\usepackage{hyperref}

\input{Macros}

\makeatother

\journal{Annals of Physics}
\begin{document}
\begin{frontmatter}
\title{Hamiltonian Approach to $1+1$ dimensional \\ Yang--Mills theory in Coulomb gauge}
\author{H.\ Reinhardt, W.\ Schleifenbaum}
\address{Institut f\"ur Theoretische Physik\\
Universit\"at T\"ubingen\\
D-72076 T\"ubingen, Germany}

\ead{hugo.reinhardt@uni-tuebingen.de}

\begin{abstract}
We study the Hamilton{ian} approach to $1+1$ dimensional Yang--Mills theory in
Coulomb gauge, considering both the {pure} Coulomb gauge and the gauge where in
addition the remaining constant gauge field is restricted to the Cartan algebra.
We evaluate the corresponding Faddeev--Popov determinants, resolve Gauss' law and
derive the Hamiltonians, which differ in both gauges due to additional zero
modes of the Faddeev--Popov kernel in the {pure} Coulomb gauge. By Gauss' law the
zero modes of the Faddeev--Popov kernel {constrain} the physical wave functionals to
zero colour charge states.
We solve the Schr\"odinger equation in the {pure} Coulomb gauge and
determine the vacuum wave functional. The gluon and ghost propagators
and the static colour Coulomb potential are calculated in the first Gribov
region as well as in the fundamental modular region, and Gribov copy
effects are studied. We explicitly demonstrate that the
Dyson--Schwinger equations do not specify the Gribov region while the
propagators and vertices do depend on the Gribov region chosen. In
this sense, the Dyson--Schwinger equations alone do not provide the full
non-abelian quantum gauge theory, but  subsidiary conditions must be
required. Implications of Gribov copy effects for lattice calculations
of the infrared behaviour of gauge-fixed propagators are discussed.
We compute the ghost-gluon
vertex and provide a sensible truncation of
Dyson--Schwinger equations. Approximations of the variational approach
to the $3+1$ dimensional theory are checked by comparison to the $1+1$
dimensional case.
\end{abstract}

\end{frontmatter}


\section{Introduction}
In recent years there has been a renewed interest in Yang--Mills theory in
the Coulomb gauge, both in the continuum approach 
{\cite{ChrLee80,Swi88,Sch85,CutWan88,Zwa03,SzcSwa02,Szc03,FeuRei04,FeuRei04a,ReiFeu04,ReiEpp07,FeuRei07,Zwa98,WatRei06,WatRei07b,WatRei07a}}
and on the lattice
\cite{CucZwa02,GreOleZwa04a,LanMoy04,diss_Moyaerts,NakSai06,GreOle07,Qua+07}. 
Being non-covariant,
Coulomb gauge is more cumbersome for perturbative calculations and, in fact, it
has not yet been proved that Yang--Mills theory in the Coulomb gauge is
perturbativ{e}ly renormalisable. However, Coulomb gauge {is
  superior to} covariant gauges {such as} Landau gauge when it comes to nonperturbative
investigations of the infrared sector of the theory. The reason is that in
Coulomb gauge, Gauss' law can be explicitly resolved (and thus the separation of
gauge dependent and gauge invariant degrees of freedom accomplished) \cite{ChrLee80}. 
This is a
particular advantage in the Hamilton{ian} approach where the resolution of Gauss' law
leads to a confining static potential between colour charges. A perturbative
calculation of this potential allows the extraction of the running coupling
constant \cite{Dre81,Khr70}. 
Furthermore, in the so-called first order formalism of the
functional integral approach, the resolution of Gauss' law in the Coulomb gauge leads
to a cancellation of the Faddeev--Popov determinant and thus avoids in principle
the introduction of ghost fields \cite{Zwa98,WatRei06}. 
The Yang--Mills Hamiltonian in the Coulomb gauge
is the starting point of a variational solution of the Yang--Mills
Schr\"odinger equation in Refs.\ \cite{SzcSwa02,Szc03,FeuRei04,ReiFeu04}. 
Using Gaussian type{s} of ans\"atze for the wave functional,
minimisation of the vacuum energy density gives rise to a set of
equations (similar to the
Dyson--Schwinger equations) for various propagators and
vertices. Imposing certain approximations, these equations have been solved analytically in the
infrared \cite{SchLedRei06} 
and numerically in the whole momentum regime
\cite{FeuRei04,EppReiSch07,FeuRei07}. If the curvature of the
space of gauge orbits expressed by the Faddeev--Popov determinant is properly
included one indeed  finds a linearly rising static potential and an infrared
diverging gluon energy, both phenomena signalising confinement. 

In the derivation of the Dyson--Schwinger equations from the variational principle
\cite{SzcSwa02,FeuRei04},
a couple of assumptions and approximations are involved which are difficult to
control in $D = 3 + 1$ dimensions. 
For this reason, we apply in the present paper the
variational approach of Ref.\ \cite{FeuRei04}
 to Yang--Mills theory in $D = 1 + 1$, which can
be solved exactly in the Coulomb gauge. With the exact solution at hand we then test
various assumptions and approximations involved in the $D = 3 + 1$ dimensional
case. 

Yang--Mills theory in $1 + 1$ dimension{s} is trivial in flat Minkowski space but
becomes non-trivial on a compact manifold. We shall consider Yang--Mills theory
on the space-time manifold {$S^1 \times \Real$} which is convenient for the Hamilton{ian}
approach. On this manifold the Yang--Mills Schr\"odinger equation has been solved
exactly {with} a (almost) complete gauge fixing \cite{HetHos89}. 
This gauge fixing consists of
the \pcg, which leaves in $D = 1 + 1$ a constant spatial gauge field,
 and an
additional gauge condition, which exploits the residual global gauge invariance
left in the \pcg to diagonalise the algebra-valued gauge field. We will
refer to this (almost complete) gauge {as the} \dcg. The additional
global gauge fixing is not implemented in the variational approach in $D = 3 +
1$ dimensions. In $D=1+1$, the {\it pure} and {\it diagonal Coulomb gauges} are here directly
compared by identifying their Gribov regions and (previously unknown) boundary
conditions on the wave functionals. The latter boundary conditions arise by
paying proper attention to those zero modes of the respective
Faddeev--Popov kernels that are not related to the field configurations on the
Gribov horizon but to the incomplete gauge fixing. Thus, the exact
vacuum wave functionals are derived in both gauges, and the
propagators and vertices are calculated within the corresponding Gribov
regions. These results for the Green functions are shown to be a
particular solution of the exact Dyson--Schwinger equations defined on
the {\it first} Gribov region. Other solutions to the Dyson--Schwinger
equations are shown to exist and to be defined on a union of
several Gribov regions, including many Gribov copies. We investigate
which kind of approximation of the Dyson--Schwinger equations leads to
which one of these solutions, for the sake of comparison to $D=3+1$
where a truncation of the Dyson--Schwinger equations in mandatory.

The calculation of the static colour Coulomb potential 
within a variational approach to $D=3+1$ yields linear quark
confinement \cite{EppReiSch07} but includes certain approximations. We
here calculate the static colour Coulomb potential in both the \pcg and
the \dcg and compare the results for the Coulomb string tension to the
gauge invariant string tension extracted from the Wilson loop. Along
these lines, the quality of approximations in the $D=3+1$ calculations can be estimated
by comparison to the $D=1+1$ case.

The organisation of this paper is as follows: In section \ref{Hamsec}, we briefly review the
Hamilton{ian} formulation of Yang--Mills theory in $1 + 1$ dimensions on the
space-time manifold {$S^1 \times \Real$}, extract the physical
configuration space from the Wilson loop and identify the gauge invariant and
gauge degrees of freedom. In section \ref{FPsec}, we discuss incomplete gauge fixing in the
Hamiltonian approach by the Faddeev--Popov method and calculate explicitly the
Faddeev--Popov determinant of the \pcg, which differs from the one in
the \dcg. In section \ref{Gribovregions}, we discuss the gauge fixing on $S^1$ in detail and
determine the Gribov regions and the so-called 
fundamental modular region. The restriction
of the configuration space to the fundamental modular region imposes boundary
conditions on the wave functionals, which are also extracted in this section. In
section \ref{Gausssec}, Gauss' law is explicitly resolved for both the
{\it pure} and {\it diagonal Coulomb gauges}. The exact solution of the Yang--Mills Schr\"odinger equation in
the \pcg is given in section \ref{statespsec}. 
With the exact vacuum wave functional at hand,
the exact propagators and vertices are calculated in section \ref{exactGreen}. 
In section \ref{potsec}, the potential between static
colour charges is determined and the Coulomb string tension is 
extracted. In section \ref{DSEsec}, the Dyson--Schwinger equations are
derived. Gribov copy effects on the Green functions are discussed in
section \ref{ManyGribov}. The quality of a common truncation of
Dyson--Schwinger equations is assessed in section \ref{truncsec}. In section \ref{variationalsec}, we apply
the variational approach of Ref.\ \cite{FeuRei04} to Yang--Mills theory
in $D = 1 + 1$ on {$S^1 \times \Real$} using the same variational wave
functional in order to check the approximations made in $D = 3 +1$. A short summary and
some concluding remarks are given in section \ref{summsec}. Some mathematical derivations
are presented in appendices.

\section{Hamilton{ian} formulation of 1+1 dimensional Yang--Mills theory on $S^1 
\times \mathbb{R}$}
\label{Hamsec}
We consider $SU (N_c)$ Yang--Mills theory in $1 + 1$ dimensions. 
In $D = 1 + 1$, the gauge field $A_\mu (x)$ is dimensionless while the gauge
coupling constant $g$ has dimension of inverse length. It is, however,
convenient to absorb the gauge coupling $g$ into the gauge field $g A_\mu \to
A_\mu$, so that $A_\mu\equiv A_\mu^a T^a$ has dimension of inverse length. We will use
anti{hermitian} generators $T_a , a = 1, \dots, N_c^2 - 1$ of the gauge group,
satisfying
\be
[T_a, T_b] = f_{abc} T_c \hk ,
\ee
where $f_{abc}$ is the structure constant. For $SU(2)$, the generators are
related to the Pauli matrices $\tau_{a = 1, 2, 3}$ by $T_a = - \frac{i}{2}
\tau_a$.

In 1+1 dimensional (flat) Minkowsk{i} space Yang--Mills theory is
trivial but becomes non-trivial on a compact manifold. The only compact space-time manifold with a canonical time is
the cylinder
\be
\label{BG1}
S^1 \, (\mbox{space}) \, \times \, \Real \,  (\mbox{time})\; .
\ee
On $S^1\times\Real$, the gauge invariant de{g}rees of freedom are the
spatial Wilson loops winding around $S^1$,
\be
\label{Wilsondef}
\overline{W}[A] = \frac{1}{N_c}\, \tr\,\mathrm{P}\, \exp \lk - \oint\limits_{S^1}
{dx^1A_1(x^0,x^1)} \rk \; ,
\ee
which represent closed electric flux lines. 
The spatial $S^1$ can be realized 
by considering a finite interval on the $x^1$-axis 
of length $L$ and imposing the {periodic} boundary condition\footnote{The
periodic boundary condition can be taken without loss of generality since the
$SU(N_c)$ bundle over $S^1$ is trivial as noticed in Ref.\ \cite{Hos89}.}
\be
\label{BG3}
A_{\mu} (x^0, x^1 = L) = A_{\mu} (x^0, x^1 = 0) \hk .
\ee
In the absence of fermions in the fundamental representation this boundary 
condition remains intact under gauge tran{s}formations 
\be
A^U_\mu  = U A_\mu U^\dagger + U \partial_\mu U^\dagger \hk 
\ee
satisfying the boundary condition
\be
\label{BG4}
U (x^0, x^1 = L) = Z_n (x^0) U (x^0, x^1 = 0) \hk ,
\ee
where $Z_n , n = 0, 1, \dots, N_c - 1$
 is an element of the centre $Z (N_c)$ of the gauge group.

Throughout the paper we will work in
 the  canonical Hamilton{ian} approach \cite{Jac80} and
 impose the Weyl 
gauge
\be
\label{Weyldef}
A_0 = 0 \hk . 
\ee
The gauge transformation required to bring a periodic gauge field $A_0
(x^0, x^1 = L) = A_0 (x^0, x^1 = 0)$ into the Weyl gauge,
\be
U^\dagger (x^0, x^1) = \mathrm{P} \exp \lk - \il^{x^0}_0 d t A_0 (t, x^1) \rk\; ,
\ee
 is also periodic and
thus within the class (\ref{BG4}).
On the flat space $\Real$, an analogous gauge transformation
\be
\label{F4A-GX}
V^\dagger (x^0, x^1) = \mathrm{P} \exp \lk - \il^{x^1}_0 d s A_1 (x^0, s) \rk
\ee
could also gauge away the field $A_1$, i.e.\ $A^V_1 = 0$. However,
the gauge transformation $V$ (\ref{F4A-GX}) is not within the class (\ref{BG4})
and therefore not allowed on $S^1$. It is the compactification of space
$\Real \to S^1$ which
makes the theory non-trivial.
\bi

\no 
The canon{i}cal quantisation is carried out at a fixed time $x^0$, so that $A_1
(x^0-\mbox{fixed}, x^1)$  is the only field ``coordinate'' left. To
simplify the notation, we will omit henceforth the spatial index $i = 1$ and write 
$x^1 \to x \hk , \hk \partial_1
\to \partial \hk , \hk 
 A^a_1(x^1) \to A^a(x)$ etc. The Hamiltonian of $1 + 1$ dimensional
Yang--Mills theory then reads
\be
\label{p4-9}
H = \frac{g^2}{2} \int d x \,\Pi^a (x) \Pi^a (x) \hk ,
\ee
where
\be
\Pi^a (x) = \frac{1}{i} \frac{\delta}{\delta A^a (x)}
\ee
is the momentum operator, which represents the electric field. 
Note in 1+1 dimensions, there is no magnetic field and hence no potential 
term 
in the Hamiltonian. 

Having quantised the theory in the Weyl gauge, the classical Gauss law
$\hat D\Pi=\rho$ is lost from the quantum equations of
motion. Enforcing Gauss' law as an operator identity contradicts the
canonical commutation relations.\footnote{Alternatively,
the Dirac bracket formalism can be used to quantise the theory after
fixing the gauge transformations generated by Gauss' law. The Gauss law can
then be imposed as an operator identity. This was shown
in $3+1$ dimensional Minkowski space to lead to the same
energy spectrum \cite{PhDSchleifenbaum}.} One therefore imposes Gauss' law as a
constraint {on} the wave functional $\Psi (A)$
\be
\label{BG6}
 \hat{D}(x) \Pi(x) \Psi (A) =  \rho (x) \Psi (A) \hk .
\ee
Here, 
\be
\label{BG7}
\hat{D}(x) = \partial +  \hat{A}(x), \hspace{0.7cm} 
\hat{A}(x) = A^a(x) \hat{T}_a, 
\ee
denotes the covariant derivative in the adjoint representation  
with
\be
( \hat{T}_a )_{bc} = f_{bac}
\ee
being the generators in the adjoint representation.
Furthermore, 
$\rho (x)$ denotes the colour density of matter fields (or external 
sources). 
The operator on the l.h.s.\ of Eq.\ \eqref{BG6}, $\hat D\Pi$, is the
generator of so-called ``small'' gauge transformations (see section
\ref{Gribovregions}) and in the absence of matter fields, $\rho (x) = 0$, Gauss' law forces 
 the wave functionals 
$\Psi (A)$  to be invariant under
 the small gauge tran{s}formations $U$,
 \be
 \Psi (A^U) = \Psi (A) \hk .
 \ee




Instead of working with gauge invariant wave functionals, it is more convenient to explicitly resolve Gauss law 
by fixing the gauge \cite{ChrLee80} and for this purpose the
Coulomb gauge
\be
\label{Couldef}
\partial \, A(x) = 0\ee
is particularly convenient. In one spatial dimension, the
Coulomb gauge
constrains the gauge field to spatially constant modes
and the field theory reduces to quantum mechanics in these constant modes. 
The gauge transformation $\Omega$ which brings a given gauge field $A$ into
the Coulomb gauge $\partial A^\Omega = 0$ can be chosen
\be
\Omega^\dagger (x) = V^\dagger (x) (V (L))^{\frac{x}{L}} \hk ,
\ee
where $V (x)$ is defined by Eq. (\ref{F4A-GX}) with the
$x^0$-dependence suppressed. Contrary to $V(x)$, $\Omega (x)$
is periodic for periodic $A (x)$, and thus within the class of allowed gauge
transformations (\ref{BG4}). The gauge-transformed field
\be
A^\Omega = \frac{1}{L} \ln V (L)
\ee
is space-independent and thus obviously satisfies the Coulomb gauge condition.


The Coulomb gauge \eqref{Couldef} (together with Weyl gauge \eqref{Weyldef}) still leaves invariance 
 with respect to global (space and time independent) gauge transformations,
  which 
 can be exploited to diagonalise the constant gauge field 
\be
\label{BG9}
A = A^aT_a\;\to\; A^{a_0}T_{a_0}=A_{diag}\; .
\ee
Here $T_{a_0}$, $a_0 = 1$, $\dots$, $N_c - 1$ denotes the generators of the 
Cartan sub-algebra which can be chosen to be diagonal. Equation (\ref{BG9}) fixes 
the global transformation $U$ up to a constant element of the Cartan sub-group, 
i.e.\ even after implementing the gauge condition $A = diagonal$ in addition 
to the \pcg, there still is a residual global abelian $U (1)^{N_c - 1}$ 
gauge symmetry.
{Moreover,} there still is a discrete symmetry left: The above gauge conditions
(\ref{Couldef}) and (\ref{BG9}) 
do not fix the so-called Weyl symmetry consisting of the $N_c!$ permutations of the
$N_c$ eigenvalues of $A^aT_a$. In the case of $SU (2)$, where the two eigenvalues have
the same modulus but opposite signs, the Weyl symmetry switches the  signs of the
eigenvalues.

Note that we can re-express the two gauge
 conditions (\ref{Couldef}) and (\ref{BG9}) as 
{\begin{subequations}
\label{BG14}
\begin{align}
f^{\bar{a}} [A] &= A^{\bar{a}} (x) = 0 \, ,\\
f^{a_0} [A] &= \partial A^{a_0} (x) = 0 \, , 
\end{align}
\end{subequations}}
\no i.e.\ the Coulomb gauge condition is imposed only on the abelian part $A^{a_0}$ 
while the 
non-abelian components of the gauge field denoted by the index 
$\bar{a} \neq a_0$ vanish. In the following we will refer to this gauge as
{\it diagonal Coulomb gauge}, while Eq.\ (\ref{Couldef}) will be called {\it
  pure Coulomb
gauge}.

To simplify the explicit calculations, we will 
confine ourselves henceforth to the gauge group $SU (2)$ where the structure constants 
coincide with the totally anti-symmetric tensor 
$f^{abc} = \varepsilon^{abc}$, $a = 1, 2, 3$, and we will choose the generator 
of the Cartan sub-group to be given by $a_0 = 3$. 
For $SU (2)$, the $A^a$ can be interpreted as the Cartesian components of a 
vector $\vA=A^a\ve_a$ in a 3-dimensional Euclidean
space with Cartesian unit vectors $\ve_{a = 1, 2, 3}$. 
In the \dcg specified by Eq.\ 
(\ref{BG14}) the colour vector $\vA$ is rotated into the
direction of the positive or negative 3-axis
\be
\label{diagdef}
\vA =  A^3 \ve_3 \hk , \hk A^3 = \pm |\vA| \; .
\ee
The two signs differ by a Weyl refle{ct}ion $\vA \to - \vA$. Thus the
gauge transformation from the \pcg into the \dcg is given by rotating
the colour vector $\vA$ into the positive 3-direction $\ve_3$, possibly
followed by a Weyl refle{ct}ion $A^3 \to - A^3$.

The modulus $|\vA |$ of the constant Coulomb gauge field 
represents the only gauge invariant degree of freedom. In fact, the only physical observable of the theory,
the spatial Wilson loop $\overline{W}$ winding (non-trivially) around the whole space
manifold $S^1$, Eq.\ \eqref{Wilsondef}, is easily
calculated in (both pure and diagonal) $SU(2)$ Coulomb gauge to be
given by
\be
\overline{W} = \frac{1}{2}\, \tr\, \exp (- AL) = \cos \vartheta \; ,
\ee
where we have introduced the dimensionless variable
\be
\label{9-30-a}
\vartheta = \frac{\abs{\vA}L}{2} \;  ,
\ee
for later convenience. $\overline{W}$ attains all possible values in $[-1,1]$ when $\vartheta$ traverses the interval
\be
\label{FMR1st}
0 \leq \vartheta \leq \pi \; .
\ee
Equation \eqref{FMR1st} defines the physical configuration space of the theory. We will later
recover the interval \eqref{FMR1st} as the so-called fundamental modular region.

To sep{a}rate the constant Coulomb gauge field $\vA$ in{to} (global) gauge 
invariant and
gauge dependent parts, it is convenient to use the spherical coordinates
$\abs{\vA}, \theta, \phi$ to write\footnote{For a $3$-vector $\vA$, the
caret ``$\hat{\ }$'' denotes as usual the unit vector
$\hat\vA=\vA/\abs{\vA}$, while for algebra- and group-valued
quantities the caret means the adjoint representation.}
\be
\label{9-28}
\vA =\abs{\vA}\, {\bf \hat{\vA}} (\theta, \phi) \equiv \frac{2}{L}\,\vth \, {\bf \hat{\vA}} (\theta, \phi) \; ,
\ee
where
\be
{\bf \hat{\vA}} (\theta, \phi) = \sin \theta \lk \cos \phi \,\ve_1 + \sin \phi \,\ve_2 \rk
+ \cos {\theta} \,\ve_3
\ee
is the radial unit vector.

The primary
 aim of the present paper is to use 1+1 dimensional Yang--Mills theory 
as testing ground for the variational approach
developed in $3 + 1$ dimensions \cite{FeuRei04}, assessing the
approximations introduced there. In studies of Coulomb gauge
Yang--Mills theory in higher dimensions, merely the \pcg is
fixed. Therefore, we will here mainly focus on the \pcg as
well. Moreover, the \dcg will be used to investigate the effect of
a complete gauge fixing on the Green functions of the theory. 

\section{Gauge fixing in the Hamilton{ian} approach by the Faddeev--Popov method}
\label{FPsec}
In the Hamiltonian approach in Weyl gauge, there is no need to fix the
residual time-independent gauge symmetry.
 In principal, one can work (in the absence of external colour charges) with 
gauge invariant wave functionals, which trivially satisfy Gauss law. 
It is only a matter of convenience that one prefers to fix the gauge. 
Furthermore, as will be explicitly shown in the context of the
resolution of Gauss' law (see section \ref{Gausssec}), the gauge fixing needs not to be complete, i.e.\ any partial 
gauge fixing is allowed in the Hamiltonian approach. In any case, the wave functionals have to
{be} invariant under the residual gauge symmetries unfixed by the gauge condition.
In the \pcg, the wave functionals have to be
invariant under global colour rotations $U = const$, which are not
fixed by that gauge. 

Gauge fixing is accomplished in the Hamilton{ian} approach by
applying the Faddeev--Popov method to
 the functional integral over the spatial gauge fields defining 
the scalar product in the Hilbert space of wave functionals.
 Consider the matrix element of a gauge invariant observable $\cO [A^U] = \cO [A]$
 \be
 \label{12ZZ}
 \langle \Psi_1 | \cO [A] | \Psi_2 \rangle = \int \cD A\, \Psi^*_1 [A]\, \cO [A]\, \Psi_2
 [A] \hk .
 \ee
  The Faddeev--Popov method amounts to inserting 
into the functional integral the identity 
\be
\label{BG32}
1 = \Det\, \cM [A] \int D \mu (\Omega) \, \delta (f^a [A^\Omega])  
\ee
where $\Omega = \e^\Theta, \Theta =
\Theta^a T_a$, denotes a gauge transformation
and
\be
\label{BG33}
\cM^{ab} (x, y) = \frac{\delta f^{a} [A^\Omega] (x)}{\delta \, \Theta^b (y)}
\ee
is the Faddeev--Popov kernel. Furthermore, for a complete
gauge fixing, 
$D \mu (\Omega)$ denotes the Haar measure of the gauge group.
 Inserting the identity (\ref{BG32}) into Eq.\ (\ref{12ZZ}) 
and exploiting the gauge invariance of both the wave functional and the 
observable,
 one finds after {a} change of {the} integration variable 
\be
\label{BG35}
\langle \Psi_1 | \cO[A] | \Psi_2 \rangle = \int \cD A \, \Det \, \cM [A] 
\, \delta (f^a [A]) {\Psi_1}^{\ast} [A] \, \cO [A] \, {\Psi_2} [A] \int D \mu (\Omega) 
\ee
where the integration of the gauge group is now explicitly separated, yielding
 a (infinite) constant which can be absorbed into {the} normalisation of the wave 
 functional.

If the gauge condition $f^a [A^\Omega] = 0$ does not fix the gauge completely, 
there are directions in the space of gauge transformations, along 
which the gauge-fixing functional $f^a [A^\Omega]$ does not change, i.e.\ 
{$\delta f^a [A^\Omega] / \delta \Theta^b = 0$}, and the tangent vectors 
corresponding to these directions represent zero 
modes of 
the Faddeev--Popov kernel (\ref{BG33}). In order that the identity (\ref{BG32}) 
holds and thus the Faddeev--Popov method works one has to exclude these zero 
modes and integrate only over the subspace of gauge transformations which are 
fixed by the gauge condition $f^a [A^\Omega] = 0$. Furthermore,
the gauge fixing is defined only in the region of gauge field
configurations $A$ where the Faddeev--Popov determinant $\Det\,\cM[A]$
is non-zero. This will be important for the 
application of the Faddeev--Popov method given below. 

The explicit calculation of the Faddeev--Popov kernel is most easily 
accomplished by noticing that the Gauss law operator $\hat{D} \Pi$ is the 
generator of (so-called small\footnote{See section \ref{Gribovregions}.}) gauge transformations 
\begin{align}
\label{BG36}
f^a [A^\Omega] &= \cG (\Theta) f^a [A] \cG^{-1} (\Theta) \, ,\\
\cG (\Theta) &= \exp \, {\lk 
- i \int d x \lk \hat{D}^{a b} (x) \Theta^b (x) \rk
\Pi^a (x) \rk}
\, .\nonumber
\end{align}
For infinitesimal gauge transformations 
$\cG (\delta \Theta) = 1 + i \int d x \, \delta \Theta^a (x) (\hat{D} 
\Pi)^a (x)$, we have accordingly
\be
\label{BG37}
f^a [A^\Omega] (x) = f^a [A] (x) + \int d y \, \delta \Theta^b (y) \, \big[ i 
(\hat{D} \Pi)^b (y), f^a [A] (x) \big]
\ee
and comparison with the Taylor expansion of $f^a [A^\Omega]$ 
in powers of $\delta \Theta^a (x)$ reveals that 
the Faddeev--Popov kernel (\ref{BG33}) can be expressed as 
\be
\label{BG38}
\cM^{ab} (x, y) 
= i \, \hat{D}^{bc} (y) \, \big[ \Pi^c (y), f^a [A] (x) \big] \; .
\ee
Below, we use this relation to determine the Faddeev--Popov kernels for the two 
gauges considered above and calculate the corresponding Faddeev--Popov 
determinants.
\subsection{Diagonal Coulomb gauge}
Applying the relation (\ref{BG38}) 
to the \dcg condition (\ref{BG14}), we obtain the Faddeev--Popov
kernel ($\bar b=1,2$)
{\begin{subequations}
\label{BGneu2}
\begin{align}
\cM^{a \bar{b}} (x, y) &= - \hat{D}^{a \bar{b}} (x) \delta (x, y) \\
\cM^{a 3} (x, y) &= - \hat{D}^{a3} (x) \partial^x \delta (x, y) 
\end{align}
\end{subequations}}
On the gauge shell (where the gauge condition is fulfilled), the Faddeev--Popov 
kernel becomes block diagonal and reads in the Cartesian basis
\begin{align}
\label{BGneu3}
\cM (x, y) =  \left( 
\begin{array}{ccc} 
- \partial_x &  A^3 & 0 \\ 
-A^3 & - \partial_x  & 0 \\ 
0 & 0 & - \partial^2 
\end{array} 
\right) \delta (x,y) 
= \left( 
\begin{array}{ccc} 
\: i\: & 0 & 0 \\ 
0 & \: i\: & 0 \\ 
0 & 0 & i \, \partial_x 
\end{array} 
\right) \; i  \hat{D}(x)\; \delta (x, y) \; .
\end{align}
The Faddeev--Popov determinant becomes 
\be
\label{BGneu4}
\Det \, \cM = \Det_{(1)} \, (-i \partial)^{33} \, \Det \, (i \hat{D})
\; .
\ee
The first factor is a (divergent) irrelevant constant, which drops out from the 
expectation values and can be absorbed into the functional integral
measure, see below. Note, 
this determinant is defined in the (one-dimensional) abelian colour sub-space $a
= a_0 = 3$  
only, indicated by the sub-script $(1)$. To evaluate the second 
factor $\Det \, (i \hat{D})$, we consider the eigenvalue equation
\be
\label{BGneu5}
i \hat{D}^{ab} [A] (x) \varphi^b (x) = \lambda \, \varphi^a (x) 
\ee
From the representation (\ref{BG33}), it is clear that the eigenfunctions of 
the Faddeev--Popov kernel have to satisfy the same boundary condition as the 
gauge angles $\Theta^a (x)$. Since the gauge transformations on the spatial 
$S^1$ have to satisfy periodic boundary conditions $\Theta (x + L) = \Theta 
(x)$, the eigenfunctions of the Faddeev--Popov kernel have to be periodic 
 as well. Since $\partial$ and $\hat{D}$ commute on the gauge-fixed
 manifold, the operators $\partial, \hat{D}$ and $\cal M$ have common eigenfunctions and we can
 impose the periodic boundary condition also on the eigenfunctions of $i
 \hat{D}$,
\be
\label{BGneu6}
\varphi^a(x + L) = \varphi^a(x) \hk .
\ee
With this boundary condition, which does not mix the different colour
components, and with the fact that the gauge field is spatially constant, the
eigenfunctions of $i \hat{D}$ factorise in space and colour dependent parts.
Using the results of appendix \ref{rots} and Eq. \eqref{diagdef}, the eigenfunctions  are given by
\be
\label{BGneu7}
\varphi^a_{n, \sigma} (x) \equiv \langle x, a | n, \sigma \rangle = 
\langle x | n \rangle \langle a | \sigma \rangle 
\ee
where $\langle a | \sigma \rangle = \e^a_\sigma$ are (the Cartesian
components of) the polar unit vectors $\ve_\sigma$ (\ref{BGneu9}) and
\be
\label{BGneu8}
\langle x | n \rangle = \frac{1}{\sqrt{L}}\, \e^{- i k_n x} \; , \quad
k_n = \frac{2 \pi n}{L}\; , \quad n\in\mathbb{Z}
\ee
are plane waves (periodic in $L$).
The corresponding eigenvalues read 
\be
\label{BGneu10}
\lambda_{n, \sigma} = k_n + \sigma A^3 \hk .
\ee
The two eigenvalues $\lambda_{n, \sigma = \pm 1}$ correspond to the 
non-abelian block $(\bar{a} = 1, 2)$ 
 in the upper left corner in the Faddeev--Popov matrix 
(\ref{BGneu3}) while the eigenvalue $\lambda_{n, \sigma = 0} = k_n$ 
corresponds to the abelian colour direction $a_0 = 3$ (see Eq.\
(\ref{BGneu9})). Furthermore, the zero mode $\varphi_{n = 0, \sigma = 0} =
 \ve_{\sigma = 0} / \sqrt{L}\equiv \ve_3/\sqrt{L}$ ($\lambda_{n = 0,
   \sigma = 0} = 0$) represents the tangent vector (to the gauge orbit) corresponding to
the infinitesimal global $U(1)$ colour rotation 
 ($\Theta^3 = const$) which is not fixed by the gauge condition (\ref{BG14}).
As discussed above, this mode has to be excluded from the spectrum of the Faddeev--Popov kernel
 (\ref{BGneu3}),
 whose eigenvalues are given by
 \be
 \label{15-66}
 \Lambda_{n, \sigma} = \left\{ \begin{array}{lcl} i \lambda_{n, \sigma} & , &\quad
 \sigma = \pm 1 \\
 k_n \lambda_{n, \sigma} = k^2_n & , &\quad \sigma = 0 \end{array} \right. \hk .
 \ee
  Note also that the eigenmodes $\varphi_{n = 0, \sigma = \pm1}$ corresponding
   to the global $SU(2)/U(1)$ gauge transformations which are fixed by the 
   gauge condition (\ref{BG14})
    do not give rise to zero eigenvalues, $\lambda_{n = 0, 
   \sigma = \pm 1} = \pm A^3 \neq 0$.

Excluding the zero mode $n = \sigma = 0$ (indicated in the following by a prime), we obtain
with $A^3 = \pm |\vA|$ (see Eq.\ \eqref{diagdef})
\begin{align}
\label{BGneu11}
\Det' \, (i \hat{D} [A]) &=\prod^{\infty}_{n = - \infty}{}' \, \prod_{\sigma = 
0, \pm 1} \lambda_{n, \sigma} = \big( \prod_{m \neq 0} k_m \big) 
\prod^{\infty}_{n = - \infty} \lambda_{n, 1} \lambda_{n, -1} \nonumber\\
&= \big( \prod_{m \neq 0} k_m \big) \, \abs{\vA}^2 \big( \prod^\infty_{n = 1} \, 
(k^2_n - \abs{\vA}^2) \big)^2 \nonumber \\
&= \big( \prod_{m \neq 0} k^3_m \big) \, \lk 
|\vA| \, \prod^{\infty}_{n = 1} 
\big( 1 - \frac{\abs{\vA}^2}{k^2_n} \big) \rk^2
\end{align}
The first factor represents $\Det (i \partial)$ with 
the zero mode $n = 0$ excluded. Using 
\be
\label{BGneu12}
\sin x = x \, \prod^{\infty}_{n = 1} \big( 1 - \big(\frac{x}{\pi n}\big)^2 \big) 
\ee
we obtain 
\be
\label{BGneu13}
\Det' \, (i \hat{D} [A]) = \Det \, (i \partial)\,\big(\frac{2}{L}\big)^2\, \sin^2 \left(\frac{|\vA|L}{2}\right)
\equiv \Det \, (i \partial)\,\big(\frac{2}{L}\big)^2\, \sin^2 \vth\; .
\ee
All field-independent factors in the Faddeev--Popov determinant \eqref{BGneu4} can be absorbed in the functional integral measure. We thus
arrive at the Faddeev--Popov determinant $\cJ_D$ of the \dcg (\ref{BG14})
\be
\label{J_Dres}
\cJ_D := \frac{\Det \cM}{\Det_{(1)} \, (-i \partial)^{33} \, \Det (i
  \partial)\,(\frac{2}{L})^2\,} = \sin^2 \vth\; .
\ee
Let us stress that it was absolutely crucial to exclude the gauge modes 
which are not fixed by the gauge condition. Otherwise the
Faddeev--Popov determinant would have vanished identically.

The Faddeev--Popov determinant $\cJ_D$ has zeros at (note that by definition $\vth \geq 0$)
\be
\label{BGneu21}
\vartheta \equiv \frac{|\vA|L}{2} = n \pi \; , \quad n = 0, 1, 2, \dots 
\ee
and thus divides up the gauge-fixed configuration space into regions
where the Faddeev--Popov method of gauge fixing is defined. These so-called {\it
  Gribov regions} are given by 
\be
\label{BGneu22}
n \pi \leq \vartheta {\, < \,} (n + 1) \pi \hk .
\ee
Recall that in the \dcg $A^3 = \pm |\vA|$, so that in this variable the
Gribov regions are given by
\be
\label{diagGribov}
\left\{A_3\;\Big|\;\frac{2 n \pi}{L} \leq  | A_3 | {\, < \,} \frac{2 (n + 1) \pi}{L}\right\} \; , \quad  n = 0,
1,  2, \dots \;\, .
\ee
The boundaries of the Gribov regions, the {\it Gribov horizons}, are given by the
discrete momenta $k_n$ (\ref{BGneu8}). We will return to the
discussion of the Gribov regions in section \ref{Gribovregions}.

Note that the Faddeev--Popov determinant $\cJ_D$ \eqref{J_Dres} vanishes also at the (classical)
perturbative vacuum $A = 0$. This is not surprising since the diagonalisation of
the gauge field, and thus the \dcg, is ill-defined for $A = 0$.
Therefore, this gauge is not suitable for perturbation theory.
\subsection{Pure Coulomb gauge}
Let us now consider the \pcg \eqref{Couldef} which leaves the global gauge 
transformations unfixed. The Faddeev--Popov kernel \eqref{BG38} is then given by
\be
\label{FPcoo}
\cM^{ab} (x, y) = (- \hat{D}^{ab} {[A]}(x) \partial_x) \, \delta (x, y)\; .
\ee
Since the gauge field in the \pcg is related to the one in the \dcg by
a global gauge transformation, we can express the covariant derivative
$\hat D^{ab}[A]$ in the \pcg by the one in the \dcg. Let $U^\dagger$
denote the global gauge transformation which rotates the colour vector
$\vA$ into the positive $3$-direction, i.e.\
\be
\label{rotier}
{A^a\hat{T}_a} =  \abs{\vA} \hat{U} \hat{T}_3 \hat{U}^T\; .
\ee
Then we have (see appendix \ref{rots}) 
\be
\label{17A*}
i \hat{D}{ [A]} = \hat{U}\, i \hat{D} [\abs{\vA}T_3]\, \hat{U}^T 
\ee
where $\hat U$ is the adjoint representation of $U$.
Here, $i \hat{D} [\abs{\vA} T_3]$ is the covariant derivative in the \dcg (with
$A^3 =\pm |\vec{A}|$) whose
eigenvalues and eigenfunctions were determined in the previous subsection. From
Eq.\ (\ref{17A*}), it follows that $i \hat{D} {[A^aT_a]}$ has the same eigenvalues $\lambda_{n,
\sigma}$ (\ref{BGneu10})
 as $i \hat{D} [\abs{\vA} T_3] $, i.e.\ 
 \be
 \label{16-75}
 \lambda_{n, \sigma} = k_n \, + \, \sigma |\vA| \hk ,
 \ee
  and that the eigenfunctions $|\tilde{\varphi} \rangle$ of $i \hat{D} [A]$ are related to the eigenfunctions $|
\varphi \rangle$ (\ref{BGneu7}) of $i \hat D [\abs{\vA} T_3]$ by
\be
\label{19-77}
\tilde{\varphi}^a_{n, \sigma} (x) = \hat{U}_{a b} \varphi^b_{n, \sigma} (x) =
\langle x | n \rangle u^a_\sigma \hk ,
\ee
where $\langle x | n \rangle$ are the periodic plane waves (\ref{BGneu8})
 and we have defined (see appendix \ref{rots})
\begin{align}
\label{20-81}
u^a_\sigma & :=  \langle a | \hat{U} | \sigma \rangle = \langle a | \hat{U} | b
\rangle \langle b | \sigma \rangle = \hat{U}_{a b} \e^b_\sigma \nonumber\\
 &=  \langle a | \tau \rangle \langle \tau | \hat{U} | \sigma \rangle =
 \e^a_\tau D^1_{\tau \sigma} (\phi, \theta, 0) \hk .
\end{align} 
Since $i \hat{D} {[A]}$ (\ref{17A*}) 
has the same eigenvalues as $i \hat{D} [\abs{\vA} {T}_3]$, and since 
$\det \, \hat{U} = 1$, one would expect
 that the Faddeev--Popov determinant in the \pcg
gauge is, up to an irrelevant constant factor $\Det \, (i \partial) / \Det_{(1)} \, (-i \partial)^{33}$, the same as in the \dcg (\ref{BG14}) 
considered above. 
However, since the \pcg does not fix the global gauge transformation 
$U$, we have to exclude the constant eigenmodes from the Faddeev--Popov kernel. 
These are given by the eigenfunctions with $n = 0$ and all $\sigma$. In 
addition to the zero mode $n = 0, \sigma = 0$ excluded already from the
Faddeev--Popov kernel of 
the \dcg, one has to exclude here also the gauge modes $n = 0, \sigma
= \pm 1$, corresponding to the non-zero eigenvalues $\lambda_{n = 0, \sigma = \pm
1} = \pm\abs{\vA}$. Although these modes correspond to non-zero eigenvalues of $i
\hat{D} [A]$ they give rise to zero modes of the full Faddeev--Popov kernel of
the \pcg,
\be
\label{18-84b}
(- \hat{D} [A] \partial ) \tilde{\varphi}_{n, \sigma}  =  {\Lambda_{n, \sigma}}
\tilde{\varphi}_{n, \sigma} \hk , \hk
\Lambda_{n, \sigma}  =  \lambda_{n, \sigma} k_n
\ee
since
$\Lambda_{n = 0, \sigma} = 0$. Note that {these zero
  modes form precisely} the global gauge transformation $\hat{U}$
which diagonalises the constant Coulomb gauge field $\hat{A} =\hat A^a {T}_a$ (cf.\ Eqs.\ 
(\ref{17A*}), (\ref{9-31})),
\be
\tilde\varphi^a_{n = 0, \sigma} (x) = \frac{1}{\sqrt{L}} \langle a | \hat{U} | \sigma
\rangle   \hk .
\ee
 Omitting these modes in calculating $\Det \, (- \hat{D} \partial)$,
 equivalently to Eq.\ \eqref{BGneu11}, we find
for the Faddeev--Popov determinant $\cJ_P$ in the \pcg
\be 
\label{J_Cres}
\cJ_P = \frac{\Det \, (- \hat{D} \partial)}{\Det \, (- \partial^2)} = 
\frac{\sin^2 \, \big(\frac{|\vA|L}{2}\big)}{\big(\frac{|\vA|L}{2}\big)^2} \equiv \frac{\sin^2
\vartheta}{\vartheta^2} \hk .
\ee
\begin{figure}
\includegraphics[scale=0.5]{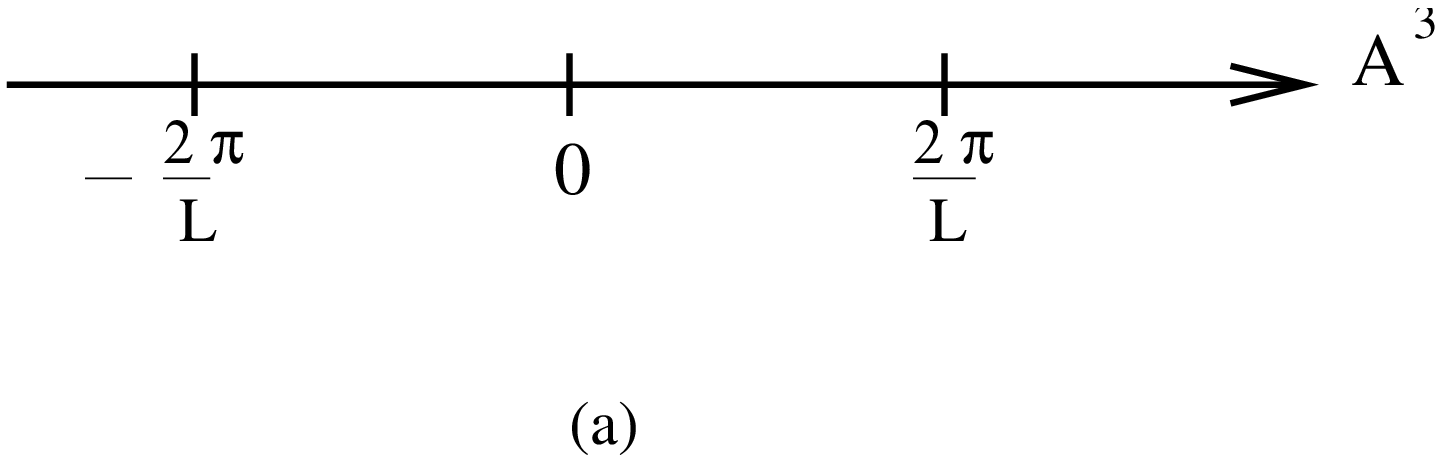}
\hspace{1.9cm}
\includegraphics[scale=0.7]{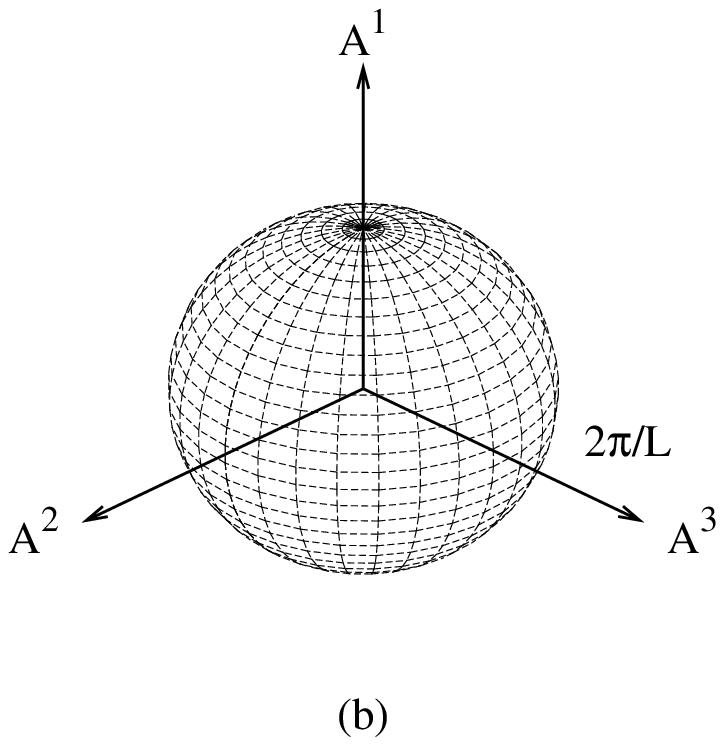}
\caption {The first Gribov {region} for  (a) the \dcg and (b) the
\pcg.}
\label {kugel}
\end {figure}
This determinant differs from the one of the \dcg (\ref{BG14}) by the
denominator (cf.\ Eq.\ (\ref{J_Dres})) and does not vanish  at the 
perturbative vacuum $A = 0$. Furthermore, since $\lim_{|\vA|\to 0}d \cJ_P / d |\vA| = 0$ 
the space of gauge {orbits} is flat near the perturbative vacuum in
this gauge. The first zero 
of the Faddeev--Popov determinant occurs at $\vartheta = |\vA|L/2 = \pi$
 which defines the 
first Gribov horizon to coincide (in the variable $\vartheta$) 
with the one in the \dcg 
(\ref{BGneu22}). However, in the \pcg the direction of the gauge field ${\vA}$
is not fixed, so the Gribov regions $\Omega_n$ are given here by the spherical shells
\be
\label{GriOmedef}
\Omega_n:=\left\{\vA\; \big|\; {(n-1)}\, \frac{2 \pi}{L} \leq | \vA |
  {\, < \,} n\, \frac{2 \pi}{L} \right\} \; , \quad n = 1, 2, 3, \dots .
\ee
In particular, the first Gribov region $\Omega_1$ is given by the sphere $| \vA | <
\frac{2 \pi}{L}$. The first Gribov regions of both the diagonal and the
pure Coulomb gauge are illustrated \mbox{in Fig.\ \ref{kugel}}. The
Gribov regions will be discussed in more detail in the upcoming section.


\section{Gribov regions and boundary conditions on the wave functionals}
\label{Gribovregions}

In this section, the configuration spaces of both the {\it pure} and
{\it diagonal Coulomb gauges} are examined. (For valuable discussions on this topic,
also see, e.g., Ref.\ \cite{PauHei98} and references therein.) The topology of
gauge transformations relating the various Gribov regions of the \pcg
and the \dcg is discussed here, the fundamental modular region is
specified, and boundary conditions on the wave functionals are given.

\subsection{Topology of gauge transformations on $S^1$}
  Since the first homotopy group $\Pi_1 (SU (N_c))$ is trivial, there 
are no ``large'' gauge transformations in one (compact) spatial dimension, i.e.\
on $S^1$. However, in the 
absence of matter fields in the fundamental representation, the gauge group is
in fact 
 $SU (N_c) / Z (N_c)$. This is because the gauge field, living in the adjoint
 representation, is invariant under centre gauge
 transformations\footnote{Since the centre $Z(N_c)$ is a discrete set,
 centre gauge transformations $Z(x)$ have to be piecewise constant. At
the jumps of $Z(x)$ (from one centre element to another) the
inhomogeneous term of the gauge transformation, $Z(x)\del
Z^\dagger(x)$, represents so-called ideal centre vortices \cite{EngRei00,Rei02}. The
centre vortices of the fundamental representation, $Z(x)\del
Z^\dagger(x)=C^a(x)T_a$, become invisible Dirac sheets (strings) in the
adjoint representation $\hat Z(x)\del \hat Z^\dagger(x)$. In
fact, a centre element $Z(x)=\exp(-\mu^a(x)T_a)\in Z(N_c)$ (in the fundamental representation, with
$\mu^a(x)$ being a co-weight vector) becomes the unit matrix in the
adjoint representation, $\hat Z(x)=\exp(-\mu^a(x)\hat T_a)=\id$, so
that the inhomogeneous term $\hat Z(x)\del \hat Z^\dagger(x)$ disappears in the adjoint representation.}  $Z(x)\in Z(N_c)$.
 Since $\Pi_1 (SU (N_c) / Z (N_c)) = Z (N_c)$, 
there are topological non-trivial gauge transformations falling in $N_c$ 
different topological classes and consequently there are $N_c$ distinct classical
vacuum configurations $A_{(n)} = U_{(n)} \partial U^\dagger_{(n)}$ between which
quantum tunnelling occurs. In the fundamental representation, the $U_{(n)}$
satisfy the boundary condition (\ref{BG4}), with $x^0$ fixed, and are specified by the
$N_c$ centre
elements $Z_n \in Z (N_c)$. The $U_{(n = 0)}$, belonging to the trivial centre
element $Z_0 = \Id$ are periodic and form the ``small'' gauge transformations. The remaining ones
are  ``large'' transformations. Since $(Z_n)^{N_c} = \Id, N_c$ successive large gauge
transformations $U_{(n)}$ belonging to the same $Z_n$ yield a small gauge
transformation. The fundamental representation of $SU(N_c) / Z(N_c)$ is the adjoint
representation of $SU(N_c)$, which we will denote in the following by a caret.
The adjoint representation 
$\hat{U} = \exp ( \Theta_a \hat{T}_a)$ is related to the
fundamental representation $U = \exp (\Theta_a T_a)$ 
(with the same $\Theta_a$ !) by
\be
\label{5-11}
U^\dagger T_a U = \hat{U}_{a b} T_b \; , \quad
\hat{U}_{ab} = - 2 \,\tr \lk U^\dagger T_a U T_b \rk \; .
\ee
From this representation it is explicitly seen that the adjoint representation $\hat{U}$ is
centre blind. Accordingly, the allowed gauge
transformations  satisfying (in the fundamental representation of $SU
(N_c)$) the boundary condition (\ref{BG4})  are periodic in the adjoint representation
\be
\hat{U} (x + L) = \hat{U} (x) \hk .
\ee
To be more specific consider the gauge group $SU(2)$ whose centre is $Z (2) = \{ 
\Id , - \Id \}$. The allowed gauge transformations satisfying Eq.\  (\ref{BG4}) are
either periodic (belonging to $Z_{{0}} = \Id$) or anti-periodic (belonging to $Z_{{1}} = -
\Id$). In the absence of matter fields in the fundamental representation,
 the true
gauge group is
$SU(2) / Z(2) = SO(3)$. Since $\Pi_1 (SO(3)) = Z(2)$, there are two 
inequivalent sets of gauge transformations allowed by Eq.\ (\ref{BG4}). The
periodic ones $U_{(0)} (L) = U_0 (0)$  belonging to the trivial centre
element $Z_{{0}} = \Id$, form the {\it small}
 gauge transformations, which can be smoothly
deformed to unity. The anti-periodic gauge transformations, corresponding to the
non-trivial centre element $Z_{{1}} = - \Id$, cannot be smoothly deformed to
unity and
 are called {\it large}.  To provide some explicit examples consider the
gauge transformation
\be
\label{gtdef}
U (x) = {\exp\left({\omega (x) {\bf \hat{n}} \cdot {\bf T}} \right)
\equiv \exp\left({- i \,\frac{\omega}{2}\,
{\bf \hat{n}} \cdot \boldsymbol{\tau}}\right)} = \cos \frac{\omega}{2} - i\, {\bf \hat{n}}\cdot \boldsymbol{\tau} \sin
\frac{\omega}{2} 
\ee
with some constant unit vector ${\bf \hat{n}}$. A small (periodic) gauge
transformation is obtained when $\omega (x)$ satisfies the
{boundary condition}
\be
\omega (x + L) = \omega (x) + 4 n \pi \hk ,
\ee
while a large (anti-periodic) gauge transformation follows for
\be
\label{antiper}
\omega (x + L) = \omega (x) + (2 n + 1) 2 \pi \hk .
\ee
The corresponding adjoint representations $\lk ( \hat{T}_a )_{bc} =
\epsilon_{bac} \rk$
\be
( \hat{U} (x) )_{ab} = \big( \e^{\omega {\bf \hat{n}} \cdot {\bf T}} \big)_{ab} =
\delta_{ab} \cos \omega + {\hat{n}}_a {\hat{n}}_b (1 - \cos \omega) + \epsilon_{acb}
{\hat{n}}_c \sin \omega
\ee
are periodic in both cases.

\subsection{Gribov and fundamental modular regions}
In section \ref{FPsec}, we found that in the \pcg the Faddeev--Popov determinant has zeros
at $\vartheta = \frac{{| {\bf A}}| L}{2} = n \pi$. Hence, in the 3-dimensional space of
constant gauge orbits 
the {\it Gribov horizons} $\del\Omega_n$
 are given by the {surfaces} of spheres around the origin with
radius $|\vA| = \frac{2 \pi n}{L} (\vartheta = n \pi)$, see Eq.\
\eqref{GriOmedef}. Note also that in the limit $L \to \infty$ the
first {Gribov} region $\Omega_1$ shrinks to the point $A= 0$, in agreement with the fact that $1
+ 1$ dimensional Yang--Mills theory becomes trivial on a flat space-time manifold.

The \pcg is not a complete gauge fixing since it leaves
 invariance with respect to global gauge
transformations, which form the zero modes of the Faddeev--Popov kernel. 
However, the global gauge transformations are not the only
symmetries left. In the following we will carefully
examine the residual symmetries left after \pcg fixing. 
Due to the existence of these residual symmetries, the first Gribov
region $\Omega_1$ cannot yet 
be
 the  {\it fundamental modular region}
 which, by definition, contains only a single
copy of each gauge orbit. Since the \pcg is contained in the \dcg, the fundamental modular region is the same in both gauges, while the
Gribov regions are, of course, different. 
The determination of the fundamental modular region is
necessary in order to identify the symmetry relations (i.e.\ boundary conditions)
to be fulfilled by the wave functionals. 

Consider the large (i.e.\ anti-periodic) gauge transformation,
{cf.\ Eqs.\ \eqref{gtdef} and \eqref{antiper}},
\be
\label{7-2}
V (\hat{\vA}) = \exp \lk \frac{2 \pi}{L} x\, \hat{\vA} \cdot \vT \rk \hk , \hk 
\hat{\vA}
= \frac{\vA}{|\vA|} 
\ee
which shifts a constant gauge field $\vA=\abs{\vA}\hat{\bf n}$ along
its direction $\hat{\bf n}$ in colour space by multiples of $\frac{2\pi}{L}$,
\be
\label{7-3}
\vA \to \vA^V = \vA - \frac{2 \pi}{L} \hat{\vA} = \lk \abs{\vA} -
\frac{2 \pi}{L} \rk \hat{\vA} \hk .
\ee
Obviously, the transformed configuration $A^V$ still satisfies the
(pure or diagonal) Coulomb gauge condition if the original does.

\begin{figure}
\originalTeX
\centerline{
\psfig{figure=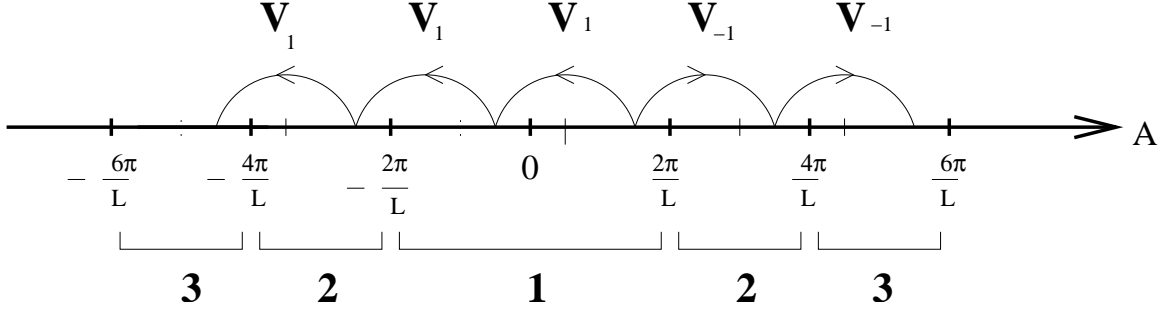,height=4cm}}
\caption {Illustration of the displacements of a field configuration $\vA$ from inside the first Gribov region to Gribov copies in the
neighbouring Gribov regions by the large gauge transformations $V_{\pm
  1} = ( V (\hat\vA) )^{\pm 1}$.}
\label {displace}
\end{figure}\no

The large gauge transformation \eqref{7-3} maps a configuration from the
$n^{th}$ Gribov region $\Omega_n$ to a configuration (on the same ray through the
origin in colour space but) within $\Omega_{n-1}$.\footnote{For a
  peculiarity of $n=1$, see below.}
In this way, any gauge configuration in one Gribov region has
a unique copy in every other Gribov region and all Gribov regions are
homeomorphic to each other. Fig.\ \ref{displace} illustrates the shifting of a
particular gauge configuration from the first Gribov region to
the neighbouring ones. 
Furthermore, the large gauge transformation (\ref{7-3}) maps a configuration of
the $n^{th}$
 Gribov {\it horizon} $\del\Omega_n$ to the configuration on the same ray through the origin
on $\del\Omega_{n-1}$.
In particular, the configurations on the
first Gribov horizon $\del\Omega_1$, where $| \vA | = \frac{2 \pi}{L}$, are mapped to the vacuum
$A = 0$. This shows that all configurations of (all) Gribov horizons are
equivalent under large gauge transformation to the vacuum $A = 0$. In
particular, since two successive large gauge transformations form a small one,
all configurations {on} a given Gribov horizon are related by small gauge
transformations and are thus gauge copies of each other. Let us stress, however,
that the configurations {on the first} Gribov horizon $\del\Omega_1$ (and on all
$\del\Omega_n$ with odd $n$) are not
equivalent to the vacuum $A = 0$ 
with respect to {\it small} gauge transformations, 
they are related to $A=0$ by a
non-trivial (large) gauge transformation. In the quantum theory tunnelling
between these vacua will occur.  This tunnelling will be entirely {accounted} for
by solving the Schr\"odinger equation.

By a large gauge transformation (\ref{7-2}), a configuration $A$
inside the first Gribov region with $|\vA| \leq \frac{\pi}{L}$ is also mapped to a
copy $A^V$ with $\frac{\pi}{L} \leq | \vA^V | \leq \frac{2 \pi}{L}$, $\hat{\vA}^V = - \hat{\vA}$ and vice versa. One could eliminate the large gauge
transformation by restricting the modulus of the gauge field to
\be
\label{7-4}
0 \leq | \vA | \,{<} \, \frac{\pi}{L} \hk .
\ee
However, Gauss' law only enforces the wave
 functionals to be invariant under small
gauge transformations while they can transform according to an arbitrary
representation of the symmetry group under large gauge transformations.
For example, in the colour singlet sector under a large gauge transformation the wave functional
can
a{c}quire a non-trivial phase $e^{i \alpha}$, which is well-known from the
$\Theta$-vacuum in $D = 3 + 1$. 
Therefore we will not remove the large gauge symmetry.\footnote{If one
restricted the configuration space
to $| \vA | \leq \frac{\pi}{L}$, from the large gauge transformations only a residual discrete symmetry on
the new Gribov horizon $| \vA | = \frac{ \pi}{L}$ would be left, which is given by 
the displacement transformations $V
(\hat{\vA})$ (\ref{7-2}) and 
 which relates the
antipodal points $\pm \frac{\pi}{L}
\hat{\vA}$. 
Identifying these antipodal points, which are equivalent by the
displacement transformation \eqref{7-3} the Gribov region, the ball $B_3$ with radius
$\frac{\pi}{L}$, becomes the group manifold of $SO(3)$, which is the gauge group
in the absence of fundamental charges.}

The first Gribov region in the \pcg, $\Omega_1$, is a ball $B_3$ around the origin with radius $| \vA | =
\frac{2 \pi}{L}$, bounded by the first Gribov horizon $\del\Omega_1$,
which is the $S_2$.
Since all configurations on this Gribov horizon are equivalent with
respect to small gauge transformations, we have to identify all points of the first
Gribov horizon, $S_2$, which
compactifies the first Gribov region to $S_3$, which is the manifold of the $SU
(2)$ group. This shows that the configuration space of Yang--Mills theory on $S^1\times\Real$ is the gauge group manifold itself. The mapping from the
configuration space into the gauge group is provided by the (untraced) 
spatial Wilson loop winding around $S^1$. In the fundamental
representation we have
\begin{align}
\label{7-5}
{W}[A]  &=  \mathrm{P} \exp \left(\;\oint_{S^1} dx  A\right) = \exp (L \vA \cdot \vT) = \id\cos \vth + {2}\hat{\vA}\cdot \vT \sin
\vth \; ,\\
&\qquad\vth=\frac{\abs{\vA}L}{2}\; ,\quad 0 \leq | \vA | \leq \frac{2 \pi}{L}\nn \hk .
\end{align}
The Gribov horizons $\del\Omega_n$ with odd $n$ (and in particular $\del\Omega_1$) are mapped onto the
non-trivial centre element
\be
\label{7-6}
W\left[ \abs{\vA} = (2n+1)\frac{2 \pi}{L} \right] = - \unit \; .
\ee
In higher dimensions, field configurations $A$,
 non-trivially linked to a closed
loop $C$ for which the corresponding Wilson loop $W [A] (C)$ equals a
non-trivial centre element, are referred to as centre {vortices}.
In this spirit, the field configurations on the Gribov horizons
$\del\Omega_n$ with odd $n$ represent centre
vortices, in agreement with {the} general observation that centre vortices are on the
Gribov horizon \cite{GreOleZwa04a}, 
due to their larger symmetry.

The first Gribov region $\Omega_1$ of the \pcg can be restricted
further by implementing the \dcg \eqref{diagdef}. In the \dcg, the first
Gribov region is given by 
\begin{equation}
  \label{7-11}
   -\pi\leq \frac{A^3L}{2}<\pi\; ,
\end{equation}
see Eq.\ \eqref{diagGribov}, and
is obviously a subset of $\Omega_1$. There are still gauge copies
within the first Gribov region of the \dcg, due to the fact that
there remains a residual discrete gauge symmetry, the Weyl reflection
\be
\label{7-12}
A^3 \to - A^3 \hk .
\ee
Removing this
symmetry by identifying configurations of opposite sign,
\be
A^3 = | \vA | = \frac{2 \vartheta}{L} \hk ,
\ee
 reduces the
configuration space (\ref{7-11}) to the {\it fundamental modular region}
\be
\label{FMRdef}
{A^3 \in \left[ 0, \frac{2 \pi}{L} \right] \; , \quad \vartheta \in [0, \pi] \hk .}
\ee
This physical configuration space was already found from the gauge-invariant spatial Wilson
loop, see Eq.\ \eqref{FMR1st}, and reduces the configuration space to
the genuinely gauge invariant degree of freedom
$\vth=\abs{\vA}L/2$. The Faddeev--Popov determinant $\cJ_D(\vth)$, see
Eq.\ \eqref{J_Dres}, is gauge invariant as well, since it corresponds to the
Faddeev--Popov determinant of a completely fixed gauge. Such a
Faddeev--Popov determinant is gauge invariant due to the invariance of the Haar measure $\cD\mu(U)$.

The gauge condition that immediately rotates the constant colour vector
$\vA$ of the \pcg into the positive 3-direction is accomplished by the gauge
transformati{on} $\hat{U}$ given by Eq.\ (\ref{8-X2}).
In this gauge the Faddeev--Popov determinant has zeros (Gribov horizons)
at $\vartheta = n \pi \hk ,{
\hk n \in \mathbb{N}}$
and  the Gribov regions are given by the one-dimensional intervals
\be
n \pi \,{\leq}\, \vartheta < (n + 1) \pi \hk , \hk n = 0, 1, 2, \dots \hk .
\ee
For such a gauge, the first Gribov region, $n=0$, coincides with the
fundamental modular region \eqref{FMRdef}.

\subsection{Boundary condition on the wave functionals}
Let us now discuss the implications 
of the residual gauge symmetries on the wave
functionals. In general,
by Gauss'
law the residual gauge symmetries which correspond to small gauge
transformations that are not fixed by the gauge considered have to be respected
by the wave functional.
In the \pcg global gauge invariance is left
unfixed and consequently
the wave functional has to respect this symmetry, i.e.
\be
\label{7-8}
\Psi (A^U) = \Psi (A) \hk , \quad U = const \; .
\ee
Since the global gauge transformations are just rotations in colour space, the
wave functionals have to be colour singlet states satisfying
\be
\label{7-9}
\vL\, \Psi (A) = 0 \hk 
\ee
where
\be
\label{26-110}
L^a = \epsilon^{abc} A^b \frac{d}{i d A^c}
\ee
is the ``orbital'' angular momentum in colour space, which is nothing but the
colour spin of the gauge field.
 In the next section, we will
obtain this constraint in the resolution of Gauss' law in the \pcg as a
consequence of the zero modes of the Faddeev--Popov kernel belonging to the global
gauge symmetry. 
Equation \eqref{7-9} implies that the wave functional is rotationally
invariant,
\begin{equation}
  \label{rotinv}
  \Psi(\vA)=\Psi(\abs{\vA})
\end{equation}
and thus depends only on the gauge invariant modulus $\abs{\vA}$ of $\vA$.

The global gauge transformations do not ex{h}aust the set of small gauge
transformations remaining unfixed in the \pcg. An even number of large gauge transformations (\ref{7-2}) forms a space-dependent
small one
\be
V^{2 n} (\hat{\vA}) = \exp \lk 2 {x} k_n\, \hat{\vA} \cdot \vT \rk \hk , \hk k_n =
\frac{2 \pi n}{L} \hk ,
\ee
which shifts the gauge field by
\be
\vA \to \vA^{V^{2 n} (\hat{A})} = \lk \vA - 2 k_n \hat{\vA} \rk \hk .
\ee
Note that the gauge transform $A^{V^{2 n}}$ still satisfies the \pcg if
the  original configuration $A$ does so. Since the wave functional has to be
invariant under small gauge transformations, it has to satisfy the condition
\be
\Psi \lk \vA - 2 k_n \hat{\vA} \rk = \Psi (\vA) \hk 
\ee
and by Eq.\ \eqref{rotinv}
\be
 \Psi \lk | 2 k_n - | \vA|| \rk =\Psi (|\vA|)\hk .
\ee
Restricting $A$ to the first Gribov region
\be
0 \leq | \vA| < k_1 \equiv \frac{2\pi}{L}\; ,
\ee
the above condition becomes
\be
\Psi \lk \frac{4 \pi}{L} - | \vA | \rk = \Psi \lk | \vA | \rk
\ee
or when expressed in terms of the dimensionless variable $\vartheta =
\frac{|\vA |L}{2}$
\be
\label{22-5.15}
\Psi (2 \pi - \vartheta) = \Psi (\vartheta) \hk .
\ee
Under large gauge transformations $V (\hat{\vA})$ (\ref{7-2}) the wave functional
needs only to be invariant up to a phase
\be
\Psi \lk A^{V (\hat{\vA})} \rk = \e^{i \alpha} \Psi (A)
\ee
and since $( V (\hat{\vA}) )^2$ is a small gauge transformation this phase
has to be $e^{i \alpha} = \pm 1$. Using Eq.\ (\ref{7-3}) and proceeding as above we
find from the effect of the large gauge transformation the boundary condition 
\be
\label{21-5.17}
\Psi (\pi - \vartheta) = \pm \Psi (\vartheta) \hk .
\ee
{The two signs correspond to two superselection sectors of the theory, which are
the discrete analog of the $\Theta$-vacuum in $D = 3 + 1$. In section
\ref{statespsec}, we will find that the ground state belongs to the sector with
the plus sign.
\bi

\no
The global gauge symmetry left in the \pcg
is used in the \dcg to diagonalise the (algebra-valued) gauge
field
\be
\label{7-10}
A^a = \delta^{a 3} A^3 \; , \quad A^3 = \pm  | \vA | \hk .
\ee
After implementing this gauge there is still the residual $SO(2) \simeq U (1)$
global
symmetry of rotations around the 3-axis. 
This abelian symmetry cannot be fixed since the gauge-fixed
configurations \eqref{7-10} are invariant under these rotations. This
implies that also the wave functional defined on the gauge-fixed
manifold automatically respects this symmetry. Therefore, the residual global
$U(1)$ symmetry can be left out in further considerations. 

A small gauge transformation consisting of two successive displacement
transformations
$
V (\ve_3)$ (\ref{7-3}) shifts $A^3$ to $A^3 - \frac{4 \pi}{L}$. By the
identification of $\pm A^3$ this configuration is equivalent to $\frac{4
\pi}{L} - A^3$. Thus $A^3$ and $\frac{4 \pi}{L} - A^3$ (or $\vartheta$ and $2
\pi - \vartheta$) represent the same configuration. This
 residual invariance under the small gauge transformations
$(V (\ve_3))^2$ left by the \dcg 
 has to be respected by the wave functional, which therefore has to
satisfy the boundary condition
\be
\label{7-15}
\Psi ( 2 \pi - \vartheta ) = \Psi (\vartheta)  \hk .
\ee
{This condition (\ref{7-15}) was already obtained above in the \pcg, see Eq.
(\ref{22-5.15}). This is not surprising:}
Since the \dcg contains the \pcg the boundary conditions
following from the residual gauge invariance in the \dcg apply also to
the \pcg.
In section \ref{statespsec} we will solve the Schr\"odinger equation thereby imposing the
boundary conditions  (\ref{7-15}) in
 the \dcg and the conditions (\ref{7-9}) and (\ref{7-15}) in the \pcg. 
\section{Resolution of Gauss' law}
\label{Gausssec}
As already discussed in section \ref{Hamsec}, in
 the Hamilton{ian} approach in Weyl gauge $(A_0 = 0)$ Gauss' law (\ref{BG6}) 
does not follow from
the Heisenberg equation of motion and has to be imposed as a constraint {on} the
wave functional.
In the following, we explicitly resolve Gauss' law in both the {\it pure} and
{\it diagonal Coulomb gauges}, thereby paying proper attention to the zero modes of
the Faddeev--Popov kernel. We
will find that the modes $n = 0, \sigma = 0 , \pm 1$ are excluded from the Coulomb
propagator in the \pcg (although the modes $n = 0, \sigma = \pm 1$ 
are not zero modes of the
Coulomb kernel $\lambda_{n = 0, \sigma \pm 1} = \sigma A \neq 0$!). In the
\dcg only the true zero mode ${n} = \sigma = 0$ is excluded from the
Coulomb propagator. This
is in accord with our discussion of the Faddeev--Popov method in
section \ref{FPsec}.
We will first outline the general strategy of resolving Gauss' law and
afterwards apply it separately to the {\it pure} and {\it diagonal Coulomb gauges}. 

First note, that even in the \pcg where only constant gauge ``fields'' are
  left, the momentum operator has  space-dependent components. 
We denote the part of the momentum operator conjugate to the modes of the gauge
field left after gauge fixing by {$\Pi_\perp $} and the remaining part by
$\Pi_{||} (x)$,
\be
\Pi (x) = {\Pi_\perp} + \Pi_{||} (x) \hk .
\ee
These components are orthogonal to each other in the sense that 
\be
\label{orth}
\int d x\, \Pi_\perp \Pi_{||} (x) = \int d x\, \Pi_{||} (x) \Pi_\perp = 0 \hk
\ee
and we refer to them here as the ``transversal'' and
``longitudinal'' components of the momentum operator, respectively, although this notation is somewhat misleading in the case of
the \dcg (see Appendix \ref{Gaussapp}). With Eq.\ \eqref{orth}, the Yang--Mills Hamiltonian (\ref{p4-9}) becomes
\be
H = {\frac{g^2}{2}} \int d x\, \Pi^2 (x) = {\frac{g^2}{2}} \int d x\, \lk \Pi^2_\perp +
\Pi^2_{||} (x) \rk \hk .
\ee
As usual, we will
solve Gauss' law for the longitudinal part {$\Pi_{||}(x)$}.
Since $\partial \Pi_\perp = 0$ (in both gauges) we can rewrite Gauss' law 
{(\ref{BG6})} as
\be
\label{15XX}
\hat{D}^{a b} (x) \Pi^b_{||} (x) \Psi (A) = \rho^a_{tot} (x) \Psi (A) \hk ,
\ee
where
\be
\label{15XX1}
\rho^a_{tot} (x) = \rho^a (x) + \rho^a_g
\ee
is the total colour charge, including the external charge $\rho^a
(x)$ and the
charge of the gauge bosons
\be
\label{19-90}
\rho^a_g(x) = - \hat{A}^{a b}(x) \Pi^b_\perp \hk .
\ee
Since Gauss' law is a constraint {on} the wave functional and not an operator
identity, one can extract from Gauss' law only $\Pi_{||} \Psi (A)$ but cannot
obtain $\Pi_{||}$ itself. For this reason, we consider the expectation value of
the Hamiltonian and perform a
partial integration with respect to the gauge field to obtain
\be
\langle \Psi | H | \Psi \rangle = \frac{g^2}{2} \int \cD A \int d x \lk \Pi (x) \Psi
(A) \rk^* \Pi (x) \Psi (A) \hk .
\ee
Implementing here the (pure or diagonal) Coulomb
gauge by the Faddeev--Popov method, splitting the
momentum operator into longitudinal and transversal parts $\Pi = \Pi_{||} +
\Pi_\perp$, expressing $\Pi_{||} \Psi$ by Gauss' law and performing a
 partial
integration with respect to the gauge-fixed field, the Hamiltonian becomes
\be
H = \frac{g^2}{2} \int  d x\,  \cJ^{- 1}_{FP}\, \Pi_\perp\, \cJ_{FP}\, \Pi_\perp + H_C \hk ,
\ee
where {$\cJ_{FP}$ is the Faddeev--Popov determinant and} $H_C$ is
the so-called Coulomb Hamiltonian, defined by
\be
\label{80-79}
 \int \cD A\, \cJ_{FP} (A) {\frac{g^2}{2}} \int d x \lk \Pi^a_{||} (x) \Psi (A) \rk^*
\Pi^a_{||} (x) \Psi (A) 
 =:  \int \cD A \,\cJ_{FP} (A) \Psi^* (A) H_C \Psi (A) \hk .
\ee
Formally, from Eq.\ (\ref{15XX}) follows
\be
\Pi^a_{||} (x) \Psi (A) = \int d y\, \langle x | \lk \hat{D}^{- 1} \rk^{a b} | y
\rangle\, \rho^b_{tot} (y)\, \Psi (A)
\ee
and the Coulomb Hamiltonian becomes
\be
\label{20-96}
H_C = \frac{g^2}{2} \int d x d y\, \cJ^{- 1}_{F P}\, \rho^a_{tot} (x) \,F^{a b} (x, y)\,
\cJ_{F P}\, \rho^b_{tot} (y) \hk ,
\ee
where
\be
\label{25-113}
{F}^{a b} (x, y) = \langle x | (- \hat{D}^{- 2} )^{a b} | y \rangle 
= \langle x | \left[ \lk - \hat{D} \partial \rk^{- 1} (-
\partial^2) \lk - \hat{D} \partial \rk^{- 1} \right]^{a b} | y \rangle 
\ee
is the so-called Coulomb kernel. However, the operator $\hat{D}$ has zero modes,
which forbid a naive inversion.
In Appendix \ref{Gaussapp}, we
 explicitly solve Eq.\ (\ref{15XX}) for $\Pi^a_{||} (x) \Psi (A)$ and
extract $H_C$  for
both the \pcg and the \dcg, thereby paying proper
attention to the zero modes. The upshot of these considerations is that the zero
modes of the Faddeev--Popov kernel, which are a consequence of incomplete gauge
fixing, give rise to additional constraints {on} the wave functionals. These
constraints basically arise from the projection of Gauss' law onto the zero modes
of the Faddeev--Popov kernel. In the \pcg these constraints read
(\ref{15-z3}) 
\be
\label{20AX}
Q^a_{tot} \Psi = 0 \hk ,
\ee
where
\be
\label{28-118}
Q^a_{tot} = \il^L_0 d x \,\rho^a_{tot} (x) = \il^L_0
 d x \lk \rho^a (x) + \rho^a_g
(x) \rk \equiv Q^a + Q^a_g 
\ee
is the total colour charge.  In this gauge the transverse momentum operator
(\ref{5-20}) reads
\be
\Pi^a_\perp = \frac{1}{L} \frac{d}{i d A^a}
\ee
 and  
the dynamical charge of the gauge bosons  $Q^a_{g} $ (\ref{19-90})
 becomes (up to a sign) the colour
angular momentum operator (more precisely the colour spin) of the
gauge field, $L^a$ (\ref{26-110}).
The residual constraint (\ref{17-74}) from Gauss' law  becomes
\be
L^a \Psi = Q^a \Psi \hk ,
\ee
where $Q^a$
is the external charge.
In the absence of external colour charges $Q^a = 0$ this constraint simplifies to
\be
\label{28-122}
\vec{L} \Psi = 0 \; ,
\ee
i.e.\ the physical wave functionals do not depend on the angle degrees of  freedom
$\hat{\vA}(\theta, \phi)$, which, in fact,
 are unphysical since they represent the residual
global colour gauge degrees of freedom, which are not fixed by the \pcg
condition. Thus, the
 physical vacuum 
 wave functionals depend only on the ``radial'' coordinate $| \vA |$ 
 which is the physical degree of freedom of the gauge field. The constraint
 (\ref{28-122}) was already found in the previous section, see Eq.\ \eqref{7-9}, and reflects the
 invariance of the wave functional under global gauge transformations.

In the \dcg, where
\be
A^a = \delta^{a 3} A^3 \hk , \quad \Pi^a_\perp  = \delta^{a 3} \Pi^3_\perp \hk ,
\ee
the dynamical charge of the gauge bosons $\rho^a_g$ (\ref{19-90}) 
vanishes and the residual constraint
from Gauss' law  implies the vanishing of the Cartan component of
the external
charge in the physical state (see Eq.\ (\ref{20-92}))
\be
\label{29-124}
Q^3 \Psi = 0 \hk .
\ee
 In absence
of external charges $Q^3 = 0$, in this gauge
there is no residual constraint {on} the wave functional from
Gauss' law. 

The \dcg rotates the constant gauge mode in the 3-direction, i.e.\ $A^3
= \pm | \vA |$ and with the restriction to the fundamental modular region \eqref{FMRdef}, in
this gauge ${A}^3$ equals the modulus of ${\vA}$. Thus, in both gauges the
physical wave functional depends only on the modulus of the constant gauge mode
${A}^a$, which is the only physical degree of freedom. Both gauges leave a
residual {\it global} gauge 
invariance: Global $SU (2)$ symmetry in the case of
the \pcg and global $U (1)$ symmetry in the case of the \dcg. By
Noether's theorem, these global symmetries imply the existence of conserved
charges: $Q^{a = 1, 2, 3}_{tot}$
 in the case of the \pcg and $Q^3$ in the \dcg. The residual constraints {on} the wave functionals obtained above from Gauss'
law are nothing but the quantum version of Noether's theorem for these global
colour symmetries.
Also in $3 + 1$ dimension{s} the \pcg fixing still leaves invariance with
respect to global colour gauge transformations and by Noether's theorem the total
colour charge has to be conserved \cite{ReiWat08}.

After resolution of Gauss' law (see Appendix \ref{Gaussapp}) one finds the following
 gauge-fixed Hamiltonian in the \pcg
\be
\label{HC}
H = {- \frac{g^2}{2L} \, \frac{1}{\cJ_P} 
\frac{d}{d {A}^a} \, \cJ_P \, \frac{d}{d {A}^a} + H_C \hk ,}
\ee
where $\cJ_P$ is defined by Eq.\ (\ref{J_Cres})
 and the Coulomb Hamiltonian  (cf.\ Eq.\ (\ref{20-96})) is given by
\be
\label{H_C}
H_C = \frac{g^2}{2} \int d x d y\, \rho^a (x) F^{a b} (x, y) \rho^b (y)
\ee
with Coulomb kernel
\be
\label{Fdef}
F^{a b} (x, y) = F^{a b} [A=A^aT_a] (x, y) 
= \sli_{n \neq 0} \langle x | n \rangle \sli_\sigma \langle a |
\hat{U} | \sigma \rangle \lambda^{- 2}_{n, \sigma} \langle \sigma | \hat{U}^T |
b \rangle \langle n | y \rangle \hk ,
\ee
from which all zero modes $n = 0$ of the Faddeev--Popov operator (\ref{FPcoo}) 
are excluded,
although $n = 0 , \sigma = \pm 1$ are {\it not} zero modes of the
operator $(- \hat{D}^2)$ in Eq.\ \eqref{25-113}!
Note also that the dynamical charge $\rho_g$ (\ref{19-90}), although being here
non-zero, has dropped out from the Coulomb Hamiltonian (\ref{H_C}). This is a
special feature of $1 + 1$ dimensions (see Appendix \ref{Gaussapp}) and is a consequence of
$\rho_g$ being space-independent in the Coulomb gauge. Similarly, the Faddeev--Popov
determinant also drops out from the Coulomb Hamiltonian.

  The first term in Eq.\ (\ref{HC}) arises from the ``transversal'' momentum operators 
  $\Pi^a_\perp$
  corresponding to the physical mode $(A^a = const)$.
   This term has the form of 
  a Laplacian in a curved space with the Faddeev--Popov determinant acting as the
    determinant of the metric. The second term of Eq.\ (\ref{HC}) arises from 
  the ``longitudinal'' (here $x$-dependent) part $\Pi_\parallel^a(x)$ of the momentum operator. 
   This 
  term gives the static potential of external static colour 
  charges, and it is considered an advantage of the \pcg that this 
  term is explicitly isolated. Note, the Hamiltonian in the \pcg
  (\ref{HC}) is still invariant under global colour rotations, which are not
  fixed in this gauge.

  Resolving Gauss' law in the \dcg (\ref{BG14}) (which does fix the
  global colour rotations) yields the gauge-fixed Hamiltonian
  \be
  \label{F22AYY}
  H = - \frac{g^2}{2 L} \frac{1}{\cJ_D} \frac{d}{d A^3} \cJ_D \frac{d }{d A^3} + H_C \hk
  ,
  \ee
  where $\cJ_D$ is defined in Eq.\ (\ref{J_Dres}). Here
  the Coulomb Hamiltonian $H_C$ is still
  given by Eq.\ (\ref{H_C}), however, with the Coulomb kernel $F^{a b} [A] 
  (x, y)$
  (\ref{Fdef}) replaced by
  \be
  \label{29-125}
  F^{a b} (x, y) = 
  F^{a b} [A=\abs{\vA}T_3] (x, y) = \sli_{n, \sigma}{}' \langle x | n \rangle \langle a | \sigma
  \rangle \lambda^{- 2}_{n, \sigma} \langle \sigma | b \rangle \langle n | y
  \rangle \hk ,
  \ee
  where the prime indicates that the mode $n = \sigma = 0$ is 
  excluded, which  is the only zero mode of the Faddeev--Popov operator in
  this gauge.

  Note, since the \pcg field $A$ is related to the field in the \dcg
  by a {global} gauge transformation, see Eq.\ \eqref{rotier}, one would expect
  that in view of Eq.\ \eqref{17A*}, the Coulomb kernels (\ref{25-113})  in these two gauges  are related by
  \be
  \label{30-130}
  F^{a b} [A = U A^3T_3 U^\dagger] (x, y) = \hat{U}_{ac} F^{c d}
   [A^3T_3] (x, y) \hat{U}^T_{d b}
  \hk .
  \ee
  This is almost the case (cf.\ Eqs.\  (\ref{Fdef}) and (\ref{29-125})) except
  for the additional zero modes $n = 0 , \sigma = \pm 1$ to be excluded from the
  kernel (\ref{Fdef}) in the \pcg. The zero modes $n = 0, \sigma = 0, \pm
  1$ are precisely given by the
  constant gauge transformation (\ref{20-81})  $u^a_\sigma = \langle a | 
  \hat{U} | \sigma \rangle (\langle x | n = 0 \rangle =
  const)$
    not fixed in the \pcg. 

  By fixing the \pcg, one switches from Cartesian to curvilinear coordinates 
and accordingly the gauge-fixed Hamiltonian a{c}quires the form of the 
Hamiltonian in curved space \cite{ChrLee80}.
 The gauge-fixed Hamiltonian is, of course, no
longer gauge invariant. In particular, the Hamiltonian (\ref{F22AYY})
 is
not even invariant under global colour rotation since the \dcg (\ref{BG14})
fixes also the global gauge transformations. Nevertheless, the \dcg
 still leaves invariance under global abelian gauge transformations (colour
rotations around the 3-axis). 

  In the absence of
  external charges $(\rho^a (x) = 0) $, the Coulomb term $H_C$ (\ref{H_C}) 
  obviously vanishes
  in both gauges,  however, for 
  different reasons, see appendix \ref{Gaussapp}. In the \pcg it vanishes because all
constant modes $n = 0, \sigma = 0, \pm 1$ are excluded from the Coulomb kernel
$F^{a b} (x, y)$, while in the \dcg it vanishes because the physical modes live all in the Cartan
algebra, resulting in a vanishing colour charge (\ref{19-90}) 
of the gauge bosons.

\section{The physical state space}
\label{statespsec}
The spectrum of the Yang--Mills Hamiltonian in $1 + 1$ dimensions can
be obtained in a gauge invariant way \cite{Raj88}. We choose
here the \dcg, with all
unphysical degrees of freedom eliminated and the physical degree
of freedom $\vth$ within the fundamental modular region $0\leq\vth\leq\pi$,
see section \ref{Gribovregions}. Thus, one can regain the gauge
invariant spectrum and simultaneously find the
physical state space. From the vacuum wave functional
$\Psi[A]$ in the \dcg, the one in the \pcg can be derived which will
be very useful in the subsequent sections.

\subsection{Diagonal Coulomb gauge}
In the \dcg (\ref{BG14}), the Yang--Mills Schr\"odinger
equation of the $1 + 1$ dimensional theory
\be
\label{4-1}
{H} \Psi_k = E_k \Psi_k
\ee
can be solved exactly \cite{HetHos89}.
In the absence of external colour charges
 the Yang--Mills Hamiltonian (\ref{F22AYY}) reads 
 in the compact variable $\vth=\frac{|{\bf A}|L}{2}$
\be
\label{BGneu23}
H = - \frac{1}{2 L} \lk \frac{g L}{2} \rk^2
 \, \frac{1}{\cJ_D(\vth)} \, \frac{d}{d \vartheta} \, 
\cJ_D(\vartheta) \, \frac{d}{d \vartheta} \hk .
\ee
To solve the Schr\"odinger equation, we introduce the ``radial'' wave functional $\phi (\vartheta)$ by
\be
\label{4-2}
\Psi(\vth) = \frac{1}{\sqrt{\cJ_D (\vartheta)}} \,\phi (\vartheta) \hk .
\ee
This eliminates the Faddeev--Popov determinant $\cJ_D(\vth)=\sin^2\vth$
in the scalar product  
\be
\label{4-3}
\langle \Psi_1 | \Psi_2 \rangle = \il^\pi_{0} d \vartheta \,\cJ_D (\vartheta) \Psi^*_1
(\vartheta) \Psi_2 (\vartheta) = \il^\pi_0 d \vartheta \,\phi^*_1 (\vartheta)
\phi_2 (\vartheta)
\ee
and reduces the Schr\"odinger equation (\ref{4-1}) to
\be
\label{4-4}
\frac{d^2\phi}{d\vth^2} =-k^2 \phi\; ,\quad k^2=1+\frac{8E}{g^2L}
\hk .
\ee
The boundary condition (\ref{7-15}) on the total wave functionals $\psi
(\vartheta)$ requires the radial wave functional $\phi (\vartheta)$ to satisfy
\be
\phi (2 \pi - \vartheta) = - \phi (\vartheta)\;  .
\ee
Thus, the solutions to Eq.\ (\ref{4-4}) read
\be
\label{4-5}
\phi_k (\vartheta) = \sqrt{\frac{2}{\pi}}\, \sin (k \vartheta)\; , \quad
k\in\mathbb{N}\; .
\ee
These are normalised with respect to the scalar product \eqref{4-3}. The corresponding energy eigenvalues are given by 
\be
\label{4-6}
E_k = \frac{g^2 L}{8}(k^2 - 1)=\frac{g^2L}{2}\, j(j+1)=\frac{g^2L}{8}\, l(l+2)\; .
\ee
Here, we have defined
\be
j=\frac{k-1}{2}=0,\,\frac{1}{2},\, 1,\,\dots
\ee
to identify the spectrum \eqref{4-6} as a rigid rotor in colour
space where the integer and half-integer $j$ correspond to the two superselection
sectors defined by the boundary condition (\ref{21-5.17}).
Alternatively, one can use the definition
\be
l = k - 1 = 2 j
\ee
to recognise in Eq.\ \eqref{4-6} the energy eigenvalues of a point particle with mass $4/(g^2 L)$ and angular momentum $l$ on a unit sphere $S_3$ in $D = 4$,
which is the group manifold of $SU(2)$. (In fact, $H$ (\ref{BGneu23}) is (up to
the constant factor) the polar angle part of the Laplacian on
$S_3$.) Either way, the eigenfunctions
\be
\psi_j (\vartheta) = \sqrt{\frac{2}{\pi}}\, \frac{\sin ((2 j + 1) \vartheta)}{\sin
\vartheta} = \sqrt{\frac{2}{\pi}} \,\chi_j (2 \vartheta)
\ee
are the characters of $SU(2)$ ($\hat{\bf n}$ -- arbitrary unit vector)
\be
\chi_j (\beta) = \sli_m \langle j m | \e^{i \frac{\beta}{2} \hat{{\bf n}}\cdot{\boldsymbol{\tau}}} |
j m \rangle
\ee
and the eigenvalues \eqref{4-6} are seen to diverge in the thermodynamic limit $L \to \infty$ except for $j = 0$.
 Therefore, all states are frozen, except the one with $j = 0$, which
 has vanishing energy $(E_{k=1} = 0)$ and the vacuum wave functional 
\be
\label{PsiD}
\Psi (\vartheta) = \sqrt{\frac{2}{\pi}} \hk .
\ee
The vacuum wave functional $\Psi$ has no dependence on the gauge ``field'' $\vartheta = \frac{\abs{\vA}L}{2}$
and describes a stochastically distributed weight of gauge field
configurations in the expectation values of the \dcg,
\be
\label{vevD}
\langle\cO(\vth)\rangle_{\textrm{diagonal gauge}} = \int_0^\pi d\vth\,\cJ_D(\vth)\cO(\vth)\abs{\Psi(\vth)}^2=\frac{2}{\pi}\il^\pi_0 d \vartheta \sin^2
\vartheta\, \cO(\vartheta) 
\hk .
\ee
It is claimed that in $3+1$ dimensions the gauge field configurations that dominate the infrared physics are located on the
common boundary of the Gribov region and the fundamental modular
region \cite{Zwa97}. In $1+1$ dimension, this is obviously not realized. The
Faddeev--Popov determinant $\cJ_D=\sin^2\vth$ actually {\it suppresses}
the contributions of $\vth=\pi$ which is the common boundary of Gribov and
fundamental modular regions. This is due to the fact that in the fully
gauge-fixed theory, i.e.\ in the \dcg, the configuration
space is merely one-dimensional and entropy does not favour boundary
contributions (although expected to do so in $D=3+1$). 


\subsection{Pure Coulomb gauge}
We are interested here in the vacuum wave functional in the \pcg. Although the \pcg Hamiltonian (\ref{HC}) apparently has a much more complicated
structure than the Hamiltonian (\ref{BGneu23}) in the \dcg
both expressions are equivalent, due to gauge invariance. In the following, we will explicitly reduce the Hamiltonian in
the \pcg to the \dcg Hamiltonian (\ref{BGneu23}). This will provide us with the
explicit form of the wave functional in the \pcg, which is needed for
subsequent considerations. 

In the absence of external charges, the Yang--Mills
Hamiltonian (\ref{HC})
 is proportional to the Laplacian $\Delta_C$ in the space of gauge orbits projected on
the hyperplane defined by the \pcg condition
\be
\label{4-12}
H=-\frac{g^2L}{8}\Delta_C\; ,\quad
\Delta_{{C}} := \frac{1}{\cJ_P } \vec{\nabla}\cdot\, \cJ_P
\vec{\nabla}\; ,\quad  {\vec{\nabla}}=\ve_a\frac{\del}{\del A^a}\; .
\ee
Using spherical coordinates of the gauge ``field'', introduced
already in Eq.\ (\ref{9-28}),
\be
\label{4-10}
\vA = A\, {\bf \hat \vA} (\theta, \phi) \; , \quad A = | \vA | = \sqrt{A^a A^a} \hk ,
\ee
and expressing
the $\vec{\nabla}$
operators in Eq.\ \eqref{4-12} by the standard form
\begin{align}
\label{4-11}
{\vec{\nabla}} & = {\bf \hat \vA}\, \frac{d}{d A} - \frac{i}{A}\,\hat{\bf
  \vA}\times\vL \; ,
\end{align}
where  $\vL$ is the colour angular momentum operator (\ref{26-110}), the Laplacian $\Delta_C$ reads
\begin{align}
\label{4-13}
\Delta_{{C}} &= \lk A^2 \cJ_P \rk^{- 1} \frac{d}{d A} \lk A^2 \cJ_P\rk \frac{d}{d A}
- \frac{\vL^2}{A^2} \nn\\
&= \frac{1}{\cJ_D } \frac{d}{dA} \cJ_D \frac{d}{dA}-
\frac{\vL^2}{A^2}\; .
\end{align}
Here, we have used the explicit forms of the Faddeev--Popov
determinants $\cJ_P$ in the \pcg \eqref{J_Cres} and $\cJ_D$ in the \dcg \eqref{J_Dres}. In the physical space of colour singlet states
where (see Eq.\ (\ref{28-122}))
\be
\label{4-15}
\vL^2 | \Psi \rangle = 0\; ,
\ee
the last term in Eq.\ \eqref{4-13} becomes irrelevant and the
Hamiltonian $H$ in Eq.\ \eqref{4-12} reduces precisely to
the one in the \dcg (\ref{BGneu23}). Its eigenfunctions in
the colour singlet Hilbert space \eqref{4-15} are
therefore the same as in the \dcg. However, in the \pcg the global gauge degrees of freedom
$\theta, \phi$ defining the orientation $\hat{\vA}(\theta, \phi)$ of the colour vector $\vA$ remain as
coordinates, which enter the definition of the scalar product in the Hilbert
space of the wave functionals,
\begin{align}
  \label{scalproC}
  \langle \Psi_1 | \Psi_2 \rangle = \frac{1}{4\pi}\int_{S_2}
  d\Omega\il^\pi_{0} d \vartheta\,\vartheta^2 \cJ_P (\vartheta)
  \Psi^*_1(\vartheta,\theta,\phi)\Psi_2(\vartheta,\theta,\phi)\; .
\end{align}
Here, $d\Omega(\theta, \phi)$ denotes the usual integration measure on $S_2$
and the Jacobian $\vartheta^2$ comes from the transformation to
spherical colour coordinates. 
Due to the factor $\frac{1}{4\pi}$ in the definition of the scalar product
\eqref{scalproC}, the normalisation of the vacuum wave functional in the
\pcg is identical to the one in the \dcg in Eq.\ \eqref{PsiD},
\begin{equation}
  \label{PsiCoul}
  \Psi(A)=\sqrt{\frac{2}{\pi}}\; .
\end{equation}
The \pcg expectation values
\begin{align}
\label{vevdef}
\langle \cO[A] \rangle & =  \int \cD
A\,\cJ_P[A]\Psi^*[A]\cO[A]\Psi[A]\nn\\ &=
\il^\pi_{0} d \vartheta\,\vth^2\,\cJ_P
(\vartheta)\frac{1}{4\pi}\int_{S_2} d\Omega\,\cO[\vth,\theta ,\phi] \,\abs{\Psi}^2\nn\\ &=
\frac{2}{\pi} \il^\pi_0 d \vartheta\, \sin^2 \vartheta \frac{1}{4 \pi}
\il_{S_2}
d \Omega \,\cO [\vartheta, \theta , \phi]  
\end{align}
average the gauge degrees of freedom $\theta , \phi$. The gauge
invariant variable $\vth$ is integrated with the same weight as in the \dcg
expectation value \eqref{vevD}. For operators $\cO(\vth)$ in the
\dcg, the expectation value \eqref{vevdef} gives the same
result as the one in the \dcg \eqref{vevD}. Hence,
we may use the definition \eqref{vevdef} exclusively.

In subsequent sections we will use the exact wave functionals to
 calculate various propagators and vertices, the colour Coulomb
 potential and derive their Dyson--Schwinger equations.
 \section{Exact Propagators and Vertices}
\label{exactGreen}
In $D=3+1$, the study of Landau gauge Dyson--Schwinger
equations (DSEs) has recently become quite popular (for a recent
review, see Ref.\ \cite{Fis06} and references therein). The covariant Landau gauge is technically convenient and the structure
of the DSEs is similar to the \pcg. In the \pcg, the variational approach to $D=3+1$ Yang--Mills theory also
results in a set of DSEs for the 
propagators and vertices \cite{FeuRei04}. In any case, a truncation of
the non-linearly coupled DSEs is unavoidable. It is difficult to assess the validity of this
approximation. In $1+1$ dimension, on the other hand, we can
calculate the {\it exact} Green functions and compare them to the solution
of approximated DSEs. In this section, we will use the exact vacuum state
(\ref{PsiCoul}) to calculate the propagators and vertices in the \pcg. We begin with the gluon propagator. After the computation of the ghost
and Coulomb propagators,  the ghost-gluon vertex is calculated in the \pcg. Finally, we discuss the propagators in the
\dcg.

\subsection{Gluon propagator in pure Coulomb gauge}
In $1+1$ dimensions, the \pcg fields are spatially independent. The gluon propagator is therefore a
constant matrix, defined as the expectation value of two field operators,
\be
\label{Ddef}
D^{a b} = \langle A^a A^b \rangle\; .
\ee
Expressing the colour vector $A^a=\frac{2}{L}\vth\,\hat \vA^a(\theta, \phi)$ in spherical 
coordinates, the \pcg expectation value (\ref{vevdef}) for the
gluon propagator \eqref{Ddef} yields
\be
\label{Dab}
D^{a b} = \frac{2}{\pi} \lk \frac{2}{L} \rk^2 \il^\pi_0 d \vartheta\,
\vartheta^2 \sin^2 \vartheta \,\frac{1}{4 \pi} \int d \Omega\, { \hat \vA}^a (\theta, \phi)
\hat{\vA}^b (\theta, \phi) \; .
\ee
By symmetry, the angular integration yields
\be
\label{AAsymm}
\frac{1}{4 \pi} \int d \Omega\, \hat{\vA}^a (\theta, \phi) \hat{\vA}^b (\theta, \phi) =
\frac{1}{3}\, \delta^{a b}
\ee
and thus for Eq.\ \eqref{Dab}
\be
\label{D_Ares}
D^{a b} =\delta^{a
  b}\frac{1}{L^2}\left(\frac{4}{9}\pi^2-\frac{2}{3}\right)=:D_A\,\delta^{ab}\; ,\quad D_A\approx 3.72\,\frac{1}{L^2}\; .
\ee
The \pcg gluon propagator has only diagonal components which
all have the same value $D_A$.
In the thermodynamic limit, $L\to\infty$, the gluon propagator is
identically zero, in agreement with the fact that the theory becomes
trivial for $L\to\infty$.
\subsection{Ghost propagator in pure Coulomb gauge}
The ghost propagator occurs in the Dyson--Schwinger equations as a consequence of the projection on the hypersurface of Coulomb gauge (or
Landau gauge). In the variational approach \cite{FeuRei04}, this
propagator is merely an auxiliary object to facilitate the
computation of the energy density. It is defined as
the expectation value of the inverse Faddeev--Popov
kernel. The Faddeev--Popov kernel $\cM$ of the \pcg is given in coordinate space
by Eq.\ \eqref{FPcoo}. In the momentum space, we have
\begin{equation}
  \label{FPmom}
  \cM^{ab}_n=\int_0^Ldx\,\cM^{ab}(x,y)\,
  \e^{-ik_nx}=\delta^{ab}k_n^2-i\hat A^{ab}k_n=:\left(G_n^{-1}\right)^{ab}\; .
\end{equation}

We refer to the inverse $\cM_n^{-1}$ as the ``ghost kernel'' denoted
by $G_n$. It is customary to consider the Cartesian colour components of the
matrix-valued ghost propagator. Let us therefore invert $\cM_n$ in the
Cartesian basis. Using the $SU(2)$ identity
\begin{equation}
  \label{dachdach}
  (\hat A\hat A)^{ab}=A^aA^b-\delta^{ab}{\bf A}^2\; ,
\end{equation}
one can verify that
\begin{equation}
  \label{FPinv}
  G_n^{ab}=\frac{1}{k_n^2}\frac{1}{1-\frac{{\bf
        A}^2}{k_n^2}}\left(\delta^{ab}-\frac{A^aA^b}{k_n^2}+i\frac{\hat A^{ab}}{k_n}\right)
\end{equation}
is indeed the inverse of $\cM_n^{ab}$ \eqref{FPmom}. The expectation value \eqref{vevdef} then defines the ghost propagator
\begin{equation}
  \label{d_ndef}
  \lla G_n^{ab}\rra=\frac{d_n^{ab}}{ k_n^2}
\end{equation}
as well as the ghost form factor $d_n^{ab}$, which measures the deviation of
$\langle G_n^{ab}\rangle$ from the tree-level behaviour $\delta^{ab}/{k_n^2}$.
The angular averages within the \pcg expectation value
\eqref{d_ndef} can be taken with Eq.\ \eqref{AAsymm} and the identity
\begin{equation}
  \label{danguint}
\frac{1}{4\pi}\int d\Omega\, \hat A^{ab}=0\; .
\end{equation}

One thus finds a colour diagonal ghost form factor,
\begin{equation}
  \label{dabres}
  d_n^{ab}=\delta^{ab}\lla\frac{1}{1-\frac{{\bf
        A}^2}{k_n^2}}\left(1-\frac{1}{3}\frac{{\bf
        A}^2}{k_n^2}\right)\rra=:\delta^{ab}\, d_n \; .
\end{equation}
Its diagonal elements $d_n$ can be rewritten as $\left(\frac{|{\bf A}|}{k_n}=\frac{\vth}{\pi n}\right)$
\begin{equation}
  \label{d_nres}
  d_n=1+\frac{2}{3}\lla\frac{\vth^2}{(\pi n)^2-\vth^2}\rra =
  1+\frac{4}{3\pi}\int_0^\pi d\vth\,\sin^2\vth \,\frac{\vth^2}{(\pi n)^2-\vth^2}\; .
\end{equation}

The above integral may be expressed by integral sine functions. The
allowed modes $n\in\mathbb{Z}\setminus\{0\}$ exclude the zero mode
$n=0$ of the Faddeev--Popov operator. In the
ultraviolet limit, the ghost form factor \eqref{d_nres} approaches
tree-level,
\begin{equation}
  \label{dnUV}
  \lim_{n\to\infty}d_n =1\; ,
\end{equation}
since the $D=1+1$ theory is super-renormalisable and there are no
anomalous dimensions. In the infrared,
the ghost form factor is enhanced, as can be seen in Fig.\
\ref{ghostfig}. This enhancement is also found in $D=3+1$ dimensions
and is understood to come from near-zero eigenvalues of the Faddeev--Popov
kernel in the vicinity of the Gribov horizon. 

\begin{figure}
  \centering
\includegraphics[scale=1.0]{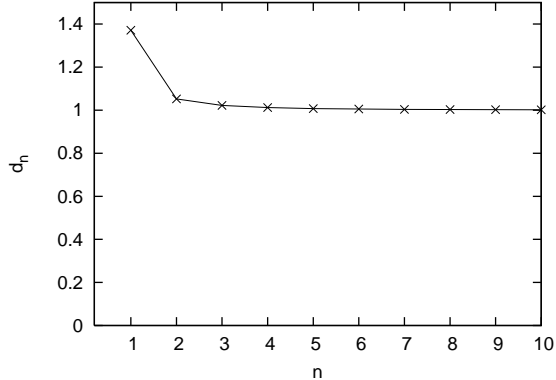} 
\caption{The exact ghost form factor,
  given by Eq.\ (\ref{d_nres}).}
\label{ghostfig}
\end{figure}

In the $3+1$ dimensional continuum theory, the inverse of the ghost
form factor, $d^{-1}$, represents the
generalised dielectric function $\epsilon$ of the Yang--Mills vacuum
\cite{Rei08}. Fig.\ \ref{ghostfig} shows that also in $D = 1 + 1$ the Yang--Mills
vacuum behaves like a dia-electric medium $(\epsilon < 1)$ in the infrared and 
becomes the ordinary vacuum $(\epsilon = 1)$ in the ultraviolet.

\subsection{Coulomb propagator in the pure Coulomb gauge}
The \pcg Hamiltonian \eqref{HC} comprises the so-called Coulomb
term $H_C$, see Eq.\ \eqref{H_C}, which accounts for the interaction energy between colour
charges. By the calculation of the expectation value $\lla H_C\rra$,
the quark potential can be found, see section \ref{potsec}. The ``Coulomb kernel'' $F$, which mediates
this interaction is defined in Eq.\ \eqref{Fdef} and reads in momentum space
\begin{equation}
  \label{Fndef}
  F_n^{ab}=G_n^{ac}k_n^2G_n^{cb}\; .
\end{equation}

Using the explicit form \eqref{FPinv} of the ghost kernel $G_n^{ab}$
and the identity \eqref{dachdach},
the Coulomb kernel $F^{ab}_n$ is cast into the form
\begin{equation}
  \label{Fnres}
  F_n^{ab}=\frac{1}{k_n^2}\frac{1}{\left(1-\frac{{\bf A}^2}{k_n^2}\right)^2}\left[\delta^{ab}\left(1+\frac{{\bf A}^2}{k_n^2}\right)+\frac{A^aA^b}{k_n^2}\left(\frac{{\bf A}^2}{k_n^2}-3\right)+2i\frac{\hat A^{ab}}{k_n}\right]\; .
\end{equation}

The expectation value \eqref{vevdef} of the operator $F_n^{ab}$ in
Eq.\ \eqref{Fnres} defines the Coulomb propagator
\begin{equation}
  \label{phindef}
  \lla F_n^{ab}\rra=\frac{\phi_n^{ab}}{k_n^2}
\end{equation}
and the form factor $\phi_n^{ab}$ which measures the deviation of
$\langle F_n^{ab}\rangle$ from tree-level, $\delta^{ab}/k_n^2$ (being the abelian case).
To evaluate the expectation value \eqref{phindef}, let us first
integrate the gauge degrees of freedom of $F_n^{ab}(A)$. Using the
identities \eqref{AAsymm} and \eqref{danguint}, one finds from
$F_n^{ab}$ in Eq.\ \eqref{Fnres}
\begin{align}
  \label{Fnaver}
  \frac{1}{4\pi}\int d\Omega\,F_n^{ab}&=\frac{\delta^{ab}}{k_n^2}{\Big(1-\big(\frac{\vth}{\pi n}\big)^2\Big)^{-2}}\left[1+\big(\frac{\vth}{\pi n}\big)^2+\frac{1}{3}\big(\frac{\vth}{\pi n}\big)^2\Big(\big(\frac{\vth}{\pi n}\big)^2-3\Big)\right]\nn\\
  &=\frac{\delta^{ab}}{k_n^2}\,\frac{1}{3}\left(1+2(\pi n)^2\frac{(\pi
    n)^2+\vth^2}{\left((\pi
      n)^2-\vth^2\right)^2}\right)\; .
\end{align}

\begin{figure}
  \centering
  \includegraphics[scale=1.0]{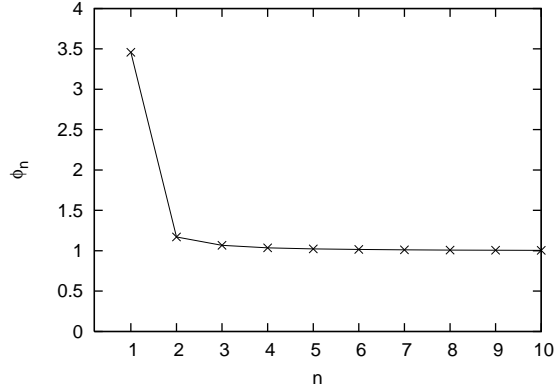}
\caption{Form factor $\phi_n$ of the
  Coulomb potential as given by Eq.\ \eqref{phinres}.}
\label{phifig}
\end{figure}

The form factor $\phi_n^{ab}$ \eqref{phindef} of the Coulomb
propagator is therefore strictly
diagonal and its diagonal elements yield
\begin{equation}
  \label{phinres}
   \phi_n^{ab}=:\delta^{ab}\phi_n\; , \quad \phi_n=\frac{1}{3}\left(1+{4}\pi n^2\int_0^\pi d\vth\,\sin^2\vth\,\frac{(\pi
    n)^2+\vth^2}{\left((\pi n)^2-\vth^2\right)^2}\right)\; .
\end{equation}
In the ultraviolet limit, the form factor $\phi_n$ approaches
tree-level. As shown in Fig.\ \ref{phifig}, $\phi_n$ is infrared
enhanced. In the $D=3+1$ theory, the infrared enhancement of the
Coulomb propagator is expected to come from the restriction of the configuration
space to the Gribov region, as claimed by Gribov in his seminal paper
\cite{Gri78}, and to lead to a confining quark potential. This will be
discussed further in section \ref{potsec}.

An issue in $D=3+1$ dimensions is whether the expectation value of
\eqref{Fndef} can be factorised, i.e.\ can the connected part be
neglected? In order to answer this question, a further form factor was
introduced in Ref.\ \cite{FeuRei04}. It measures the deviation from
the factorisation,
\begin{equation}
  \label{fdef}
  \lla G_n\, k_n^2 G_n \rra = : \lla G_n\rra\, k_n^2\, f_n\,\lla
  G_n\rra\; ,\quad \Rightarrow\; f_n=\frac{\phi_n}{d_n^2}\; .
\end{equation}
and can be expressed by the ratio of the form factor $\phi_n$ of the
Coulomb propagator and the ghost form factor squared. Following Ref.\
\cite{FeuRei04}, we refer to
$f_n$ as the ``Coulomb form factor''. In $1+1$ dimensions, where the
exact solutions for $d_n$ and $\phi_n$ are available, we can calculate
$f_n$ exactly. In Fig.\ \ref{ffig}, the result for $f_n$ is depicted. It
shows an infrared enhancement. A further discussion in the context of
the $D=3+1$ theory will follow in section \ref{variationalsec}.

\begin{figure}
  \centering
  \includegraphics{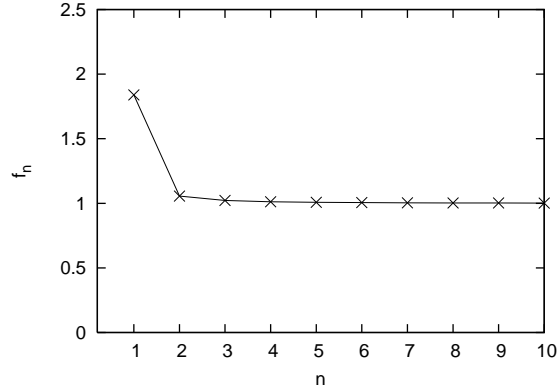}
  \caption{Coulomb form factor $f_n$, as defined in Eq.\ \eqref{fdef}.}
  \label{ffig}
\end{figure}

\subsection{Ghost-gluon vertex in pure Coulomb gauge}
In the variational approach in the \pcg \cite{FeuRei04} and the
Dyson--Schwinger approach in Landau gauge, the ghost-gluon vertex is of
particular interest. In these approaches the {proper} ghost-gluon vertex is usually replaced by the bare one with the
argument that this vertex is not renormalised \cite{Tay71}. In fact, recent lattice calculations performed in $D = 1 +
1$ Landau gauge provide little evidence for a dressing of this vertex
\cite{Maa07}.  However, although larger lattices are available in $D = 1 + 1$, there are significant statistical
errors. The lattice results are also plagued by the
existence of Gribov copies. The $1 + 1$ dimensional continuum theory,
on the other hand, has full control of the Gribov problem, see section \ref{Gribovregions}. In the following we will {calculate} the
proper (one-particle irreducible) ghost-gluon vertex {in the \pcg}.

The bare ghost-gluon vertex is defined by \cite{FeuRei04}
\be
\label{5a-2}
\Gamma^{0,a}(x, x') = - \frac{d {G}^{- 1} (x, x')}{{d A^a}} =  \hat{T}^a
\partial_x \delta (x, x') \hk ,
\ee
where ${G}^{-1}\equiv\cM$ is the Faddeev--Popov kernel (\ref{FPcoo}). Fourier expansion 
\be
\label{5a-3}\Gamma^{0,a}(x, x')  =  \frac{1}{L} \sli_{n} \e^{i k_n (x -
x')} \Gamma^{0,a}_n 
\ee
yields the momentum space representation
\be
\label{57-278}
\Gamma^{0,a}_n  =  i k_n \hat{T}^a \hk .
\ee
In the $D=1+1$ Coulomb gauge the ghost-gluon vertex depends only on
a single momentum for there is only the zero momentum mode of the gauge field.
The {proper} ghost-gluon vertex $\Gamma_n^a$ is defined via the
expectation value for the connected ghost-gluon vertex\footnote{For
  the wave functional chosen in Ref.\ \cite{FeuRei04}, the definition
  \eqref{ggzconn} coincides with the one used there, $\lla
  G\Gamma^{0}G\rra=:\lla G\rra\Gamma\lla G\rra$.}
\begin{equation}
  \label{ggzconn}
  \lla A^a G^{bc}_n\rra =
  D^{aa'}\,\lla G_n^{bb'}\rra\,(\Gamma_n^{a'})^{b'c'}\,\lla G_n^{c'c}\rra = D_A\,\frac{d_n}{k_n^2}\,(\Gamma_n^{a})^{bc}\,\frac{d_n}{k_n^2}
\end{equation}
which is the proper vertex with propagators attached on its legs. We
have used the fact that the exact ghost and gluon propagators in
the \pcg are colour diagonal.
Using the explicit form \eqref{FPinv} of the ghost kernel $G_n^{ab}$,
we can write the expectation value \eqref{ggzconn} as
\begin{equation}
  \label{ggzop}
  \lla A^aG^{bc}_n\rra =\frac{1}{k_n^2}\lla \frac{1}{1-\frac{{\bf
        A}^2}{k_n^2}}\left(A^a\delta^{bc}-A^a\frac{A^bA^c}{k_n^2}+iA^a\frac{\hat A^{bc}}{k_n}\right)\rra\; .
\end{equation}
In the \pcg expectation value \eqref{vevdef}, the angular integration
renders the first two terms in Eq.\ \eqref{ggzop} zero, while the last
one yields
\begin{align}
  \label{ggz1}
   \lla A^aG^{bc}_n\rra &= \frac{1}{k_n^2}\, ik_n(\hat T^d)^{bc}\lla
   \frac{A^aA^d}{k_n^2-{\bf A}^2}\rra\nn\\
   &=
   \frac{1}{k_n^2}\,(\Gamma_n^{0,d})^{bc}\,\frac{1}{3}\,\delta^{ad}\lla\frac{\vth^2}{(\pi n)^2-\vth^2}\rra\nn\\
   &=
   \frac{1}{k_n^2}\,(\Gamma_n^{0,a})^{bc}\,\frac{1}{2}\left(d_n-1\right)
\end{align}
where in the last line we have used the expression \eqref{d_nres} for
the ghost form factor $d_n$. It is helpful to define the form factor
$\gamma_n$ of the ghost-gluon vertex by
\begin{equation}
  \label{gamman}
  \Gamma_n^a=:\gamma_n\,\Gamma_n^{0,a}\; .
\end{equation}
This form factor can now be expressed by Eq.\ \eqref{ggz1} using the definition
\eqref{ggzconn} of $\Gamma_n^a$,
\begin{equation}
  \label{gammanres}
  \gamma_n=\frac{k_n^2(d_n-1)}{2\,D_A\,d_n^2}\; .
\end{equation}

\begin{figure}
  \centering
  \includegraphics[scale=1.0]{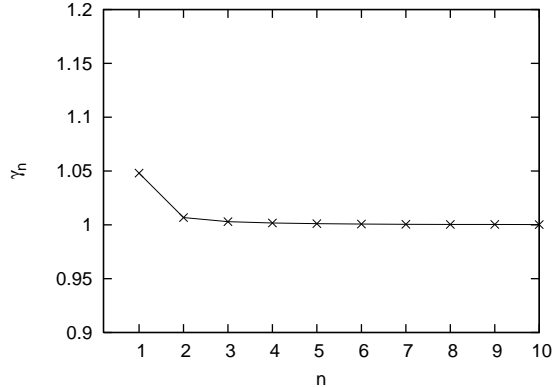}
  \caption{Form factor $\gamma_n$ of the proper ghost-gluon vertex, as
    given by Eq.\ \eqref{gammanres}.}
  \label{ggzfig}
\end{figure}

Let us check the ultraviolet limit, $n\to\infty$. From Eqs.\
\eqref{d_nres}, \eqref{Dab} and \eqref{AAsymm}, one can see that
\begin{equation}
  \label{auxlim}
  \lim_{n\to\infty}k_n^2(d_n-1)=\frac{16}{3\pi
  L^2}\int_0^{\pi}d\vth\,\vth^2\sin^2\vth\equiv 2D_A
\end{equation}
where $D_A$ is the gluon propagator defined in Eq.\ \eqref{D_Ares}. It then follows immediately from
Eqs.\ \eqref{gammanres} and \eqref{dnUV} that the ghost-gluon vertex approaches
tree-level asymptotically,
\begin{equation}
  \label{gammalim}
  \lim_{n\to\infty}\gamma_n=1\; .
\end{equation}
In Fig.\ \ref{ggzfig}, one can see that the form factor $\gamma_n$,
given by Eq.\ \eqref{gammanres}, hardly deviates from tree-level for
the entire momentum range. Deviations are in the range of $5\%$. This strongly supports the popular
approximation of truncating Dyson--Schwinger equations by choosing a
tree-level ghost-gluon vertex. Further discussion will follow in
section \ref{truncsec}.

\subsection{Propagators in diagonal Coulomb gauge}
\label{propasec}
For a comparison to the \pcg, which does not completely eliminate the gauge
degrees of freedom, we here calculate the propagators in the \dcg with
restriction to the fundamental modular region. The gluon propagator $D^{ab}$ in the \dcg is
most easily found by letting $A^a\to\frac{2}{L}\vth\,\delta^{a3}$ before taking
the expectation value \eqref{Ddef}. We can then express $D^{aa}$ in the \dcg by
the \pcg result \eqref{D_Ares} for $D_A$,
\be
D^{33} =  \frac{2}{\pi} \lk \frac{2}{L} \rk^2 \il^\pi_0 d \vartheta\,
\vartheta^2 \sin^2 \vartheta= 3D_A\; ,\quad D^{11}=D^{22}=0
\ee
The colour trace of the gluon propagator is evidently the same in both
gauges. From the observation that the \dcg can be reached
from the \pcg by the unitary transformation (see appendix \ref{rots})
\begin{equation}
  \label{globrot}
  A\to U\, A\, U^\dagger
\end{equation}
of the matrix-valued variable $A=A^aT^a$, the invariance of the colour
trace of the gluon propagator is seen to be an immediate consequence.

The Faddeev--Popov kernel $\cM_n$ in the \dcg is substantially different from
the one in the \pcg, see Eq. \eqref{BGneu3} (cf.\ Eq.\ \eqref{FPcoo}). It is
convenient to expand $\cM_n$ in the spherical basis \eqref{5-22} where
$\cM_n$ is diagonal,
\begin{equation}
  \label{Mnsph}
  \bra{\sigma '}\cM_n\ket{\sigma}=\Lambda_{n,\sigma}\delta_{\sigma
    '\sigma}=:\big(G_n^{-1}\big)^{\sigma}\delta_{\sigma '\sigma}
\end{equation}
and the eigenvalues $\Lambda_{n,\sigma}$ are given by Eq.\ \eqref{15-66}. We
use the vacuum expectation value of the ghost kernel $G_n^\sigma$ in
the \dcg \eqref{Mnsph}
to define the ghost form factor components $d_n^\sigma$,\footnote{For
  the components $\sigma=\pm 1$, we related the form factor to the
  propagator differently from the $\sigma=0$ component, in order for all $d_n^\sigma$ to have the same dimension and complex phase. This circumstance
which does not occur in the \pcg can here be traced back to the
structure of the gauge
condition \eqref{BG14} of the \dcg.}
\begin{equation}
  \label{dnsidef}
  \lla G_n^{\sigma}\rra =: \left\{
      \begin{array}{cl}
        \frac{id_n^\sigma}{k_n}\; ,&\qquad \sigma=\pm 1\\
        \frac{d_n^\sigma}{k_n^2}\; ,&\qquad \sigma=0
      \end{array}
      \right.\; .
\end{equation}
The vacuum expectation values \eqref{vevdef} then yield for $n\neq 0$
\begin{equation}
  \label{dnsires}
  d_n^{\sigma=\pm 1}=\lla\frac{1}{1+\sigma\frac{\vth}{\pi n}}\rra\;
  ,\quad d_n^{\sigma=0}=1\; .
\end{equation}
Taking the colour trace $\sum_\sigma d_n^\sigma$ of the ghost form
factor in the \dcg gives the same result as summing the diagonal
elements $d_n^{aa}$ of the ghost form factor in the \pcg (see Eq.\ \eqref{d_nres}),
\begin{equation}
  \label{dntr}
  \sum_{\sigma}d_n^\sigma = 1+2\lla\frac{1}{1-(\frac{\vth}{\pi
      n})^2}\rra = \sum_a d_n^{aa}\; .
\end{equation}
While the abelian component $\sigma=0$ of the \dcg ghost form factor
is at tree-level, see Eq.\ \eqref{dnsires}, the other diagonal
components are larger than the \pcg result \eqref{d_nres} for $d_n$, such that the colour trace
is invariant. 

The same scenario occurs for the Coulomb propagator $\langle
F_n\rangle$. With the the definition \eqref{29-125} of $F_n$ and
$\lambda_{n,\sigma}$ given by Eq.\ \eqref{BGneu10}, we have
with $F_n\ket{\sigma}=F_n^\sigma\ket{\sigma}$ 
\begin{equation}
  \label{phisidef}
  \lla F_n^\sigma\rra=\lla\frac{1}{\lambda_{n,\sigma}^2}\rra=\lla\frac{1}{(k_n+\sigma A^3)^2}\rra = \frac{\phi_n^\sigma}{k_n^2}
\end{equation}
and find for the form factor $\phi_n^\sigma$ of the Coulomb propagator
for $n\neq 0$
\begin{equation}
  \label{phisires}
  \phi_n^{\sigma=\pm 1}=\lla\frac{1}{(1+\sigma\frac{\vth}{\pi n})^2}\rra\;
  ,\quad \phi_n^{\sigma=0}=1\; .
\end{equation}
Noting that the Coulomb kernel $F_n$ in the \dcg actually follows from
a rotation in colour space from the one in the \pcg, see Eq.\
\eqref{30-130}, the invariance of the colour trace is clear. 

The above results exclude the $n=0$ modes which are the zero modes of the Faddeev--Popov operator in the
\pcg. In the \dcg, however, the constant $n=0$ modes with $\sigma=\pm
1$ are allowed. These give the results
\begin{align}
  \label{Gzero}
  \lla G_{n=0}^{\sigma=\pm 1}\rra &=
  i\sigma\frac{L}{2}\lla\frac{1}{\vth}\rra\approx 0.39\, i\sigma L\\
  \label{zerom}
  \lla F_{n=0}^{\sigma=\pm 1}\rra &=\frac{L^2}{4}\lla\frac{1}{\vth^2}\rra\approx 0.23\, L^2
\end{align}
It turns out that these values are of no importance for the
considerations below and are just given here for completeness.

\section{The static quark potential}
\label{potsec}
The gauge invariant potential energy of a quark-antiquark pair in
$D=1+1$ is well-known from the calculation of the temporal Wilson loop. A
linearly rising potential emerges and the corresponding string tension $\sigma_j$ shows
strict Casimir scaling \cite{BlaTho93},
\be
\sigma_j = \frac{g^2}{2} j ( j + 1) \; .
\ee
In the fundamental representation we have with ${j = \frac{1}{2}}$ the
string tension 
\begin{equation}
\label{sigma_j}
\sigma:=\sigma_{\frac{1}{2}}= \frac{3}{8}\, g^2 \; .  
\end{equation}
In the Hamiltonian approach,
the static colour potential can be obtained by taking the expectation value
of the Hamiltonian if the charge
distribution $\rho^a(x')$ is chosen to be a pair of
opposite point charges of strength $q^a$ and $- q^a$, localised at the
positions $x$ and $y$, respectively,
\begin{equation}
\label{rhoext}
\rho^a (x') = q^a \left(  \delta (x' - x) - \delta (x' - y)\right)\; .
\end{equation}
This calculation is shown here to also give rise to a linear potential
which is, however, a gauge dependent quantity. Therefore, the string
tension $\sigma_C$ defined by this linear potential gives only an upper bound \cite{Zwa97,Zwa03b} on the gauge
invariant string tension $\sigma$ in Eq.\
\eqref{sigma_j}. Nevertheless, the Coulomb potential is calculated
here for comparison to the potential calculated in the $D=3+1$ theory.

\subsection{Pure Coulomb gauge}
We first fix only the \pcg. For the colour charge density \eqref{rhoext}, the Coulomb Hamiltonian (\ref{H_C}) reduces to
\be
\label{37-163}
H_C = \frac{g^2}{2} q^a q^b \left[ F^{a b} (x, x) + F^{a b} (y, y) - F^{a b}
(x, y) - F^{a b} (y, x) \right] \hk ,
\ee
with the Coulomb kernel $F^{ab}(x,y)$ \eqref{Fdef} in the \pcg. The first two terms of Eq.\ \eqref{37-163} represent the self-energy of the static point charges.
Only the charges belonging to the generators of the Cartan subalgebra can be
specified and the expectation value of (\ref{37-163}) in the Yang--Mills vacuum
state for  abelian (Cartan) unit charge{s} 
defines the static quark potential. With $q^a = \delta^{a 3}$ for $SU (2)$, the
static quark potential $V_C(r)$ with $r=|x-y|$ becomes
\begin{align}
\label{V_Cdef}
V_C (r) &=  \left. \langle H_C \rangle \right|_{q^a = \delta^{a 3}} = 
 g^2 \,\langle F^{33}(x,x) - F^{33}(x,y) \rangle
\end{align}
To obtain the above expectation value, we first take the angular
average of the momentum space Coulomb kernel $F_n^{ab}$, see Eq.\ \eqref{Fnaver},
\begin{align}
  \frac{1}{4 \pi}
\il_{S_2}
d \Omega \, &\left(F^{33}(x,x)-F^{33}(x,y)\right)\nn\\ 
  &=  \frac{1}{L}\sum_{n\neq 0}\, \frac{1}{4 \pi}
\il_{S_2}
d \Omega \,F_n^{33}\,\left(1-\e^{-ik_nr}\right)\nn\\
 &=  \frac{2}{L}\sum_{n=1}^\infty\,\frac{1}{3}\left(\frac{L}{2\pi n}\right)^2\left(1+2(\pi n)^2\frac{(\pi
    n)^2+\vth^2}{\left((\pi
      n)^2-\vth^2\right)^2}\right)\left(1-\cos\left(\textstyle\frac{2\pi r}{L}\, n\right)\right)\; .
\end{align}
With $1-\cos(2\alpha)=2\sin^2(\alpha)$, the Fourier transformation yields
\begin{align}
  \label{Fdiff}
  \frac{1}{4 \pi}
\il_{S_2}
d \Omega \, &\left(F^{33}(x,x)-F^{33}(x,y)\right)\nn\\ 
  &=
  \frac{1}{3}\,\frac{L}{\pi^2}\,\sum_{n=1}^\infty\,\left(\frac{1}{n^2}+2\,\frac{\del}{\del\vth}\,\frac{\vth}{n^2-\left({\vth}/\pi\right)^2}\right)\,\sin^2\left(\textstyle\frac{\pi r}{L}\, n\right)\nn\\
  &=\frac{1}{3}\,\left(\frac{1}{2}r-\frac{1}{2}\frac{r^2}{L}+\frac{1}{2\sin^2\vth}\left(L-r \cos \left(2\, \textstyle\frac{L-r}{L} \,\vth\right)-(L-r) \cos \left(2\,\textstyle\frac{r}{L}\, \vth \right)\right)\right)
\end{align}
Here, we have used the formulae in Ref.\ \cite{Gradshteyn} to obtain
for the above sums
\begin{equation}
  \label{sum}
  \sum_{n=1}^\infty\,\frac{\sin^2\left(\frac{\pi r}{L}\, n\right)}{n^2-(\vth/\pi)^2} =
  \frac{\pi^2}{2}\,\frac{\sin(\frac{L-r}{L}\vth)\,\sin(\frac{r}{L}\vth)}{\vth\,\sin\vth}\;\stackrel{\vth\to 0}{\longrightarrow}\;   \frac{\pi^2}{2}\,\left(\frac{r}{L}-\big(\frac{r}{L}\big)^2\right)\; .
\end{equation}
The (elementary) $\vth$-integration of the expression \eqref{Fdiff} as defined in the
expectation value \eqref{vevdef} yields the static quark potential
\eqref{V_Cdef} 
{\be
\label{V_Cpure}
 V_C(r) = \frac{g^2}{3}\,\left( L + \frac{1}{2} r - \frac{r^2}{2 L} + L\, \frac{2 r - L}{L - r}\,
\frac{\sin (2 \pi \frac{r}{L})}{2 \pi \frac{r}{L}}
\right)=\frac{g^2}{2}\,r\,\left(1 + \cO\big(\frac{r}{L}\big)\right)\; .
\ee}
In the thermodynamic limit, the string tension $\sigma_C$ is defined by
\begin{equation}
  \label{sigmaC}
  \lim_{L\to\infty}V_C(r)=\sigma_C\, r\; ,\quad\Rightarrow\,\sigma_C=\frac{g^2}{2}>\sigma
\end{equation}
and is found to be larger than the gauge invariant string tension
$\sigma$ \eqref{sigma_j}, as expected.

\subsection{Diagonal Coulomb gauge}
Let us now apply the global gauge transformation from the pure to the
diagonal Coulomb gauge and recalculate the string tension of the
potential. The $33$-component of the Coulomb kernel $F_n^{ab}$ can be
found by setting $A^a=\frac{2}{L}\vth\,\delta^{a3}$ in Eq.\
\eqref{Fnres},
\begin{equation}
  \label{F33dia}
  F_n^{33}=\frac{1}{k_n^2}\; .
\end{equation}

It apparently mediates only the tree-level (abelian) part of the
Coulomb interaction. Since $F_n^{33}$ is thus field independent, the vacuum
expectation value needs not to be taken and we directly arrive at
the Coulomb potential \eqref{V_Cdef} by Fourier
transformation,\footnote{Strictly speaking, the sum of momentum modes $n$ must
  be changed to ${\sum_n}'$ which excludes the zero mode
  $n=\sigma=0$ of the Faddeev--Popov operator in the \dcg, but includes the modes
  $n=0$ and $\sigma=\pm 1$, see section \ref{FPsec}. The result for
  these zero modes is given in Eq. \eqref{zerom}. However, in the subtraction of
  eigenenergies in the potential $V_C(r)$, see Eq.\ \eqref{V_Cdef}, all modes with $n=0$ cancel
and it makes no difference whether $\sum_{n\neq 0}$ or ${\sum_n}'$ is used.}
 \begin{align}
   \label{V_Cdia}
   V_C(r) 
     &=   g^2\,\frac{1}{L}\,     \sum_{n\neq 0}\,F_n^{33}\,\left(1-\e^{-ik_nr}\right)=   g^2\,\frac{L}{\pi^2}\, \sum_{n=1}^\infty\,\frac{\sin^2\left(\frac{\pi r}{L}\, n\right)}{n^2}\nn\\
     &=   \frac{g^2}{2}\,\Big(r-\frac{r^2}{L} \Big) = \frac{g^2}{2}\,r\,\left(1 + \cO\big(\frac{r}{L}\big)\right)\; .
 \end{align}

\begin{figure}
  \centering
  \includegraphics[scale=0.8]{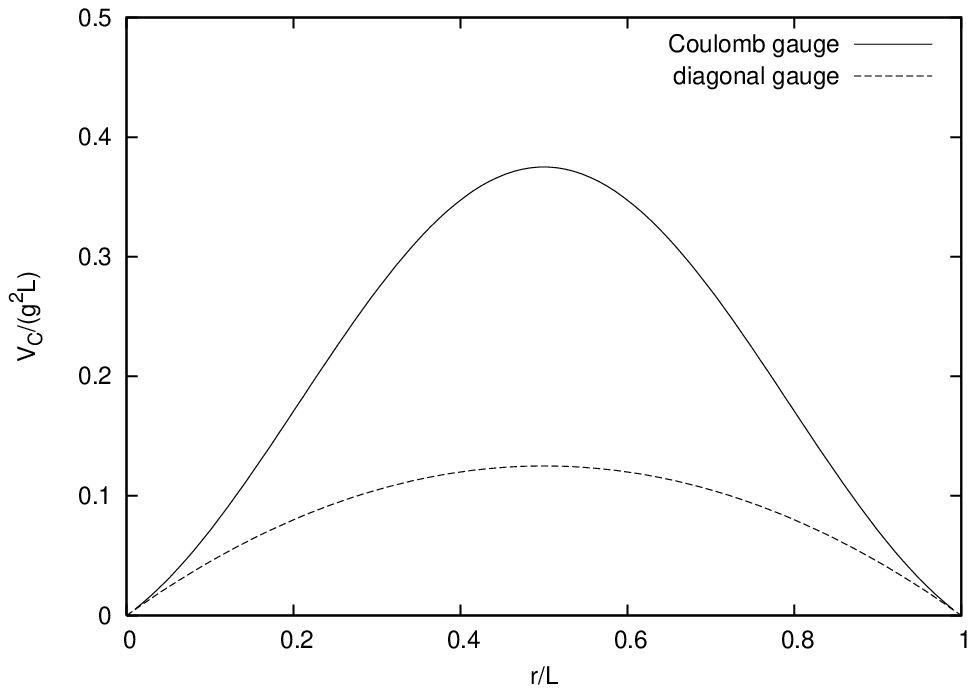}
  \caption{Coulomb potential $V_C(r)$ for the {\it pure} and the \dcg.}
  \label{V_Cfig}
\end{figure}

The string tension of the \dcg is obviously identical to the
one in the \pcg, see Eq.\ \eqref{sigmaC}. It can
actually be seen by taking the limit $L\to\infty$ in the \pcg
expression \eqref{Fdiff} that there Coulomb kernel also turns out
field independent (i.e.\ $\vth$ independent) and coincides with the
Fourier transform of the \dcg kernel \eqref{F33dia}.

The potential $V_C(r)$ is shown in Fig.\ \ref{V_Cfig} for the result
in the \pcg \eqref{V_Cpure} and in the \dcg
\eqref{V_Cdia}. It is clear that in either gauge the function $V_C(r)$ must be
symmetric about the axis $r=\frac{L}{2}$, since a separation of
charges by $r$ is identified with a separation by $L-r$ on the spatial
manifold $S^1$. For $\frac{r}{L}\ll 1$, the potentials $V_C(r)$ are seen
in Fig.\ \ref{V_Cfig} to have the same slope in both gauges, i.e.\ the
string tensions have the same value $\sigma_C$ given by Eq.\
\eqref{sigmaC}.

\section{Dyson--Schwinger equations}
\label{DSEsec}
In this paper, it is intended to test the approximations made in the
study of Dyson--Schwinger equations (DSEs) in higher dimensions by
considering the $D=1+1$ case. The exact DSEs for the ghost and gluon
propagators are usually derived from the partition function of
Yang--Mills theory. In the
present case, we can simply use the definition \eqref{vevdef} of vacuum
expectation values to come by this set of equations. We restrict
ourselves to the exact ground state of the \pcg, leaving aside the global rotation
to the \dcg. It will be shown how Gribov copies affect
the Dyson--Schwinger equations and their solution.


Let us start with the derivation of the DSE for the gluon propagator. It follows
directly from the expectation value \eqref{vevdef} that ($\Psi(A)=const$)
\begin{equation}
  \label{DSE}
  0=\int_{\Omega_1}\cD A\,\frac{\delta}{\delta
    A^a}\cJ_P(A)\,\abs{\Psi(A)}^2\,\e^{j^cA^c}=\lla\left(\frac{\delta\ln\cJ_P}{\delta A^a}+j^a\right)\e^{j^cA^c}\rra
\end{equation}
holds, 
since the integral of the total derivative in Eq.\ \eqref{DSE} is
proportional to the Faddeev--Popov determinant $\cJ_P$ evaluated at
the first Gribov horizon $\del\Omega_1$, where it vanishes. 
Applying a derivative $\delta/\delta j^b$ and setting the sources
$j^c$ to zero gives
\begin{equation}
  \label{wDSE1}
  0=\lla A^b\,\frac{\delta\ln\cJ_P}{\delta A^a}+\delta^{ab}\rra =
  -\lla A^b\,\Tr\,
   G\, \Gamma^{0,a}\rra + \delta^{ab}\; .
\end{equation}
The trace ``$\Tr$'' in Eq.\ \eqref{wDSE1} sums up the diagonal
elements in colour space as well as all modes $k_n$ with $n=\pm 1,\pm
2\,\dots$, excluding the zero mode $n=0$ of the Faddeev--Popov
determinant 
\begin{equation}
\cJ_P=\exp\Tr\ln G^{-1}  
\end{equation}
 in the \pcg. 
We recognise in Eq.\ \eqref{wDSE1} the connected ghost-gluon vertex
$\langle A^bG\rangle$. With its decomposition \eqref{ggzconn} into the proper vertex
$\Gamma^a_n$ and the attached propagators in momentum space, Eq.\ \eqref{wDSE1} can be written after
contracting with $\delta^{ab}$ as
\begin{equation}
  \label{wDSE}
  D_A^{-1}=\sum_{n\neq 0}\left(\frac{\tr\,\Gamma_n^{0,a}\Gamma_n^a}{(N_c^2-1)k_n^2}\right)\frac{d^2_n}{k_n^2}
\end{equation}

This Dyson--Schwinger equation holds for any $SU(N_c)$ but we
eventually set $N_c=2$ for comparison with the exact $SU(2)$ results in
chapter \ref{exactGreen}. Using the definition \eqref{gamman} of the form factor
$\gamma_n$ for the ghost-gluon vertex and $\tr\, \hat T^a\hat T^a=-N_c(N_c^2-1)$, the DSE for the gluon propagator \eqref{wDSE} can be written more
concisely,
\begin{equation}
  \label{wDSEshort}
  D_A^{-1}=N_c\sum_{n\neq 0}\gamma_n\,\frac{d_n^2}{k_n^2}\; .
\end{equation}

\begin{figure}
  \centering
  \includegraphics[scale=1.0, bb= 140 650 430 700, clip=]{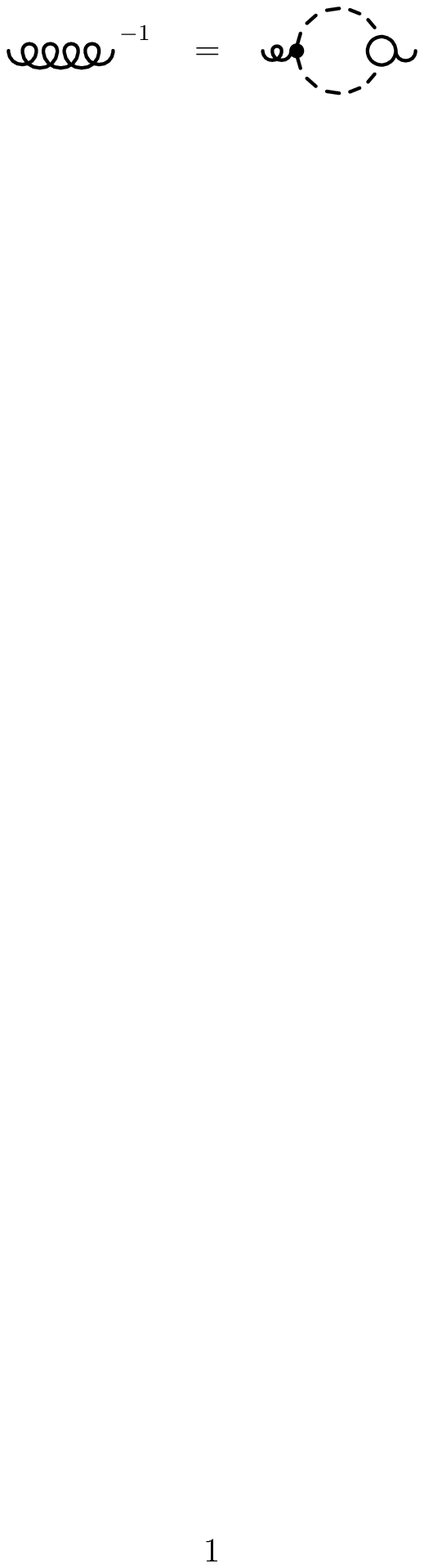}
  \caption{DSE for the gluon propagator. Curly lines represent the
    gluon propagator, dashed lines the ghost propagator. The empty
    blob represents a proper ghost-gluon vertex, the dot a tree-level
    ghost-gluon vertex.}
  \label{wDSEfig}
\end{figure}

Diagrammatically, the inverse gluon propagator is given by a ghost
loop as shown in Fig.\ \ref{wDSEfig}.\footnote{Since in $D=1+1$
  \pcg the
  gauge field is constant, the
  convolution integral of the loop breaks down into a simple product
  in momentum space, see Eq.\ \eqref{wDSEshort}.} Inserting into the right-hand side
of Eq.\ \eqref{wDSEshort} the exact expressions for the ghost form
factor $d_n$ \eqref{d_nres} and the ghost-gluon vertex $\gamma_n$ \eqref{gammanres}, one
can explicitly show that the ghost loop (r.h.s.\ of Eq.\ \eqref{wDSEshort}) equals the expression for
$D_A^{-1}$ obtained in Eq.\ \eqref{D_Ares}. In Ref.\
\cite{FeuRei04}, the ghost loop is referred to as the ``curvature''
since it incorporates the curvature of the space of gauge-fixed variables. It
is found that the curvature governs the infrared behaviour of the
gluon propagator such that the exact DSE of $1+1$ dimensions in Fig.\
\ref{wDSEfig} is the infrared limit of the corresponding DSE in $3+1$
dimensions. 


\begin{figure}
  \centering
  \includegraphics[scale=1.0, bb= 170 660 430 700, clip=]{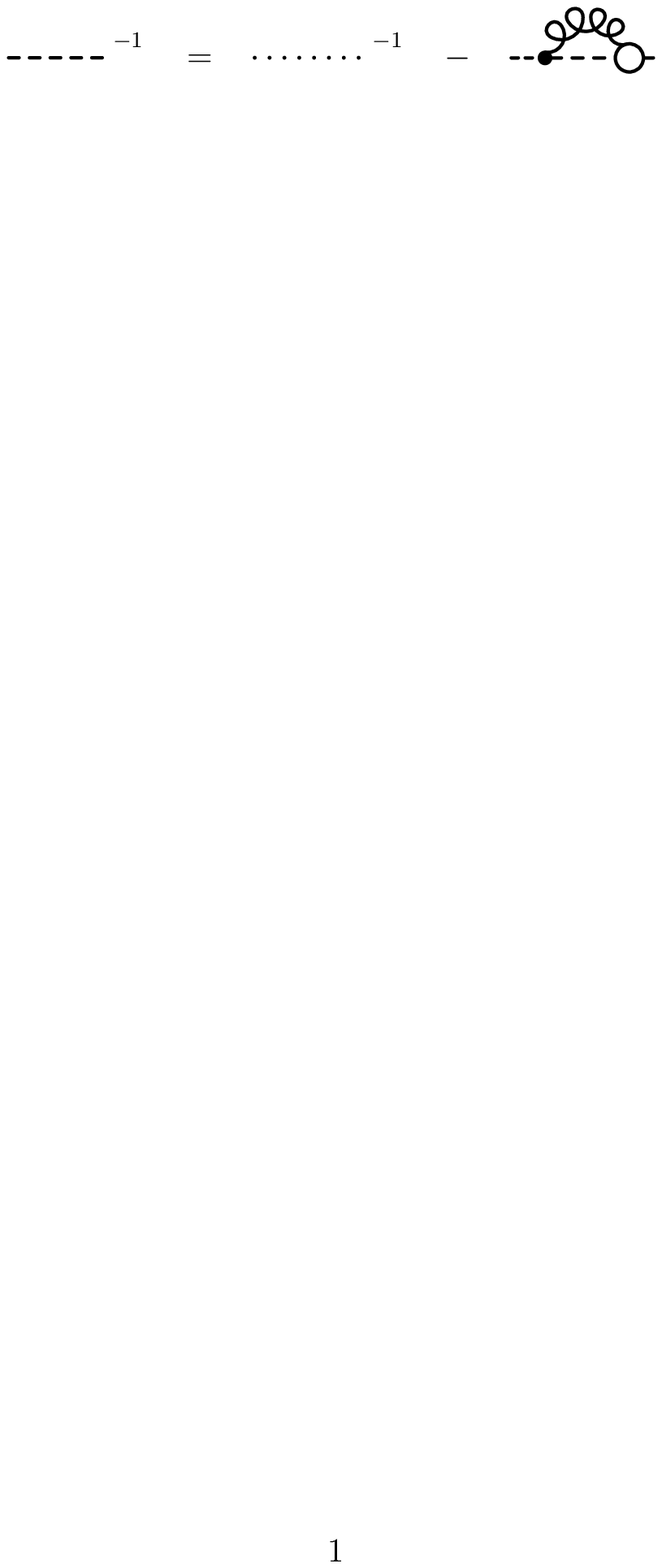}
  \caption{DSE for the ghost propagator. The dotted line is the
    tree-level propagator.}
  \label{dDSEfig}
\end{figure}

The DSE for the ghost propagator can be derived from the following operator
identity\footnote{Alternatively, one can introduce ghost fields and
  proceed similarly as for the gluon propagator.} for the ghost kernel $G$:
\begin{equation}
  \label{gDSEop}
  G_n^{ab}=k_n^{-2}\delta^{ab}+k_n^{-2}(\Gamma_n^{0,d})^{ac}A^dG_n^{cb}\; 
\end{equation}
which follows from definition \eqref{FPmom} and Eq.\ \eqref{57-278}. 
Taking the expectation value of Eq.\ \eqref{gDSEop}, we find with the decomposition \eqref{ggzconn} of the
connected ghost-gluon vertex
\begin{equation}
  \label{gDSE3}
  d_n\delta^{ab}=\delta^{ab}+(\Gamma_n^{0,d})^{ac}D_A\frac{d_n}{k_n^2}(\Gamma_n^d)^{cb}\frac{d_n}{k_n^2}\; .
\end{equation}
After contraction with $\delta^{ab}$ and using the definition
\eqref{gamman} for the form factor $\gamma_n$ of the ghost-gluon vertex, Eq.\ \eqref{gDSE3} turns into
\begin{equation}
  \label{gDSE}
  d_n=1+\left(\frac{\tr\,\Gamma_n^{0,a}\Gamma_n^a}{(N_c^2-1)k_n^2}\right)D_A\frac{d_n^2}{k_n^2}=1+N_c\gamma_nD_A\frac{d_n^2}{k_n^2}
\end{equation}
In Fig.\ \ref{dDSEfig}, the ghost Dyson--Schwinger equation
\eqref{gDSE} is depicted. It is equivalent to the exact
ghost DSE in $3+1$ dimensions. This is due to the fact that Eq.\ \eqref{gDSE} follows from the
operator identity \eqref{gDSEop} and not from the details of the wave
functional.\footnote{On the other hand, the gluon DSE as it stands in \eqref{wDSE} is only true for
the actual constant wave functional.}
The ghost DSE \eqref{gDSE} can be solved for $\gamma_n$ which confirms
the relation \eqref{gammanres} found in section \ref{exactGreen} for $N_c=2$.

In the derivation of the DSEs \eqref{wDSEshort} and \eqref{gDSE} the
integrated configuration space was set to be the first Gribov region
$\Omega_1$ of the \pcg, given by $\abs{\vA}<\frac{2\pi}{L}\equiv k_1$
(see section \ref{Gribovregions}). We now point at an important
property of the Dyson--Schwinger equations. If the configuration
space, here being the first Gribov regions $\Omega_1$, is replaced by
the union of the first two Gribov regions, $\Omega_1\cup\Omega_2$,
the DSEs \eqref{wDSEshort} and \eqref{gDSE} do not change. This is
most readily seen for the ghost DSE \eqref{gDSE} which follows from
the operator identity \eqref{gDSEop} and therefore is not affected by
the choice of the configuration space. The gluon DSE \eqref{wDSEshort}
is derived from the path integral identity \eqref{DSE} which makes use of
the fact that the Faddeev--Popov determinant $\cJ_P$ vanishes at the
first Gribov horizon $\del\Omega_1$. Changing the configuration space
to $\Omega_1\cup\Omega_2$, the path integral identity \eqref{DSE} still
holds true since by definition, the Gribov horizon is where the
Faddeev--Popov determinant vanishes,
\begin{equation}
  \label{horizdef}
  \cJ_P[A\in\del\Omega_n]=0\quad\forall n\; .
\end{equation}
We are therefore led in $\Omega_1\cup\Omega_2$ to the same gluon DSE as in Eq.\ \eqref{wDSEshort}.
More generally speaking: Regardless of the configuration space, so
long as it is a union of Gribov regions, $\bigcup_n\Omega_n$, the
Dyson--Schwinger equations stay form invariant. This form invariance
also applies to the DSEs in $3+1$ dimensions.

\begin{figure}
  \centering
  \includegraphics[scale=0.36]{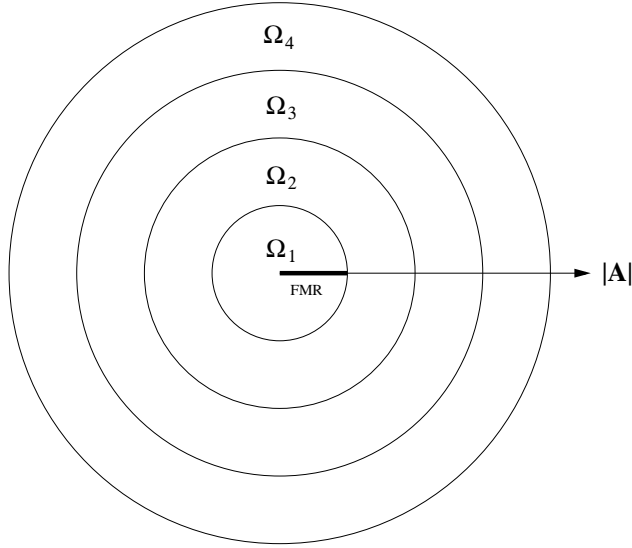}
  \caption{Gribov regions of the \pcg, sketched as spherical shells
    defined by the magnitude of $|{\bf
      A}|=k_1,\,k_2,\,k_3,\,k_4,\,\dots\;$. The location of the
    fundamental modular region (FMR) is indicated by a fat line.}
  \label{Omegasfig}
\end{figure}

In Fig.\ \ref{Omegasfig}, the various Gribov regions $\Omega_n$ of the \pcg are
sketched. For definiteness, we put to our disposal the union
\begin{equation}
  \label{setunion}
  \Gamma_{N_G}:=\Omega_1\cup\Omega_2\cup\dots\Omega_{N_G}\; ,\quad
  N_G\in\mathbb{N}\; 
\end{equation}
of $N_G$ Gribov regions. 
By the choice of $N_G$ the configuration space is restricted to
$\abs{\vA}<\frac{2\pi}{L}N_G\equiv k_{N_G}$. 
Despite the form invariance of the DSEs with respect to
$\Gamma_{N_G}$, their solution cannot be expected to be the same for
all ${N_G}$. As a matter of fact, the gluon
and ghost propagators strongly depend on $N_G$, as shown in the
upcoming section. Thus, the set of DSEs does not have a unique
solution, but for every $N_G$ there exists (at least) one separate solution. The
information on the configuration space \eqref{setunion} is missing in
the set of DSEs and must be provided by subsidiary conditions, i.e.\
put in by hand. In this sense, the DSEs alone do not provide the full
non-abelian quantum gauge theory.
In the $D=3+1$ dimensional case, some approximation is used to
solve the DSEs. Having obtained an approximative solution, there
is no means of deciding on the value of $N_G$, i.e.\ whether this solution
approximates the exact solution in the first Gribov region
$\Gamma_1\equiv\Omega_1$, or rather in $\Gamma_2$, or any other
$\Gamma_{N_G}$, remains unknown. The effect of truncating the set of DSEs in $D=1+1$
will be studied in section \ref{truncsec}.


\section{Many Gribov copies}
\label{ManyGribov}
In this section, the configuration space is extended from the first
Gribov region to a union $\Gamma_{N_G}$ \eqref{setunion} of several Gribov regions,
thus including many Gribov copies. Using the exact constant wave
functional, the propagators, vertices and the colour Coulomb potential
are calculated in $\Gamma_{N_G}$. Moreover, we extend the
configuration space to $N_G=\infty$, damping large $N_G$ contributions
with a Gaussian wave functional. This will illustrate the effect of
insufficient gauge fixing on the infrared features of the theory.

\subsection{Exact Green functions}
\label{NGdependsec}
The calculation of the Green functions in the extended configuration
space $\Gamma_{N_G}$, defined in Eq.\ \eqref{setunion}, is
identical to the one in section \ref{exactGreen} with the exception
that the constant wave functional \eqref{PsiCoul} needs to be normalised differently. This can be
accounted for by letting
\begin{equation}
  \label{psiNG}
  \int_0^{\pi}d\vth\;\dots\quad\to\;\frac{1}{N_G}\int_0^{\pi N_G} d\vth\;\dots
\end{equation}
in the expectation values \eqref{vevdef}. 

For the gluon propagator in
the \pcg we thus find from Eq.\ \eqref{Dab}
\begin{equation}
  \label{D_AresNG}
  D_A(N_G)=\frac{1}{L^2}\left(\frac{4}{9}N_G^2\pi^2-\frac{2}{3}\right)
\end{equation}
and it reduces to the result \eqref{D_Ares} for $N_G=1$. Note that there is a strong dependence on the parameter $N_G$,
i.e.\ a strong Gribov copy effect.

The result for the ghost
form factor $d_n$ in the \pcg follows from making the replacement
\eqref{psiNG} in the expectation value \eqref{d_nres},
\begin{equation}
  \label{dnresNG}
  d_n(N_G)=1+\frac{4}{3\pi}\frac{1}{N_G}\int_0^{\pi
    N_G}d\vth\,\frac{\vth^2\sin^2\vth}{(n\pi)^2-\vth^2}\; .
\end{equation}
Obviously, the ghost form factor still approaches tree-level for
the ultraviolet limit $n\rarr\infty$. There are, however, substantial changes in
the infrared. The result \eqref{dnresNG} for the exact ghost form factor $d_n$ can be seen in Fig.\
\ref{ghost+ggz}. In the first Gribov region, $N_G=1$, the ghost form
factor shows an infrared enhancement, as already shown in section \ref{exactGreen}. Including further gauge copies
makes $d_n$ peak for intermediate momenta. This peak resembling a
resonance appears at $n=N_G$ where the momentum $k_n$ equals the
radius of the configurations space.\footnote{In the thermodynamic
  limit, $L\to\infty$, the peak of the ghost form factor in the first
  Gribov region is asymptotically at $k=0$. This corresponds to the
  ``horizon condition'' in $D=3+1$ dimensions.} For $n<N_G$, the ghost
form factor drops below
tree-level, due to negative eigenvalues of the Faddeev--Popov
operator. In the limit $N_G\rarr\infty$, the lowest momentum mode $d_{n=1}$ approaches a definite value,
\begin{equation}
  \label{one3rd}
  \lim_{N_G\to\infty} d_1(N_G) = \frac{1}{3}\; .
\end{equation}

\begin{figure}
  \centering
  \includegraphics[scale=1.0]{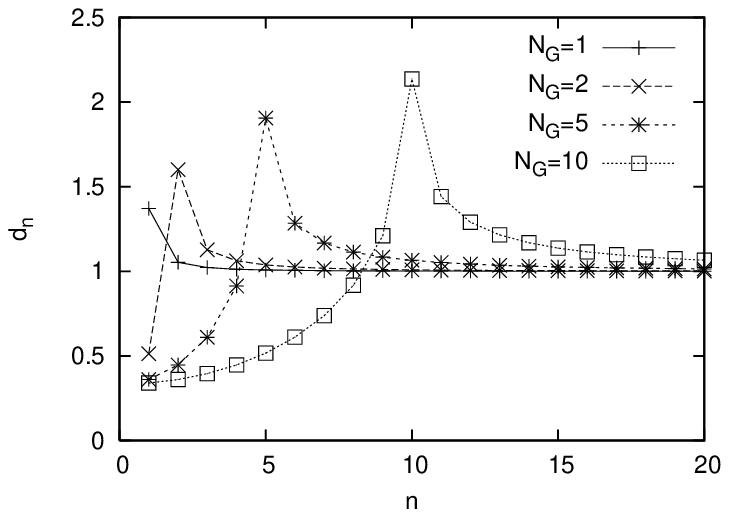}
  \includegraphics[scale=1.0]{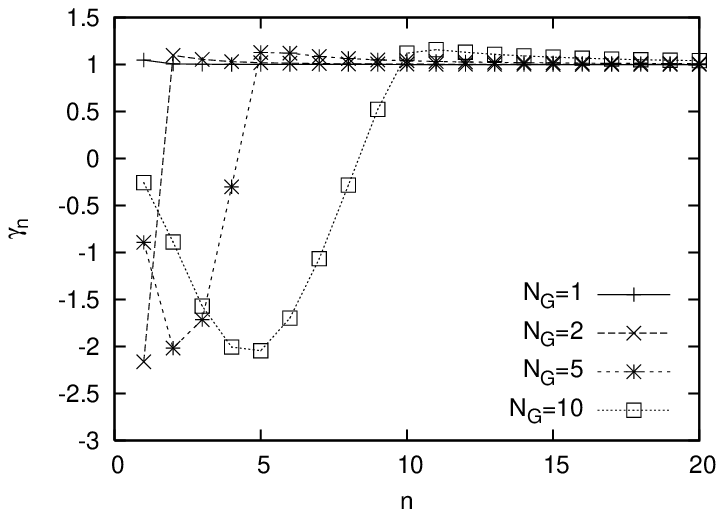}
  \caption{{\it Left:} Exact ghost form factor, including $N_G$ Gribov
    regions. {\it Right:} The form factor $\gamma_n$ of the ghost-gluon vertex for various choices of $N_G$.}
  \label{ghost+ggz}
\end{figure}

It is interesting to note that the deviation of the ghost form factor
$d_n$ from tree-level, when summed over all modes, gives a constant
independent of $N_G$,
\begin{equation}
  \label{const}
  \sum_{n\neq 0}(d_n-1)=\frac{4}{3\pi}\frac{1}{N_G}\int_0^{\pi
      N_G}d\vth\,\sum_{n\neq 0}\frac{\vth^2\sin^2\vth}{(n\pi)^2-\vth^2}= \frac{4}{3\pi}\frac{1}{N_G}\int_0^{\pi
      N_G}d\vth\,\sin\vth(\sin\vth-\vth\cos\vth)=1\; .
\end{equation}
This is illustrated in Fig.\ \ref{ghost+ggz} and can be understood as
being a consequence of the gluon propagator DSE. Solving the ghost DSE \eqref{gDSE} for the ghost-gluon vertex $\gamma_n$,
\begin{equation}
  \label{gammanDSE}
    \gamma_n=\frac{k_n^2\,(d_n-1)}{N_c\,D_A\,d_n^2}\; ,
\end{equation}
 and plugging it
into the gluon DSE \eqref{wDSEshort} directly leads to
Eq.\ \eqref{const}. 

Relation \eqref{gammanDSE} holds for any $SU(N_c)$ gauge group and,
more importantly, for any choice $\Gamma_{N_G}$ of the configuration
space. Having calculated the gluon propagator $D_A$ and the ghost form
factor $d_n$ as functions of $N_G$, Eq.\ \eqref{gammanDSE} gives the
solution for the form factor $\gamma_n$ of the ghost-gluon vertex. The result is shown in
Fig.\ \ref{ghost+ggz}. While within the first Gribov region, $N_G=1$,
there is hardly any deviation from tree-level, this deviation is quite
pronounced for $N_G>1$. Let us note here that an approximation of the
proper ghost-gluon vertex by the tree-level vertex is good if and only
if the configuration space is restricted to the first Gribov region
$\Omega_1$. Working with this approximation in solving the
Dyson--Schwinger equations has an important effect on the
propagators. This will be discussed in the next section. 

\begin{figure}
  \centering
  \includegraphics[scale=1.1]{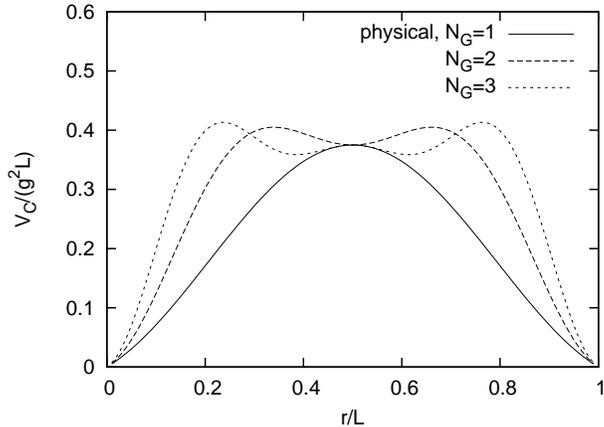}
  \caption{Coulomb potential $V_C(r)$.}
  \label{potfigNG}
\end{figure}

The form factor  $\phi_n$ for the  Coulomb propagator with the result
\eqref{phinres} for $N_G=1$ is found for general $N_G$ by making the
replacement \eqref{psiNG} in Eq.\ \eqref{phinres}. By Fourier transformation, the Coulomb potential $V_C(N_G,r)$ for external
charges separated by $r$ can be obtained in the same manner as
presented for $N_G=1$ in section \ref{potsec}. The result is
\begin{align}
V_C(N_G,r)  &=\frac{g^2}{3}\left(\frac{1}{2}r
   -\frac{r^2}{2L}+L-\,L\,\frac{ L-2 r}{L-r}\,\frac{ \sin
     \left(2 \pi N_G \frac{r}{L}\right)}{2\pi {N_G}\frac{r}{L}}\right)\\
\label{VcresNG}
 &\stackrel{L\to\infty}{\longrightarrow}\;\;\frac{g^2}{2}\, r\quad\forall N_G
\end{align}
and can be seen in Fig.\ \ref{potfigNG}. While the string tension
remains the same for all values of $N_G$ (cf.\ Eqs.\ \eqref{V_Cpure}
and \eqref{VcresNG}), visible effects occur for
large ratios $\frac{r}{L}$. For $N_G>1$, there are locally
stable minima of the potential near $r=\frac{L}{2}$, an unphysical
gauge copy effect.

\subsection{Insufficient gauge fixing}
\label{insuff}
If lattice calculations use a gauge fixing, the common technique is to
minimise a suitable functional along the gauge orbit. The restriction to the fundamental
modular region is achieved only at the absolute minimum of this
functional. If this absolute minimum cannot be reached exactly, extra gauge copies will alter the
result. In order to mimic the situation that the influence of gauge
copies cannot be strictly excluded but only suppressed, we choose here a
Gaussian damping of the gauge field configurations,
\begin{equation}
  \label{psia}
  \psi[A]=\cN\,\sqrt{\frac{2}{\pi}}\,\exp\left(-\frac{\vth^2}{4\alpha^2}\right)\; ,\quad \cN^2=\textstyle{\frac{\sqrt{2\pi}}{\alpha}}\left(1+\coth(\alpha^2)\right)\; ,
\end{equation}
extending the configuration space, see Eq.\ \eqref{setunion}, to
$\Gamma_\infty$. Expectation values in the state \eqref{psia} are most
readily obtained by replacing
\begin{equation}
  \label{psiarepl}
  \int_0^{\pi}d\vth\;\dots\quad\to\;\cN^2\int_0^{\infty} d\vth\,  \e^{-\frac{\vth^2}{2\alpha^2}}\;\dots
\end{equation}
in the corresponding integrals. 
By adjustment of the (free) parameter $\alpha\in\Real$, it can be controlled
how many Gribov copies have a considerable weight in an expectation value. A large
value $\alpha$ will take along many Gribov copies while a small value localises
the weight of field configurations around $A=0$, suppressing Gribov
copies.

\begin{figure}[p]
  \centering
  \includegraphics[scale=1.0]{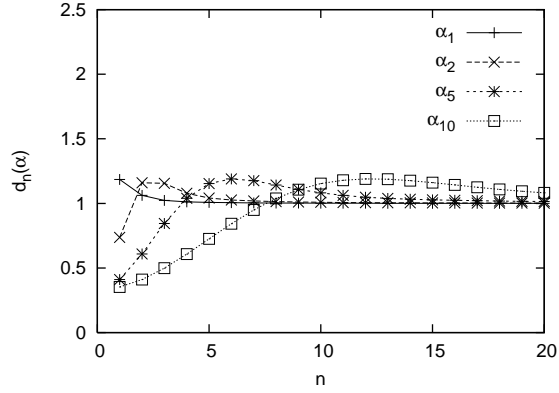}
  \caption{
Ghost form factor $d_n(\alpha)$ in the Gaussian wave functional
\eqref{psia} with those choices $\alpha_1,\alpha_2,\alpha_5,\alpha_{10}$
that yield the gluon propagator in the exact vacuum state with $N_G=1,2,5,10$.}
  \label{dnafig}
\end{figure}

\begin{figure}
  \centering
    \includegraphics[scale=1.0]{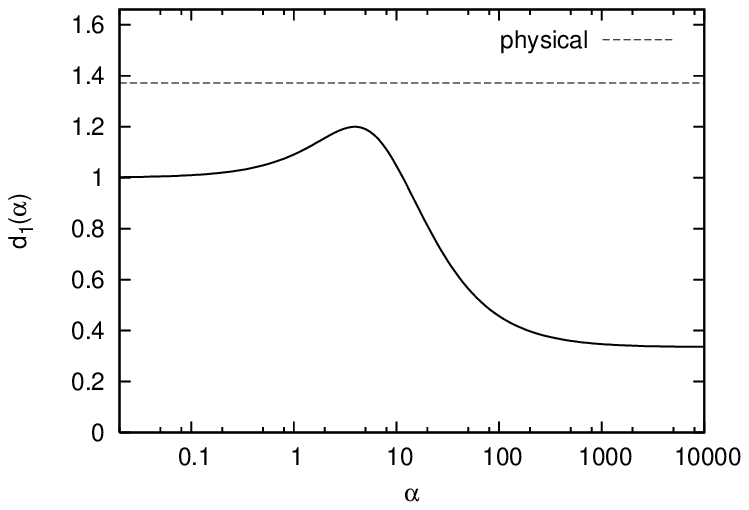}
    \includegraphics[scale=1.0]{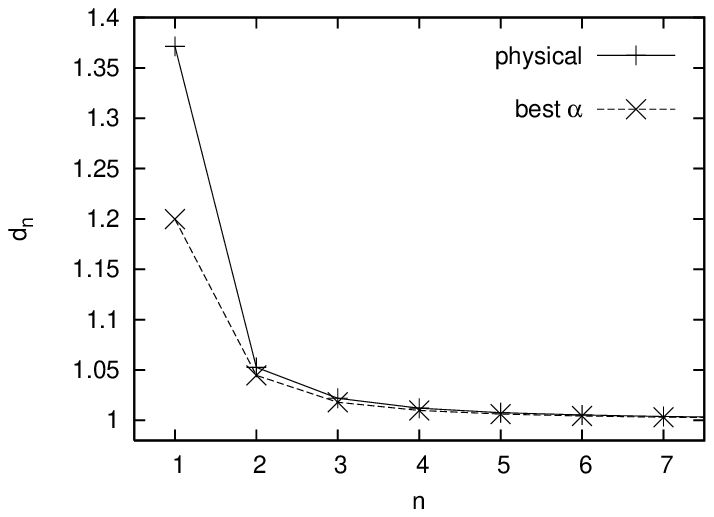}
  \caption{{\it Left:} The infrared enhancement, $d_1(\alpha)$, is
    smaller than the physical value, $d_1\approx 1.37$ for all values
    of $\alpha$. {\it Right:} The ghost form factor in the state
    \eqref{psia} for the value $\alpha\approx 1.40$ that gives the
    strongest infrared enhancement, compared to the exact physical result.}
\label{dncompfig}
\end{figure}

Let us calculate the constant gluon propagator $D_A$ \eqref{Ddef} in
the Gaussian wave functional \eqref{psia} as
a function of $\alpha$. Using the replacement \eqref{psiarepl} in the
expectation value \eqref{Dab} gives
\begin{equation}
  \label{DAares}
  D_A(\alpha)=\frac{2}{3\pi}\,\cN^2\left(\frac{2}{L}\right)^2\int_0^\infty
  d\vth\,\sin^2\vth\,\vth^2\,\e^{-\vth^2/2\alpha^2}=
  \frac{1}{L^2}\,\frac{4}{3}\left(\alpha^2+\frac{4\alpha^4}{\e^{2\alpha^2}-1}\right)\; .
\end{equation}

Obviously, the
larger the width $\alpha$ of the Gaussian \eqref{psia}, the larger $D_A(\alpha)$. This
agrees with the result \eqref{D_AresNG} of the exact wave
functional where $D_A(N_G)$ rises quadratically with
the number $N_G$ of Gribov copies. One can now adjust $\alpha$ in Eq.\
\eqref{DAares} such that it equals the result \eqref{D_AresNG} for the
exact vacuum state, given a specific $N_G$. For the values $N_G=1,2,5,10$,
respectively, the corresponding values for $\alpha$ are
\begin{equation}
  \label{alphDA}
  \alpha_1\approx 
1.63
\; ,\quad \alpha_2\approx 
3.56
\; ,\quad
  \alpha_5\approx 
9.04
\; ,\quad \alpha_{10}\approx 
18.1   \; .
\end{equation}
This procedure simulates the inclusion of $N_G$ Gribov regions in the
expectation values by a choice of the Gaussian wave functional \eqref{psia}.

We now proceed to calculate the ghost form factor $d_n(\alpha)$ in the
state \eqref{psia}. The replacement \eqref{psiarepl} in the integral
\eqref{d_nres} yields
\begin{equation}
  \label{dna}
  d_n(\alpha)=1+\frac{4}{3\pi}\,\cN^2\,\int_0^\infty d\vth\,\sin^2\vth \frac{\vth^2}{(\pi n)^2-\vth^2}\,\e^{-\vth^2/2\alpha^2}\; .
\end{equation}

The limits of $\alpha\to 0$ and $\alpha\to\infty$ produce results that
can be anticipated. As
$\alpha\to 0$, the Gaussian picks out the point $A=0$ from
configuration space. Hence the tree-level behaviour of the ghost form
factor appears:
\begin{equation}
  \label{da0}
  \lim_{\alpha\to 0} d_n(\alpha)=1\; ,\quad \forall n\; .
\end{equation}
The other extreme, $\alpha\to\infty$, takes infinitely many Gribov
copies into account and therefore must resemble the case
$N_G\to\infty$ for the exact vacuum state. Indeed, for $d_1(\alpha)$,
we find
\begin{equation}
  \label{d1inf}
  \lim_{\alpha\to\infty} d_1(\alpha)=\frac{1}{3}\; ,
\end{equation}
in agreement with Eq.\ \eqref{one3rd}.

In Fig.\ \ref{dnafig}, the result \eqref{dna} for $d_n(\alpha)$ is shown for the four values of $\alpha$ in
Eq.\ \eqref{alphDA} which yield the exact gluon propagator for
$N_G=1,2,5,10$. This should be compared to the ghost propagator in the
exact wave functional for
$N_G=1,2,5,10$ in Fig.\ \ref{ghost+ggz}. The effect visible in the
exact wave functional, that taking more Gribov copies into account
damps the infrared enhancement of $d_n$ and produces a spurious peak
(here weakened) at
intermediate momenta, can indeed be mimicked by the wave functional
\eqref{psia} with the appropriate Gaussian damping $\alpha$. 

However, if one tries to quantitatively achieve the infrared
enhancement of the {\it physical} solution, i.e.\ the exact $d_n$
within the first Gribov region $\Omega_1$, the wave functional
\eqref{psia} fails. There exists no value for $\alpha$ such that the infrared
enhancement of the exact physical solution, $d_n$ in Fig.\
\ref{ghostfig}, is realized. The value $d_{n=1}=1.37$ of the exact form
factor \eqref{d_nres} is larger than any
choice of $\alpha$ can produce for the mode $d_{n=1}(\alpha)$ in
the Gaussian wave functional \eqref{psia}. 
In Fig.\ \ref{dncompfig}, it is shown how $d_1(\alpha)$ varies with $\alpha$. For the value $\alpha_{max}$ where the
infrared enhancement of the ghost propagator in the Gaussian wave functional \eqref{psia} is the strongest,
\begin{equation}
  \label{alphmax}
\alpha_{max}\approx 1.40\; ,\quad  d_1(\alpha_{max})\approx 1.20\; ,
\end{equation}
 it is seen how it still underestimates the physical result, $d_1(\alpha_{max})<1.37$. This indicates that if a lattice calculation is unable
to exclude all Gribov copies, the genuine infrared physics cannot be
described. In higher dimensions, this would mean that the infrared
enhancement of the ghost form factor on the lattice is weaker than expected from
continuum studies, an effect that is indeed observed \cite{LanMoy04,CucMen07}.

\begin{figure}
  \centering
  \includegraphics{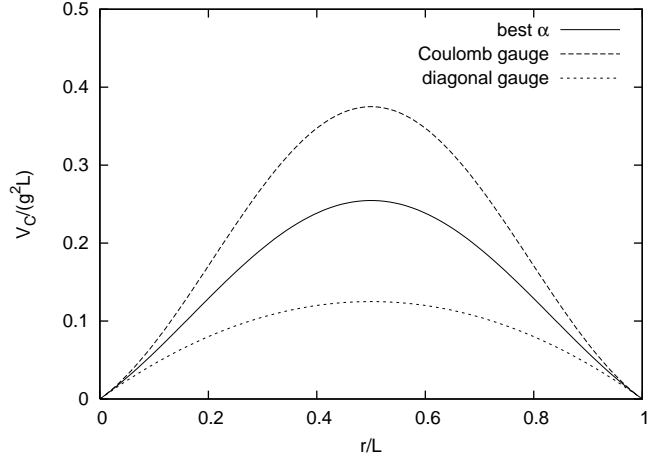}
    \caption{Coulomb potential $V_C(\alpha,r)$ in the Gaussian wave functional
    \eqref{psia} with the ``best'' value \eqref{alphmax} for the width
    $\alpha$. Also shown are the results in the exact ground state for
  the \pcg and the \dcg from Fig.\ \ref{V_Cfig}.}
  \label{potfiga}
\end{figure}

Expecting that with the choice $\alpha=\alpha_{max}$, most (though not all) of the infrared
features can be carried along,\footnote{This choice is reminiscent of
  the ``horizon condition'' in higher dimensions.} we go on to compute
other expectation values of interest. For instance, the Coulomb
potential $V_C(\alpha,r)$ between two external static colour charges separated by $r$ can
be computed by taking the \pcg expectation value of the operator
\eqref{Fdiff} in the Gaussian wave functional \eqref{psia},
\begin{align}
  \label{Vca}
  V_C(\alpha,r)&=g^2\frac{2}{3\pi}\,\cN^2\,\int_0^\infty
  d\vth\,\sin^2\vth \,\e^{-\frac{\vth^2}{2\alpha^2}}
\nn\\ &\qquad\qquad\times\left(\frac{1}{2}r-\frac{1}{2}\frac{r^2}{L}+\frac{1}{2\sin^2\vth}\left(L-r \cos \left(2\, \textstyle\frac{L-r}{L} \,\vth\right)-(L-r) \cos \left(2\,\textstyle\frac{r}{L}\, \vth \right)\right)\right)\nn\\
&=\frac{g^2}{6}\left(r-\frac{r^2}{L}+\left(1+\coth(\alpha^2)\right)\left(L-r\,\e^{-2\alpha^2(1-\frac{r}{L})^2}-(L-r)\,\e^{-2\alpha^2(\frac{r}{L})^2}\right)
\right)\nn\\
 &\stackrel{L\to\infty}{\longrightarrow}\;\;\frac{g^2}{2}r
\quad\forall \alpha
\end{align}
In the thermodynamic limit $L\to\infty$, the potential \eqref{Vca}
approaches the same behaviour as $V_C(r)$ in the exact ground state. It was already shown above that
the Coulomb string tension $\sigma_C$, defined in the thermodynamic limit,
turns out independent of the wave functional and the configuration space
$\Gamma_{N_G}$. For the choice \eqref{alphmax} of $\alpha$, the potential
$V_C(r)$ in Eq.\ \eqref{Vca} is shown in Fig.\ \ref{potfiga}. Varying
$\alpha$, the potential is seen to vary between the exact solutions of
the \pcg and the \dcg. As a matter of fact, the result \eqref{V_Cdia} of the \dcg is
reached in the limit $\alpha\to 0$,

\begin{equation}
  \label{Vca0}
  \lim_{\alpha\to
    0}V_C(\alpha,r)=\frac{g^2}{2}\left(r-\frac{r^2}{L}\right)\equiv\left. V_C(r)\right|_{\textrm{diagonal gauge}}\; .
\end{equation}

Recall that the limit $\alpha\to 0$ turns the Gaussian wave functional \eqref{psia} into
a delta distribution, peaked at $A=0$. The limit \eqref{Vca0}
therefore emphasises that in $1+1$ dimensions, the Coulomb potential
$V_C(r)$ is a quantity that is independent of quantum fluctuations,
when properly evaluated in the fundamental modular region.

\begin{figure}
  \centering
  \includegraphics{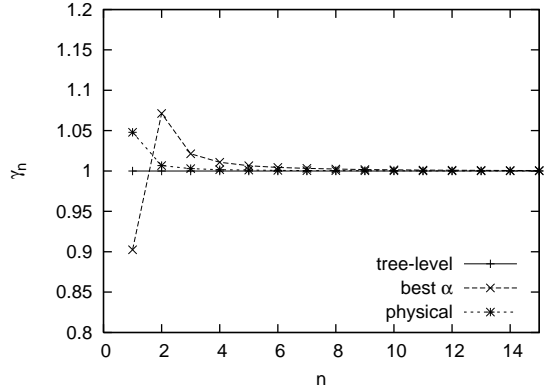}
  \caption{Ghost-gluon vertex form factor $\gamma_n$ in the Gaussian wave functional
    \eqref{psia} with the ``best'' value \eqref{alphmax} for the width
    $\alpha$. Also shown is the physical solution, see Fig.\ \ref{ggzfig}, and the tree-level value.}
  \label{ggzlatt}
\end{figure}

In the opposite limit, $\alpha\to\infty$, the potential $V_C(\alpha,r)$
coincides with $V_C(N_G,r)$ in the exact vacuum state,
see Eq.\ \eqref{VcresNG}, having taken the limit $N_G\to\infty$. This is intuitively clear since the Gaussian
state \eqref{psia} becomes constant for $\alpha\to\infty$ (and not
normalisable). Note, however, that the limits $\alpha\to\infty$ and
$L\to \infty$ are not interchangeable. In order to find the right
string tension, one is to first evaluate the energy in a normalisable
state ($\alpha<\infty$), take the thermodynamic limit $L\to\infty$,
and then determine the string tension as a function of $\alpha$. In
the opposite order of limits, one finds a different result for $\sigma_C$.

Finally, we calculate the ghost-gluon vertex in the state
\eqref{psia}. It can be found using
the expression \eqref{gammanDSE} which holds for any wave functional
(and any $\Gamma_{N_G}$); in the evaluation we simply use the
results \eqref{DAares} and \eqref{dna} for the gluon and ghost propagators. The result is shown in Fig.\
\ref{ggzlatt}. The deviation of $\gamma_n$ from the exact physical
result is small, although it is enhanced for the infrared
modes. Anyhow, the tree-level vertex is a better approximation of the
exact ghost-gluon vertex than the one in the Gaussian wave functional \eqref{psia}.

\section{Truncation effects}
\label{truncsec}

In higher dimensions, it is not possible to obtain exact nonperturbative expressions
for the Green functions. A common approximation that enables us to
find the infrared asymptotic solutions of the Dyson--Schwinger equations is to render the
ghost-gluon vertex tree-level. We will make this approximation here in
$1+1$ dimensional \pcg and investigate the effect on the
propagators. To be explicit, we set
\begin{equation}
  \label{bareGGZ}
  \Gamma_n^a\to\Gamma_n^{0,a}\; ,\quad \Rightarrow\;\gamma_n=1\; .
\end{equation}
Recall it was shown in the preceding sections that despite the form
invariance of the exact DSEs with respect to the configuration space
$\Gamma_{N_G}$, the true Green functions (calculated with the exact
vacuum wave functional) do depend on $\Gamma_{N_G}$. With the approximation \eqref{bareGGZ}, the DSEs for the gluon and ghost propagators,
Eqs.\ \eqref{wDSEshort} and \eqref{gDSE}, turn into
\begin{equation}
  \label{DSEapp}
   d_n=1+N_cD_A\frac{d_n^2}{k_n^2}\; ,\quad D_A^{-1}=N_c\sum_{n\neq
     0}\frac{d_n^2}{k_n^2}\; .
\end{equation}

Diagrammatically, these equations are depicted in Figs.\ \ref{wDSEfig} and
\ref{dDSEfig}, with the blobs replaced by dots. We note that due to the
approximation \eqref{bareGGZ}, this set of equations is closed and can
be solved. The crucial point is to realize that the solution will no longer
depend on the choice of the configuration space $\Gamma_{N_G}$. Since
the ghost-gluon vertex was chosen to be independent of $\Gamma_{N_G}$,
see Eq.\ \eqref{bareGGZ}, the gluon and ghost propagators are now also
independent of $\Gamma_{N_G}$. The original exact set of DSEs holds
within any of the Gribov regions. However, 
solving the approximated set of DSEs, the information is lost
in which Gribov region the Green functions are evaluated in. This
problem also occurs in the infrared ghost dominance model of the
$D=3+1$ theory. 

Let us now solve Eq.\ \eqref{DSEapp} explicitly. Since it is quadratic
in $d_n$, we find two solutions,
\begin{equation}
  \label{dnsolpm}
  d_n=\frac{1\pm\sqrt{1-4\frac{N_cD_A}{k_n^2}}}{2\frac{N_cD_A}{k_n^2}}\; .
\end{equation}
In order for the limit $d_n\rarr 1$ for $k_n=\frac{2\pi
  n}{L}\rarr\infty$ to be fulfilled, we need the lower sign. Plugging
this solution into the second DSE in Eq.\ \eqref{DSEapp}, we find
\begin{equation}
  \label{DAsum}
  D_AN_cL^2=\sum_{n\neq 0}\, (\pi
  n)^2\left(1-\sqrt{1-\frac{D_AN_cL^2}{(\pi n)^2}}\right)^2\; \,\quad
  \Rightarrow  D_AN_cL^2 \approx 7.74
\end{equation}
The numerical solution yields with $N_c=2$ for the gluon propagator
\begin{equation}
  \label{DAsol}
  D_A\approx 3.87\,\frac{1}{L^2}\; .
\end{equation}
This value can be plugged into Eq.\ \eqref{dnsolpm}, with the lower sign, to immediately get the
result for the ghost form factor.

The gluon propagator \eqref{DAsol} is evidently independent of the Gribov region. One can
choose $\Gamma_{N_G}$ with any $N_G$, the approximation
\eqref{bareGGZ} will always yield the gluon propagator given by Eq.\
\eqref{DAsol}. On the other hand, we know the {\it exact} result for
the gluon propagator with a given value of $N_G$, see Eq.\
\eqref{D_AresNG}. The latter can now be compared with the
approximative result \eqref{DAsol}, for each $N_G$.

\begin{figure}
  \centering
  \includegraphics[scale=1.]{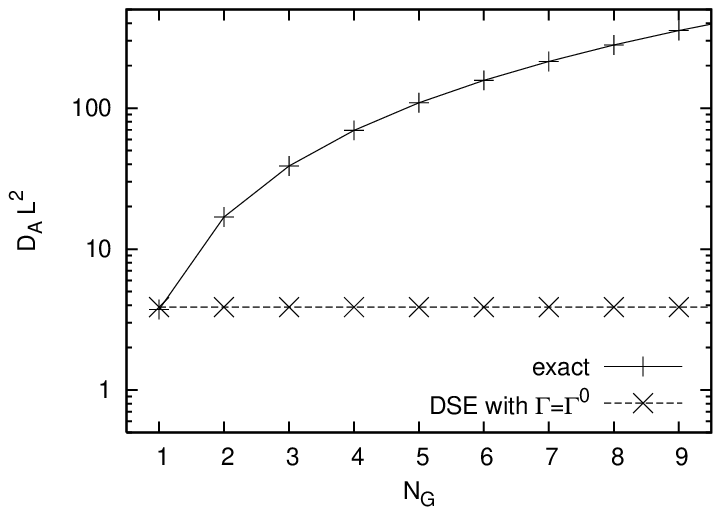}
  \includegraphics[scale=1.]{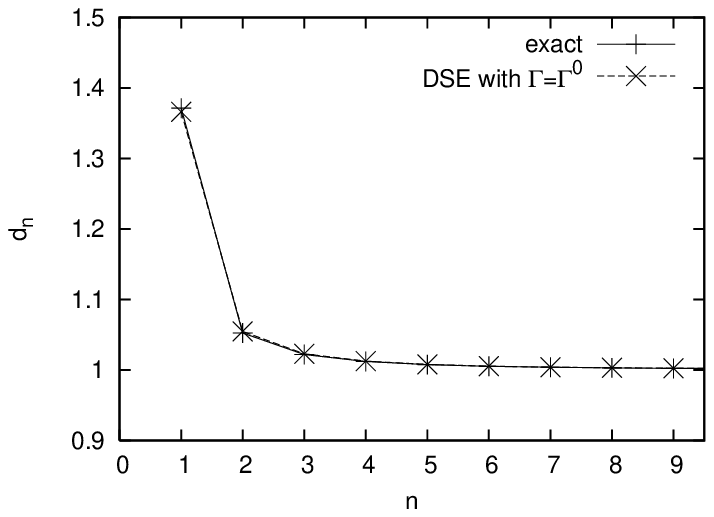}
  \caption{{\it Left:} Comparison of the exact gluon propagator to the
    one from the DSEs with a tree-level ghost-gluon vertex, shown as a
    function of $N_G$. {\it Right:} Comparison of the exact ghost form
    factor $d_n$ in the first Gribov region ($N_G=1$) to one
    calculated from the truncated DSEs, both shown as functions of
    the momentum mode $n$.}
  \label{DAcomp}
\end{figure}

In Fig.\
\ref{DAcomp}, it is seen how for large $N_G$, the exact and
approximative results differ dramatically. However, for $N_G=1$, i.e.\ in the first Gribov region, there is a good
agreement. The tree-level approximation for the ghost-gluon vertex
yields a value for the gluon propagator that is very close to the
exact result in the first Gribov region. The same occurs for the ghost propagator. 
As can be seen in on the right
panel of Fig.\ \ref{DAcomp}, 
the approximative (with $\gamma_n=1$) DSE result for the ghost form factor $d_n$ hardly
deviates from the exact result for $N_G=1$. At the same time, the
approximative and exact solutions of $d_n$ for other values of $N_G$
disagree both quantitatively and qualitatively, cf.\ Fig.\
\ref{ghost+ggz}. Also shown in Fig.\ \ref{ghost+ggz} is the ghost-gluon
vertex in the exact vacuum state for different values of
$N_G$. One can realize that the tree-level approximation \eqref{bareGGZ} is only
for $N_G=1$ a good one.

We infer that while the exact Dyson--Schwinger equations are form invariant with respect to the
configuration space $\Gamma_{N_G}$, the approximation of the
ghost-gluon vertex by its tree-level value effectively puts the approximative
solution of the propagators into the first Gribov region $\Gamma_1\equiv\Omega_1$.
This supports the approach of solving DSEs in $D=3+1$, truncated by means
of a tree-level ghost-gluon vertex.

\section{Variational approach}
\label{variationalsec}

In order to test the variational approach to
Yang--Mills theory in $3+1$ dimensions, the same ansatz for the wave
functional as in Ref.\ \cite{FeuRei04} is here applied to solve the Yang--Mills
Schr\"odinger equation in $D = 1 + 1$ in the \pcg. The variational
calculation is performed using the same approximations as in Ref.\ \cite{FeuRei04}. 

The variational wave functional is given by
\be
\label{psiw}
\Psi (A) = \frac{1}{\sqrt{\cJ_P (A)}} \,\widetilde{\Psi} (A) \; , \quad
\widetilde{\Psi} (A) = \cN \e^{- \frac{1}{2} A^a \omega A^a} := \langle A |
\omega \rangle \hk ,
\ee
where $\omega$ is a variational parameter determined by minimising the vacuum
energy density.  The
ansatz (\ref{psiw}) is mainly motivated by simplicity. It removes the
Faddeev--Popov determinant $\cJ_P$ from the integration measure in the
expectation values of the \pcg,
\be
\label{vevw}
\langle \Psi |\cO(A) |\Psi \rangle = \int_{\Gamma_\infty} \cD A\, \cJ_P (A)\, \cO(A)\abs{\Psi(A)}^2 =
\int_{\Gamma_\infty} \cD A\, \cO(A)\abs{\widetilde{\Psi} (A)}^2 =: \langle \cO(A) \rangle_\omega \; .
\ee
 and thus allows for an immediate
application of Wick's theorem. Here and 
in the following, $\langle \dots \rangle_\omega$ denotes the expectation value in
the state $| \omega \rangle$ (\ref{psiw}),
for which the scalar product is defined with
 the flat
integration measure $\cD A\equiv \pli_a d A^a$.

The configuration space in the expectation value \eqref{vevw} is not
restricted to the first Gribov region $\Omega_1$ but is extended to
the union $\Gamma_\infty$ of all Gribov regions, see Eq.\
\eqref{setunion}. This is motivated from the $D = 3 + 1$ case where only
little is known about the Gribov
horizon and a restriction to $\Omega_1$ is technically cumbersome.\footnote{The so-called
  Gribov--Zwanziger action is used in a few studies \cite{Dud+08c} to realize a
  restriction to $\Omega_1$.} In section \ref{insuff}, we have seen that
within $\Gamma_\infty$, any Gaussian damping of gauge copies will fail
to recover the exact infrared behaviour of the ghost form
factor. Note, however, that the wave functional in Eq.\ \eqref{psiw} is
supplemented by a Faddeev--Popov determinant. We will show below that
with appropriate approximations, the correct infrared behaviour of the
ghost form factor can thus still be maintained.


The normalisation constant $\cN$ in the ansatz \eqref{psiw} for the
wave functional is chosen such that $\lla 1\rra_\omega=1$ and is given by
\be
\ln\cN = {\frac{N^2_c - 1}{4}}\ln\lk \frac{\omega}{\pi}
\rk \; .
\ee
Thus, the  static (equal-time)
 gluon propagator reads
\be
\label{6-3}
D^{ab}=\langle \Psi | A^a A^b | \Psi \rangle = \langle A^a A^b \rangle_\omega =
\delta^{a b} (2 \omega)^{- 1} =:\delta^{ab}\,D_A\; .
\ee
In $D = 3 + 1$ the kernel 
$\omega$ has the meaning of the gluon energy. In the present
$1 + 1$ dimensional case $\omega$ has dimension $mass^{-2}$ and is
required by normalisability of the wave functional to be positive,
$\omega>0$. Let us emphasise that whatever the variational principle
yields for $\omega$, it will determine the gluon propagator $D_A$ by
Eq.\ \eqref{6-3}.


The vacuum energy $E(\omega)$ is calculated by taking the expectation value of the
Yang--Mills Hamiltonian in the \pcg (\ref{HC}) in the absence of external charges.
After a partial integration, $E(\omega)$ yields
\begin{align}
\label{6-5}
E (\omega) & =  \langle \Psi | H | \Psi \rangle = \int \cD A\,
{\cJ_P(A)}\Psi^*
(A) H \Psi (A) \nonumber\\
& =  \frac{g^2 L}{2} \int \cD A\, \lk \widetilde{\Pi}_\perp^a \widetilde{\Psi} \rk^* \lk
\widetilde{\Pi}_\perp^a \widetilde{\Psi} \rk \hk ,
\end{align}
where
\be
\label{6-6}
\widetilde{\Pi}_\perp^a = {\cJ_P^{1/2}} 
\Pi_\perp^a {\cJ_P^{-1/2}} 
= \Pi_\perp^a - \frac{1}{2} \lk \Pi_\perp^a \ln \cJ_P \rk \hk .
\ee
Using
\be
\label{6-7}
\widetilde{\Pi}_\perp^a \widetilde{\Psi} =\frac{i}{L} \left[ \omega A^a + \frac{1}{2} \lk \frac{d \ln \cJ_P}{d
A^a} \rk \right] \widetilde{\Psi}
\ee
we find
\begin{align}
\label{Evonw}
E (\omega) & =  \frac{g^2}{2 L} \lla \lk \omega A^a + \frac{1}{2} \frac{d \ln \cJ_P}{d
A^a} \rk^2 \rra_\omega \nonumber\\
& =  \frac{g^2}{2 L} \left[ \omega^2 \lla A^a A^a \rra_\omega + \omega
\lla A^a \frac{d\ln \cJ_P }{d A^a} \rra_\omega + \frac{1}{4} \lla \lk
\frac{d\ln \cJ_P}{d A^a}  \rk^2 \rra_\omega \right] \hk .
\end{align}
Following Ref.\ \cite{FeuRei04} 
we will explicitly calculate the first two terms and
then find the last term by completing the result to a total square, which is
correct up to two loops. 
The first term in Eq.\ \eqref{Evonw} can obviously be expressed by the
gluon propagator \eqref{6-3}. For the second term, we use the abbreviation
\be
\label{chidef}
\chi^{ab}:=-\,\omega\lla A^a \frac{d\ln \cJ_P}{d A^b}  \rra_\omega 
\hk ,
\ee
which was referred to as the ``curvature'' in Ref.\
\cite{FeuRei04}. Using the definition \eqref{ggzconn} of the proper
ghost-gluon vertex and its form factor $\gamma_n$ \eqref{gamman}, the curvature $\chi^{ab}$ can be written as\footnote{The definition \eqref{chidef} of the curvature is equivalent to
  the one in Ref.\ \cite{FeuRei04} within the wave functional
  \eqref{psiw}. The same holds for the proper ghost-gluon vertex.} 
\begin{align}
  \label{curvghostloop}
  \chi^{ab}&=+\,\omega\lla
  A^a\,\Tr\,G\,\Gamma_n^{0,b}\rra_\omega=\frac{1}{2}\sum_{n\neq 0}\,\gamma_n\,\frac{d_n^2}{k_n^4}\,\tr\left(\Gamma_n^{0,a}\Gamma_n^{0,b}\right)\nn\\
  &=\frac{N_c}{2}\sum_{n\neq 0}\gamma_n\,\frac{d_n^2}{k_n^2}\,\delta^{ab}=:\chi\,\delta^{ab}
\end{align}
and its diagonal elements define the scalar curvature
$\chi$. Performing the quadratic completion, the
expression \eqref{Evonw} for the vacuum energy can be cast into the form
\be
\label{psiw2}
E (\omega) = g^2\,\frac{N^2_C - 1}{4 L} \frac{(\omega - \chi)^2}{\omega} \hk 
\ee
and it is obviously minimised for the choice
\be
\label{psiw3}
\omega =  \chi 
\ee
of the variational kernel $\omega$. Equation \eqref{psiw3} is called
the gap equation and it gives rise to an infrared divergent gluon
energy $\omega(k)$ in $3+1$ dimensions \cite{FeuRei04,EppReiSch07}. 

The gap equation \eqref{psiw3} states that the gluon propagator
\eqref{6-3} can be related to the curvature $\chi$ and we may use the
definition \eqref{chidef} of $\chi^{ab}$ to calculate the gluon propagator
exactly. However, approximations have been made
and---more importantly---the configuration space was not properly
restricted to $\Gamma_1$.  Let us look at the task of determining the
solution for the gluon and ghost propagators differently. We note that
plugging the gap equation \eqref{psiw3} into Eq.\ \eqref{6-3} yields
with Eq.\ \eqref{curvghostloop}
\begin{equation}
  \label{DSEvaria}
  D_A^{-1}=N_c\sum_{n\neq 0}\gamma_n\,\frac{d_n^2}{k_n^2}\; .
\end{equation}
Turning back to Eq.\ \eqref{wDSEshort}, we recognise that relation
\eqref{DSEvaria} is identical to the gluon propagator DSE in the exact vacuum
state. Moreover, we can use the ghost propagator DSE \eqref{gDSE} from
the exact vacuum state since it follows from an operator identity,
independent of the wave functional or configuration space. The set of
equations the variational calculation above resulted in is equivalent
to the set of Dyson--Schwinger equations derived in the exact
vacuum state. The difference is that here the expectation values
(i.e.\ $d_n$, $D_A$, $\gamma_n$) are evaluated in the state
\eqref{psiw} and in the configuration space $\Gamma_\infty$, yielding
different results. It was shown in section \ref{ManyGribov} that choosing a set
of several ($N_G>1$) Gribov regions for $\Gamma_{N_G}$ results in
drastic changes for the Green functions. However, if we use the tree-level
approximation for the ghost-gluon vertex, it is clear from the
discussion in section \ref{truncsec} that the solution to the DSEs so obtained
is very close to the exact solution. With a conspiracy of approximations, namely
the quadratic completion in Eq.\ \eqref{Evonw} and the vertex
approximation $\gamma_n=1$, the
variational state \eqref{psiw} in $\Gamma_\infty$ yields the same
propagators as the exact vacuum state \eqref{PsiCoul} in the first
Gribov region $\Gamma_1\equiv\Omega_1$.

Identifying the variational wave functional \eqref{psiw} for the
solution $\omega=\chi$ with the exact wave functional
$\Psi=const$ implies that the Gaussian must cancel the
Faddeev--Popov determinant $\cJ_P$,
\be
\label{psiw4}
\cJ_P \;\longrightarrow\; \exp \lk -  A^a \chi^{a b} A^b \rk \hk .
\ee
In Ref.\ \cite{ReiFeu04} it was shown that up to two-loop
order in the energy the replacement \eqref{psiw4} is exact and thus
results in the correct DSEs.

In $D = 3 + 1$ both $\omega$ and $\chi$ are momentum
dependent and the cancellation of $\cJ_P^{- \frac{1}{2}}$ against the Gaussian in the wave functional is
obtained in the infrared limit $k \to 0$ only. We thus observe that {in} 
the infrared limit {the wave functional}
in $D = 3 + 1$ reduces to the exact wave functional in $D = 1 + 1$ dimensions.
As discussed in Ref.\ \cite{ReiFeu04} the 
 constant wave functional does
not constrain the infrared modes of the gauge field and thus describes a
stochastic vacuum where the infrared modes can {arbitrarily} fluctuate.

As shown in Ref.\ \cite{ReiFeu04}, the cancellation of
Gaussian and Faddeev--Popov determinant persists in $D = 3 + 1$ in the infrared even if the
more general ansatz is used,
 \be
 \label{6-4}
 \Psi (A) = \cJ^{- \alpha}_P  (A) \widetilde{\Psi} (A)\; .
 \ee
In this state with a arbitrary exponent $\alpha$ of the Faddeev--Popov
determinant, the gluon propagator becomes
\be
\langle A^a A^b \rangle = \delta^{a b} (2 \widetilde{\omega})^{- 1} \hk ,
\ee
where
\be
\label{43-X2}
\widetilde{\omega} = \omega - (2 \alpha - 1) \chi \hk .
\ee
The gap equation is the same as above, see Eq.\ (\ref{psiw3}), except that
$\omega$ is replaced by $\widetilde{\omega}$,
\be
\widetilde{\omega} = \chi \hk .
\ee
We therefore find from (\ref{43-X2})
\be
\label{47-242}
\omega = 2 \alpha \chi \hk .
\ee
For $\alpha = \frac{1}{2}$ we recover, of course, the previous result
(\ref{psiw3}), while $\alpha = 0$ yields $\omega = 0$ and the variational wave
functional (\ref{6-4}) becomes the exact one 
\be
\Psi (A) = \cN = const. 
\ee
and thus yields also the exact results for the propagators, provided the range of
the field $A$ is properly restric{ted} to the first Gribov {region.}

\bigskip

Finally, let us turn to the Coulomb form factor $f_n$ which measures
the deviation of the Coulomb propagator $\lla G(-\del^2)G\rra$ from
the factorised form $\lla G\rra(-\del^2)\lla G\rra$ and was calculated
in the exact wave functional at the end of section \ref{propasec}. In
$3+1$ dimensions, the (momentum-dependent) form factor $f(k)$ is set
to unity since it fails to satisfy the corresponding integral
equation within the approximations made \cite{Epp+07}. The form factor
$f(k)$ requires a higher-order calculation (as pointed out in Ref.\ \cite{FeuRei04,PhDSchleifenbaum}), and for this reason it is investigated
here in $1+1$ dimensions where approximations are not necessary.

The integral equation that is derived for the Coulomb form factor
$f_n$ follows from the identity \cite{MarSwi84}
\be
\label{MSrel}
F(A) = G(A)(-\del^2)G(A)=\frac{\partial}{\partial g} (g G(g\bA)) \; .
\ee
Here, we have scaled the gauge field by the coupling constant $g$,
\begin{equation}
  \label{Abar}
  A=g\,\bA
\end{equation}
 so that with 
 \begin{equation}
G^{-1}(g\,\bA)=-\,\del^2-g\,\hat\bA\,\del   
 \end{equation}
we can derive Eq.\ \eqref{MSrel} by differentiation. Following Ref.\ 
\cite{Swi88}, in the variational approach \cite{FeuRei04} the
vacuum expectation value of the relation \eqref{MSrel} was taken, thereby ignoring
the implicit $g$-dependence of the wave functional to 
obtain the approximative relation
\be
\label{test}
\langle {F} \rangle \approx \frac{\partial}{\partial g} \lk g  \langle
{G} \rangle \rk \; 
\ee
which is the so-called Swift relation \cite{Swi88}. In the present $1 +
1$ dimensional case the exact vacuum wave functional is independent of $g$ but
the Faddeev--Popov determinant $\cJ_P(g\bA)$ in the integration measure is $g$-dependent and
this $g$-dependence is ignored in Eq.\ (\ref{test}). Expressing $\langle
{G} \rangle $ and $\langle
{F} \rangle $ in terms of the ghost and Coulomb form factors, $d_n$
\eqref{d_ndef} and
$f_n$ \eqref{fdef}, the Swift relation \eqref{test} becomes\footnote{This relation differs from the one
given in Ref.\ \cite{FeuRei04}, where an 
extra factor of $g$ was included in the ghost form
factor.}
\be
\label{fSwi}
f_n \approx d_n^{-2}\frac{\del}{\del g}(g\, d_n)=d_n^{-1}- g\frac{\partial}{\partial g} d^{- 1}_n \hk .
\ee

Using the inverse form of the DSE \eqref{gDSE} for the ghost form
factor,
\begin{equation}
  \label{gDSEinv}
  d_n^{-1}=1-N_c\,\gamma_n\,D_A\,\frac{d_n}{k_n^2}\; ,
\end{equation}
an integral equation for $f_n$ can be found,
\begin{equation}
  \label{fDSE}
  f_n\approx 1+N_c\,\gamma_n\,D_A\,\frac{d_n^2 f_n}{k_n^2}\; .
\end{equation}

\begin{figure}
  \centering
  \includegraphics{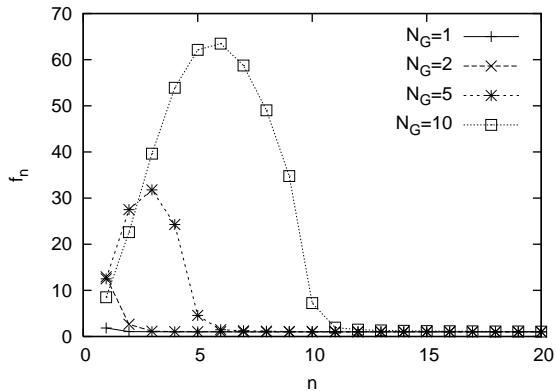}
  \caption{Coulomb form factor $f_n$, defined by Eq.\ \eqref{fdef}, in the exact vacuum state depending on the number $N_G$ of included Gribov regions.}
  \label{flattfig}
\end{figure}

Note that this integral equation is not exact, due to the Swift
approximation \eqref{test}. 
In the exact wave functional of the $1+1$ dimensional \pcg, the ghost
propagator is independent of $g$ \cite{PhDSchleifenbaum}\footnote{The same is true for the
  ghost propagator in $3+1$ dimensions if we consider the stochastic
  vacuum $\Psi[A]=const$ of the Coulomb gauge Hamiltonian approach or
  the ghost dominance model $S_{YM}=0$ in the Landau gauge to study
  the infrared asymptotics. }
\begin{align}
  \label{gindep}
  \frac{\del}{\del g}\lla G\rra& = \frac{\del}{\del
    g}\,\frac{\int_{\Omega_1}\cD \bA\, G(g\,\bA)\,\Det\,
    G^{-1}(g\,\bA)}{\int_{\Omega_1}\cD \bA\, \Det\,
    G^{-1}(g\,\bA)}\nn\\
 &= \frac{\del}{\del
    g}\,\frac{\int_{\Omega_1}\cD \bA\, G(\bA)\,\Det\,
    G^{-1}(\bA)}{\int_{\Omega_1}\cD \bA\, \Det\, G^{-1}(\bA)}\; =0
\end{align}
and hence the ghost form factor also is, $\del d_n/\del g=0$. The
Swift relation \eqref{fSwi} thus simplifies to
\be
\label{fisdinv}
f_n = d^{- 1}_n \; .
\ee
This relation implies that $f_n$ is infrared suppressed if $d_n$ is
infrared enhanced, which is in contradiction to the true behaviour of
$f_n$ and $d_n$, see Figs.\ \ref{ghostfig} and \ref{ffig}. Even more
directly, the contradiction can be seen by plugging Eq.\ \eqref{fisdinv}
into the approximative integral equation \eqref{fDSE} and comparing to
the exact ghost form factor DSE \eqref{gDSEinv}.

The contradiction arises from the approximation made to arrive at the  Swift relation 
(\ref{test}). In the infrared limit of the $D=3+1$
theory which is correctly described by the $D=1+1$ wave functional \cite{ReiFeu04},
the Swift relation therefore must lead to inconsistencies \cite{PhDSchleifenbaum,Epp+07}.

Recent lattice calculations in $D=3+1$ of the Coulomb form factor $f(k)$ seem to indicate that in the infrared $f(k)$ is
enhanced. This enhancement gets weaker the better the Coulomb gauge is
fixed. In $1+1$ dimensions, the form factor $f_n$ can be calculated
exactly, see section \ref{propasec}. The effect of including several Gribov
regions, i.e.\ choosing $\Gamma_{N_G>1}$ as done in section \ref{NGdependsec},
leads to an interesting observation. In Fig.\ \ref{flattfig}, it is
shown how the exact result for $f_n$ varies with the number $N_G$ of
included Gribov regions. The larger $N_G$, the more pronounced a
(spurious) infrared enhancement. This gauge copy effect is in
agreement with the findings on the lattice in $3+1$ dimensions and
indicates that in order to get the exact result for $f(k)$,
gauge fixing on the lattice has to be performed very carefully.

\section{Summary and Conclusions}
\label{summsec}
In this paper, we have considered $1 + 1$ dimensional $SU (2)$ 
Yang--Mills theory in canonical quantisation in the \pcg as a testing ground for
 Yang--Mills theory studies in higher dimensions. The investigations were carried out in the \pcg
and in the \dcg, 
where the residual global gauge invariance, left unfixed in
the \pcg, is fixed by diagonalising the constant spatial gauge
field.  {Although the} two
gauges differ only by a global $SU(2)/U(1)$ gauge-fixing con{straint}, they
have different Faddeev--Popov determinants due to additional zero modes of
the Faddeev--Popov kernel in the \pcg. While the \pcg is perfectly
suitable for perturbation theory, the \dcg is ill-defined for the
{pertur}bative vacuum $A  = 0$, for which the Faddeev--Popov determinant vanishes.
The occu{rr}ence of such gauge-fixing defects is a characteristic feature of
so-called abelian {gauges} the \dcg belongs to. 
In higher
dimensions gauge-fixing defects of abelian gauges manifest themselves as
magnetic monopoles in the corresponding abelian projection 
\cite{Hoo81a} (see also Ref.\ \cite{Rei97}).

We have expli{citly} demonstrated that the Faddeev--Popov method does not require
complete gauge fixing but works for any partial gauge fixing, provided that 
the zero
modes of the Faddeev--Popov kernel arising from the residual gauge symmetry
(left unfixed by the partial gauge fixing) are properly treated. 
In the resolution of
Gauss' law,  these zero
modes give rise to residual constraints {on} the wave functional, which express
the invariance of the wave functional under the residual gauge symmetry: The
Noether charges corresponding to these residual symmetries must vanish in
physical states. The {const}raints on the wave functionals arising from the
residual unfixed gauge symmetry exist also in higher dimensions but have not
been explicitly identified so far, except for space-independent gauge
transformations \cite{ReiWat08}. They also naturally emerge in the functional
integral approach in the so-called first order formalism where the temporal
gauge field can be explicitly integrated to leave a 
$\delta$-functional, which enforces Gauss' law \cite{WatRei06}. In
the \pcg, the Gauss' law constraint can be worked out analogously to the
Hamiltonian approach and the $\delta$-functional can be used to integrate out
the longitudinal components of the momentum field. When the resolution of
Gauss' law is properly done, i.e.\ the zero modes of the Faddeev--Popov kernel
properly treated, from the $\delta$-functional some ordinary $\delta$-function
{survives}, which precisely enforces the vanishing of the Noether charges
corresponding to the residual unfixed gauge symmetries \cite{ReiWat08}.

The exact spectrum of the Yang--Mills Hamiltonian was obtained within both
 the {\it diagonal} and the \pcg, having implemented the constraints {on} the wave functional arising in the 
resolution of Gauss' law from the zero modes of
the Faddeev--Popov kernel. In the thermodynamic
limit, we recovered the well-known spectrum that leaves only the vacuum
state at zero energy, freezing out all excited states. The exact
vacuum state was used to calculate the ghost and gluon propagators,
the ghost-gluon vertex and the static colour Coulomb potential. We
compared the results in the \pcg restricted to the first Gribov region
to those in the \dcg restricted to the fundamental modular region. For
the propagators, the colour trace was found to be left invariant when transforming
from the \pcg to the \dcg. We found that the ghost propagator is
infrared enhanced, in agreement with the horizon condition widely used in
Dyson--Schwinger studies of $D=3+1$. This infrared enhancement is the
strongest when the configuration space is properly restricted to the
first Gribov region. We studied the effect of including several Gribov
copies, either by extending the configuration space to a union of
Gribov regions, or by using all Gribov regions with a Gaussian
damping. It was seen that the quantitative infrared enhancement of
the ghost propagator cannot be realized by any of these calculations
that include gauge copies from outside the first Gribov region. This
indicates that lattice calculations of the ghost propagator require a
very accurate gauge fixing and explains the shortcomings of the
infrared enhancement of the ghost propagator on the lattice when compared to
continuum studies \cite{LanMoy04,CucMen07}.

The Coulomb string tension
 yielded  the same results for both gauges which is a fortunate result,
 considering that in $D=3+1$ calculations the gauge is not completely
 fixed. The quantitative result of the static colour Coulomb potential,
 away from the thermodynamic limit, differs for both gauges. The result of the
 \pcg can be sort of artificially deformed into the result in the \dcg
 by suppressing Gribov copies with a Gaussian wave functional of width
 zero, as discussed in section \ref{ManyGribov}. The investigations showed that on
 $S^1\times\Real$, the Coulomb string tension arises from the abelian
 part of the Coulomb interaction and actually is identical to the
 string tension of the abelian theory, thus providing an upper bound
 of the gauge invariant string tension \cite{Zwa03b}. The effect of gauge
 copies on the static colour Coulomb potential was studied by taking
 several Gribov regions into account, and resulted in spurious locally
 stable minima for large separations of external colour charges.

The Dyson--Schwinger equations for the propagators and
vertices in the \pcg  were derived. It was shown that the exact solution within the
first Gribov region satisfies the Dyson--Schwinger equations, but that
these solutions are not the only ones. Changing the configuration
space from any union of Gribov regions to another leaves the
Dyson--Schwinger equations form invariant. This persists in the $D=3+1$
case. Therefore, it is legitimate to ask: In which union of Gribov
regions are the Green functions given when the set of Dyson--Schwinger
equations is solved by means of a truncation? We found in $D=1+1$ that
choosing the ghost-gluon vertex at tree-level, effectively puts
the solutions for the propagators and vertices into the first Gribov region. As for
the colour traces of the propagators, even the result within the fundamental modular
region is attained. The tree-level ghost-gluon vertex approximation for solving
Dyson--Schwinger equations---which was advocated by several
investigations before \cite{FeuRei04,CucMenMih04,Sch+05,Ste+05}---thus receives further strong support. 

The variational approach to Coulomb gauge Yang--Mills theory in
$D=3+1$ dimensions \cite{FeuRei04} integrates over all Gribov regions
for technical reasons. It was shown in section \ref{variationalsec} by
using the same ansatz for the vacuum wave functional in $D=1+1$,
that with the appropriate approximation (quadratic completion in
kinetic energy expression) the variational principle yields a set of
Dyson--Schwinger equations that is the exact one. With the tree-level
ghost-gluon vertex, the exact solution within the first Gribov region
is thus very well approximated. It can be expected that the infrared limit of the
$D=3+1$ theory, which is described by a stochastic wave functional as
in $D=1+1$, is thus also well-approximated. Furthermore, we found that the
Coulomb form factor that is necessary for the calculation of the
static colour Coulomb potential is not too far from tree-level in the
exact $D=1+1$ calculation and that the inclusion of many Gribov copies
simulate a spurious infrared enhancement of the Coulomb form
factor. We infer there are no indications that the choice of a
trivial Coulomb form factor is worse than any other approximation
made in the $D=3+1$ calculations.



\no
{\Large \bf Acknowledgements}

Useful discussions with  D. Campagnari, M. Quandt, P. Watson are
greatly acknowledged.  This work was supported in part by DFG under
contract no.\ DFG-Re856/6-1
and DFG-Re856/6-2.

\newpage

\begin{appendix}
\makeatletter
\def\@seccntformat#1{\csname Pref@#1\endcsname \csname the#1\endcsname\quad}
\def\Pref@section{Appendix~}
\makeatother

\section{$SU(2)$ colour rotations}
\label{rots}
The unitary matrix $U$, which
rotates the colour vector $
{\bf \hat{n}}
(\theta, \phi)$ 
 into the 3-direction  (and thus, in particular, diagonalises the gauge field
 (\ref{BG9})) is defined by
 \be
 \label{9-31}
 U^\dagger {\bf \hat{n}} (\theta, \phi) \cdot \vT\, U = T_3 \hk .
 \ee
This matrix is defined up to an abelian gauge transformation $U \to U \omega \hk
, \hk \omega = \exp (\varphi T_3) \in U (1) \subset
 SU (2)$, i.e.\ it  is defined on
the coset $SU (2) / U (1)$. The adjoint representation $\hat U$,
defined by
\begin{equation}
  \label{X1}
  U^\dagger T_aU=\hat U^{ab}T_b
\end{equation}
is related to the fundamental representation $U$ by
\begin{equation}
  \label{X2}
  \hat{U}_{ab} = - 2 \,\tr \lk U^\dagger T_a U T_b \rk \; , \quad \tr \lk
T_a T_b \rk = - \frac{1}{2} \delta_{ab} \hk .
\end{equation}
Equation (\ref{X2}) is most easily proved by
Taylor expanding $U$ in terms of $\Theta = \Theta_a T_{a}$ and using
\be
[T_a, \Theta] = \hat{\Theta}^{a b} T_b \hk .
\ee
Furthermore, since (\ref{X1}) is based only on the
 algebra of the generators it is  valid
in any representation, in particular the adjoint representation (where
$U^\dagger = U^T$),
\be
\label{13-42}
\hat{U}^T \hat{T}_a \hat{U} = \hat{U}_{ab} \hat{T}_b \hk .
\ee
The matrix $U$ can be realized by
\be
\label{5-19}
U (\theta, \phi) = \e^{{\displaystyle \theta
 \ve_\phi\cdot \vT}} \hk , \quad \vT = - \frac{i}{2}
\boldsymbol{\tau} \hk ,
\ee
where
\be
\label{5-20}
\ve_\phi = - \sin \phi \,\ve_1 + \cos \phi \,\ve_2 \hk 
\ee
is the unit vector in the direction of the azimuthal angle $\phi$. The matrix
$U$ (\ref{5-19}) can be alternatively expressed in terms of Euler angles as
\be
\label{5-21}
U (\theta, \phi) = \e^{\phi T_3} \e^{ \theta T_2}  \hk .
\ee
Equations \eqref{9-31}, \eqref{5-19}, \eqref{5-20} are valid in any
representation of $SU(2)$ and thus also in the adjoint representation
\be
\label{8-X2}
\hat{U} (\theta, \phi)\, := \e^{{\displaystyle \theta \ve_\phi \cdot \hat{\vT}}} = \e^{\phi \hat{T}_3} \e^{\theta
\hat{T}_2}  \hk .
\ee
From the defining equations (\ref{9-31}), \eqref{X2} and 
$\hat{U}^T = \hat{U}^{- 1}$ also follows
that the components of the unit colour vector ${\bf \hat{n}} (\theta, \phi)$ are given
by 
\be
\label{11XX}
{\bf \hat{n}}^a (\theta, \phi)  =  \hat{U}_{a3} (\theta, \phi) \hk .
\ee
It is convenient to use the bracket notation
\be
\hat{U}_{a b} \equiv \langle a | \hat{U} | b \rangle
\ee
and to express $\hat{U}$ in the basis of the eigenvectors $| \sigma  = 0 ,
 \pm 1 \rangle $ of the
spin 1 operators 
\be
\hat{S}^a = i \hat{T}^a \hk , \hk ({\hat T}_a)^{b c} = \epsilon^{bac}
\ee
satisfying
\begin{align}
\label{5-22}
\hat{S}^2 |  \sigma \rangle & = 1 (1 + 1) | \sigma \rangle \nonumber\\
\hat{S}_3 | \sigma \rangle & = \sigma |  \sigma \rangle \hk .
\end{align}
The transition elements
\be
\label{11-47}
\langle a | \sigma \rangle =: \e^a_\sigma
\ee
are the Cartesian components of the spherical unit vectors (in 
colour space)
\be
\label{BGneu9}
\ve_{\sigma = 1} = - \frac{1}{\sqrt{2}} \, \left( \begin{array}{c} 1 \\ i \\ 
0 \end{array} \right), \hspace{0.7cm} \ve_{\sigma = -1} = \frac{1}{\sqrt{2}} 
\left( 
\begin{array}{c} 1 \\ - i \\ 0 \end{array} \right), \hspace{0.7cm} \ve_{\sigma 
= 0} = \left( \begin{array}{c} 0 \\ 0 \\ 1 \end{array} \right) \hk .
\ee
Here, Greek letters $\sigma, \tau, \dots$
denote spherical colour components $\left\{ 1, 0, - 1 \right\}$, while Latin
letters $a, b, \dots$ denote the Cartesian colour components $\{1, 2, 3\}$.
The matrix elements of the adjoint representation $\hat{U}$ 
(\ref{8-X2}) in the spherical basis 
\be
\langle  \sigma | \hat{U} (\theta, \phi) |  \sigma' \rangle  = 
\langle \sigma | a \rangle \langle a | \hat{U} | b \rangle \langle b | \sigma'
\rangle =
\e^{a^*}_\sigma \hat{U}_{a b} (\theta, \phi) \e^b_{\sigma'} 
\ee
are related to the Wigner D-function by
\be
\label{5-23}
\langle \sigma | \hat{U} (\theta, \phi) | \sigma' \rangle 
 =  D^{J = 1}_{\sigma \sigma'} (\phi,  \theta, 0)
\hk .
\ee
Using $\langle \sigma | 3 \rangle = \e^{3^*}_\sigma = \delta_{\sigma 0}$,
the colour unit vector (\ref{11XX}) can be expressed as
\be
{\hat{n}}^a (\theta, \phi) = \langle a | \sigma \rangle \langle \sigma |
\hat{U} | \tau \rangle \langle \tau | 3 \rangle
= \e^a_\sigma D^1_{\sigma 0} (\phi, \theta, 0) \hk .
\ee

\section{Explicit resolution of Gauss' law}
\label{Gaussapp}
To identify $\Pi^a_{||} (x)$  we  Fourier expand the periodic
gauge field $A (x + L) = A (x)$
\be
\label{5-13}
A (x) = \frac{1}{L}\,\sli_n \e^{i k_n x} A (n) \hk , \quad k_n =
\frac{2 \pi n}{L} \hk , \quad n\in\mathbb{Z}\; .
\ee
The inverse transformation reads
\be
A (n) = \il^L_0 d x\, \e^{- i k_n x} A (x) \hk 
\ee
and the continuum limit $L \to \infty$ is obtained by the replacement
\be
\frac{1}{L} \sli_n \to \int \frac{d k}{2 \pi} \hk .
\ee
For later use we also quote the completeness and orthogonality relations
\be
\label{72-348}
\delta (x) =  \frac{1}{L} \sli_n \e^{i k_n x} \hk , \quad
\delta_{m, n}  =  \frac{1}{L} \int_0^L d x\, \e^{i \lk k_m - k_n\rk x} \hk ,
\ee
where $\delta (x)$ denotes the periodic $\delta$-function, satisfying
\be
\delta (x + L) = \delta (x) \hk .
\ee
From (\ref{5-13}) we find for the momentum operator 
\be
\label{20-100}
{\Pi^a(x)=\frac{\delta}{i\delta A^a (x)} = \frac{1}{L} \sli_n \e^{- i k_n x} \frac{d}{id A^a
(n)} \hk .}
\ee
\subsection{Pure Coulomb gauge}
In momentum space the \pcg (\ref{Couldef}) reads
\be
A (n) = \delta_{n, 0} A (0) \hk ,
\ee
where 
\be
A (0) = \frac{1}{L} \il^L_0 d x A (x) =: A
\ee
is the constant part of the gauge field, which is left after gauge fixing. 
From Eq.\ (\ref{20-100}) we read off the
transversal ($x$-independent) and longitudinal ($x$-dependent) parts of
 the momentum operator 
to be given by 
\be
\label{5-20app}
\Pi^a_\perp = \frac{1}{L} \frac{d}{i d {A}^a} \hk , \quad \Pi^a_{||} (x) =
\frac{1}{L} \sli_{n \neq 0} \e^{- i k_n x} \frac{d}{id A^a (n)} \hk .
\ee
In the \pcg, where the degrees of freedom are $A^{a = 1, 2, 3}$, the
charge of the gauge bosons (\ref{19-90}) 
is space independent but non-zero. Using (cf.\ Eq.\ (\ref{17A*}))
\be
i {\hat{D}^{a b}} \e^{- i k_n x} = \e^{- i k_n x} \langle a | \hat{U} |
\sigma \rangle \lambda_{n, \sigma} \langle \sigma | \hat{U}^T | b
\rangle \; ,
\ee
we obtain
\be
i {\hat{D}^{a b}} \Pi^b_{||} (x) = \frac{1}{L} \sli_{n \neq 0} \e^{- i k_n x}
\langle a | \hat{U} | \sigma \rangle \lambda_{n, \sigma} \langle \sigma |
\hat{U}^T | b \rangle \frac{d}{i d A^b (n)} \hk .
\ee
Inserting this  relation into Gauss' law (\ref{15XX}) 
and {multiplying} the resulting equation by
$\e^{i k_m x}$, and integrating over $x$ thereby using Eq.\ (\ref{72-348})
 we obtain
\be
\label{15-a}
 \sli_{n \neq 0} \delta_{n, m} \langle a | \hat{U} | \sigma
\rangle \lambda_{n, \sigma} \langle \sigma | \hat{U}^T | b \rangle {\frac{d}{id A^b
(n)}} \Psi (A)
 =  i  \il^L_0 d x \e^{i k_m x} \rho^a_{tot} (x) \Psi (A) \hk .
\ee
For $m = 0$ the l.h.s. of Eq.\ (\ref{15-a}) vanishes and we find that the wave
functional has to satisfy the following constraint
\be
\label{17-74}
Q^a\,\Psi(A)\equiv \il^L_0 d x\, \rho^a_{tot} (x)\, \Psi (A) = 0 \hk .
\ee
For $m \neq 0$ the summation over $n$ on the l.h.s. collapses to the term $m =
n$ and Eq.\ (\ref{15-a}) becomes after multiplying it by $\langle \sigma |
{\hat{U}^T} | a \rangle $ and summing over $a$
\be
\label{15-y}
 \lambda_{m, \sigma} \langle \sigma | \hat{U}^T | b \rangle \frac{d}{i d A^b
(m)} \Psi (A) 
 =  i  \langle \sigma | \hat{U}^T | a \rangle \il^L_0 d x\, \e^{i k_m x}
\rho^a_{tot} (x) \Psi (A) \hk .
\ee
Multiplying Eq.\ (\ref{15-y}) by ${\frac{1}{L}}\e^{-i k_m y} \langle c | \hat{U} | \sigma \rangle
\lambda^{- 1}_{m, \sigma}$ and summing over $m \neq 0$ and $\sigma$ we get
\begin{align}
\Pi^c_{||} (y) \Psi (A) & =  \frac{1}{L} \sli_{m \neq 0} \e^{- i k_m y}
\frac{d}{i d A^c (m)} \Psi (A) \nonumber\\
& =  i \frac{1}{L} \sli_{m \neq 0} \sli_\sigma \e^{- i k_m y} \il^L_0 d x\, \e^{i
k_m x} \langle c | \hat{U} | \sigma \rangle \lambda^{- 1}_{m, \sigma}
 \langle \sigma | \hat{U}^T | a \rangle \rho^a_{tot} (x) \Psi (A) \\
\label{15-y1}
& =  i \il^L_0 d x \sli_{m \neq 0} \sli_\sigma \tilde{\varphi}^c_{m, \sigma}
(y) \lambda^{- 1}_{m, \sigma} \tilde{\varphi}^{a^*}_{m, \sigma} (x) \rho^a_{tot}
(x) \Psi(A)\hk ,
\end{align}
where we have used the explicit form of the {eigen}functions
$\tilde{\varphi}^a_{n, \sigma} (x)$ (\ref{19-77}) 
of the covariant derivative $i
\hat{D}^{a b}$. Note if the modes $m = 0, \sigma = \pm 1$ were included in
Eq.\ (\ref{15-y1}), the sum would produce the inverse kernel
\be
\langle y, b | (i \hat{D})^{- 1} | a, x \rangle = \sli_{m, \sigma} {}'
\tilde{\varphi}^b_{m, \sigma} (y) \lambda^{- 1}_{m, \sigma}
\tilde{\varphi}^{a^*}_{m, \sigma} (x)\hk ,
\ee
where the prime indicates that the mode $m = \sigma = 0$ is excluded (while $m = 0,
\sigma = \pm 1$ is included).

It is now straightforward to calculate the Coulomb Hamiltonian {$H_C$}
defined by Eq.\ (\ref{80-79}).
With Eq.\ (\ref{15-y1}) we obtain after straightforward manipulations
\be
\label{18-80}
H_C = \frac{g^2}{2} \int d x d y\, \rho^a_{tot} (x) F^{a b} (x, y) {\rho^b_{tot}}
(y)
\hk ,
\ee
where
\begin{equation}
\label{15-x1}
F^{a b} (x, y)  =  \sli_{n \neq 0} \sli_\sigma \langle x, a | \hat{U} | n,
\sigma \rangle \lambda^{- 2}_{n, \sigma} \langle n, \sigma | \hat{U}^T | y, b
\rangle  =  \sli_{n \neq 0} \sli_\sigma \tilde{\varphi}^a_{n, \sigma} (x) \lambda^{-
2}_{n, \sigma} \tilde{\varphi}^{b^*}_{n, \sigma} (y)
\end{equation}
is the so-called Coulomb kernel. Let us stress the mode $n = 0, \sigma = \pm 1$
is here not included although it is not a zero mode $\lambda_{n = 0, \sigma =
\pm 1} \neq 0$. Since this mode is also excluded from the ghost kernel
\be
\hat{G}^{a b} (x, y)  =  \langle x, a | \hat{G}^{- 1} | y, b \rangle
 =  \langle x, a | (- \hat{D} \partial)^{- 1} | y, b \rangle 
 =  \sli_{n \neq 0} \sli_\sigma \tilde{\varphi}^a_{n, \sigma} (x) \lk k_n
\lambda_{n, \sigma} \rk^{- 1} \tilde{\varphi}^{b^*}_{n, \sigma} (y)
\ee
the Coulomb kernel (\ref{15-x1}) can be represented as
\be
F^{a b} (x, y) = \langle x, a | (- \hat{D} \partial)^{- 1} (- \partial^2) (-
\hat{D} \partial)^{- 1} | y, b \rangle \hk ,
\ee
which is the usual representation. In the abelian case, this kernel
 reduces to the usual Coulomb potential. In the non-abelian theory, this is a 
 dynamical object depending on the field variables via the covariant derivative.
  \bi
  
 \no
  The above derivation of $H_C$ has shown that
Gauss' law in the \pcg does not only give rise to the Coulomb Hamiltonian
$H_C$ (\ref{18-80}) but in addition yields the constraint (\ref{17-74}) 
on the wave
functional. This constraint arises from the zero modes of the Faddeev-Popov
kernel, which are a consequence of the incomplete gauge fixing. In a complete
gauge fixing such residual constraints would not arise\footnote{ 
When the constraint (\ref{17-74}) 
is obeyed by the wave functional the mode $m = 0 ,
\sigma = \pm 1$ can safely be included in the Coulomb kernel (\ref{15-x1}) since
it does not contribute when the Coulomb Hamiltonian acts on the wave functional.
Thus with the constraint (\ref{17-74}) 
satisfied we can use the alternative kernel
\be
F^{a b} (x, y)  =  \sli_{n, \sigma}{}' \langle x, a | \hat{U} | {n, \sigma} \rangle
\lambda^{- 2}_{n, \sigma} \langle {n, \sigma} | \hat{U}^T | y, b \rangle 
 = \langle x, a | (i \hat{D})^{- 2} | y, b \rangle  \hk 
\ee
in the Coulomb Hamiltonian. 
It is precisely this kernel (but with $A$ restricted to the {hyperplane} of the
\dcg) which arises as ``Coulomb kernel'' in the \dcg derived
in the next subsection.}.

Due to the fact that constant modes $k_n = 0$ are excluded from the Coulomb
kernel, the dynamical charge of the gauge bosons, $\rho_g$, being
space-independent, drops out from the Coulomb term (\ref{18-80}). In fact, with
the explicit form of eigenfunctions $\tilde{\varphi}^a_{n, \sigma} (x)$ 
(\ref{19-77})
we have
\be
\int d x F^{ab} (x, y) = \frac{1}{L} \int d x \sli_{n \neq 0} \e^{ i k_n
(x - y)} F^{a b}_n 
 =  \sli_{n \neq 0} \delta_{n, 0} F^{a b}_{n} = 0 \hk .
\ee
Thus we can replace in {$H_C$} (\ref{18-80}) the total charge $\rho^a_{tot}$ by
the external charge $\rho^a$,
\be
\label{19ZZ}
H_C = \frac{g^2}{2} \int d x d y \rho^a (x) F^{a b} (x, y) \rho^b (y) \hk ,
\ee
and in the absence of external charges $\rho^a (x) = 0$ the Yang-Mills
Hamiltonian reduces to the transversal part \eqref{4-12}.
\subsection{Diagonal Coulomb gauge}
In the \dcg (\ref{BG14}) the remaining physical {degree} of freedom of the
gauge field is ${A}^3 \equiv A^3 (n = 0)$ and the corresponding physical
momentum reads
\be
\label{15-z1}
\Pi^a_\perp = \delta^{a 3} \Pi^3_\perp \hk , \hk \Pi^3_\perp = \frac{1}{L}
\frac{d}{id A^3 } \hk .
\ee
We will keep here the same notation as in the \pcg and denote the remaining
unphysical part of the momentum operator by
\be
\Pi^a_{||} = \Pi^a - \Pi^a_\perp \hk .
\ee
This part is given here by
\be
\label{15-z2}
\Pi^a_{||} = \frac{1}{L} \sli_n {}' \e^{- i k_n x} \frac{d}{i d A^a (n)} \hk ,
\ee
where the prime indicates that the term $n = 0$ is excluded for the generator of
the Cartan algebra $a = a_0 = 3$ 
only.
Note that contrary to the \pcg, the ``transverse'' components $d / d A^{a =
\bar{a}} (0)$  belonging to the generators $a = \bar{a}$ of the coset $SU(N) /
U(1)^{N - 1}$ are here parts of $\Pi_{||}$.

With the explicit form of the gauge-fixed field (\ref{BG14}) and the corresponding
momentum $\Pi_\perp$ (\ref{15-z1}) 
one notices that the charge of gauge boson (\ref{19-90}) vanishes in this case,
\be
\rho^a_{g} = - \hat{A}^{a b}_\perp \Pi^b_\perp = - \hat{A}^{a 3}_\perp
\Pi^3_\perp  = - f^{a 33} A^3_\perp \Pi^3_\perp = 0 \hk .
\ee
Inserting the explicit
form of   $\Pi^a_{||}$,
given by Eq.\  (\ref{15-z2}), into Gauss' law
(\ref{15XX}) and using furthermore
\be
i \hat{D}^{a b} [A^3T_3] \e^{- i k_n x} = \e^{- i k_n x} \sli_\sigma \langle a | \sigma
\rangle \lambda_{n, \sigma} \langle \sigma | b \rangle \hk ,
\ee
Gauss' law becomes
\be
\frac{1}{L} \sli_n {}' \e^{- i k_n x} \sli_\sigma \langle a | \sigma \rangle
\lambda_{n, \sigma} \langle \sigma | b \rangle \frac{d}{i d A^b (n)} \Psi(A) = i
\rho^a (x) \Psi(A) \hk .
\ee
Multiplying this equation by $\e^{i k_m x} {\langle  \sigma | a \rangle}$,
integrating over $x$ and summing {over $a$}, we obtain
\be
\label{15-z3}
\sli_n {}' \delta_{n, m} \lambda_{n, \sigma} \langle \sigma | b \rangle \frac{d}{i
d A^b (n)} \Psi (A) = {i}\langle \sigma | a \rangle \il^L_0 d x \e^{i k_m x} \rho^a
(x) \Psi (A) \hk .
\ee
Recall that the prime indicates that the term $n = 0$ is excluded from the sum 
for $b = 3$. 
Since $\langle \sigma | b = 3 \rangle = \e^{3^*}_\sigma \equiv \delta_{\sigma
0}$ on the l.h.s. the term $n = 0$ is excluded for $\sigma = 0$. Thus for $m =
\sigma = 0$  the l.h.s. vanishes and we find the following
constraint on the wave functional
\be
\label{20-92}
Q^3 \Psi \equiv \il^L_0 d x \rho^3 (x) \Psi = 0 \hk ,
\ee
which should be compared with the constraint (\ref{17-74}) in the \pcg.
Since in the present case the charge of the gauge bosons vanishes, Eq.
(\ref{20-92}) is the restriction of the constraint (\ref{17-74}) to the charge
of the Cartan subgroup.
For $m = 0, \sigma \neq 0$ and for $m \neq 0, \sigma$-arbitrary, Eq.\ (\ref{15-z3}) becomes
\be
\label{15-z4}
 \lambda_{m, \sigma} \langle \sigma | b \rangle \frac{d}{i d A^b (m)} \Psi
(A) 
 = {i} \langle \sigma | a \rangle \il^L_0 d x \, \e^{i k_m x} \rho^a (x) \Psi (A) \hk
.
\ee
Note that since $\langle \sigma \neq 0 | b = 3 \rangle = 0$ the
summation over $b$ is for $m = 0, \sigma \neq 0$ restricted to $b = 1, 2$.
Multiplying the last equation by $\langle c | \sigma \rangle \lambda^{- 1}_{m,
\sigma}$ and summing over $\sigma$ we obtain 
\be
\frac{d}{i d A^c (m)} \Psi (A) ={i} \sli_\sigma {}' \langle c | \sigma \rangle
\lambda^{- 1}_{m, \sigma} \langle \sigma | a \rangle \int_0^L d x\, \e^{i k_m x} \rho^a
(x) \Psi (A) \hk ,
\ee
where the prime indicates again that the term $\sigma = 0$ is excluded for $m = 0$.
Multiplying this equation by $\e^{- i k_m y} / L$ and summing over $m$ and using
(\ref{15-z2}) we obtain
the desired representation
\be
\label{15-z5}
\Pi^c_{||} (y) \Psi (A) = {i}\il^L_0 d x \langle y, c | (i {\hat{D} [A^3T_3]})^{- 1} | {x, a}
\rangle \rho^a (x) \Psi (A) \hk ,
\ee
where
\be
\langle y, c | (i {\hat{D} [A^3T_3]})^{- 1} | {x, a} \rangle  =  \sli_{m, \sigma} {}'
\varphi^c_{m,\sigma} (y) \lambda^{- 1}_{m, \sigma} \varphi^{a^*}_{m,\sigma}
(x)\; ,\quad  \varphi^a_{m, \sigma} (x)  =  \frac{1}{\sqrt{L}} \e^{- i k_m x} \e^a_\sigma \hk
.
\ee
With Eq.\ (\ref{15-z5}) one finds for the Coulomb Hamiltonian $H_C$
  defined by  Eq.\ (\ref{80-79}) in this gauge with $\cJ_{FP} = \cJ_D$
  the following expression
\be
\label{20-102}
H_C = \frac{g^2}{2} \int d x d y \rho^a (x) F^{a b} (x, y) \rho^b (y)
\ee
with the Coulomb kernel given by
\be
F^{a b} (x, y) = \langle {x, a} | (i {\hat{D} [A^3T_3]}))^{- 2} | {y, b} \rangle
 =  \sli_{n, \sigma}{}' \langle x | n \rangle \langle a | \sigma
  \rangle \lambda^{- 2}_{n, \sigma} \langle \sigma | b \rangle \langle n | y\rangle \hk .
\ee
Contrary to the Coulomb kernel in the \pcg (\ref{15-x1}) 
here only the zero mode $m =
\sigma = 0$ is excluded, as indicated by the prime, while the mode $m = 0 ,
\sigma = \pm 1$ is included. The above considerations show that the Coulomb
Hamiltonian depends on the details of the gauge fixing and is thus {a priori} not a
physical quantity.
\end{appendix}

\newpage
 \bibliography{biblio}{}
 \bibliographystyle{utcaps}

\end{document}

%% file: Macros.tex
\newcommand{\be}{\begin{equation}} 
\newcommand{\ee}{\end{equation}}
\newcommand{\bea}{\begin{eqnarray}} 
\newcommand{\eea}{\end{eqnarray}}
\newcommand{\Det}{\hbox{Det}}
\newcommand{\hbo}{\hbox to 1 true cm {\hfill } } 
\newcommand{\tr}{\hbox{tr}}
\newcommand{\Tr}{\hbox{Tr}}

\newcommand{\Id}{ \mathbbm{1} }

\newcommand{\unit}{\mathbbm{1}}

\newcommand{\Real}{\mathbbm{R}}

\newcommand{\id}{{\mathbbm 1}}

\newcommand{\bra}[1]{\langle #1|}
\newcommand{\ket}[1]{|#1\rangle}

\renewcommand{\vec}[1]{\mbox{\boldmath$#1$\unboldmath}}

\newcommand{\cN}{{\cal N}}

\newcommand{\cG}{{\cal G}}
\newcommand{\cD}{{\cal D}}

\newcommand{\cO}{{\cal O}}

\newcommand{\cM}{{\cal M}}
\newcommand{\cJ}{{\cal J}}

\newcommand{\bA}{{\bar A}}

\newcommand{\vA}{{\mathrm{\bf A}}}

\newcommand{\ve}{{\vec{e}}}

\newcommand{\vT}{{\mathrm{\bf T}}}
\newcommand{\vL}{{\mathrm{\bf L}}}

\newcommand{\bi}{\bigskip}
\newcommand{\no}{\noindent}
\newcommand{\hk}{\hspace{0.1cm}}

\newcommand{\dslash}{\partial\hskip -0.5em/}

\newcommand{\Dslash}{D \hspace{-1.55em}|\hspace{.5em}}

\newcommand{\rk}{\right)}
\newcommand{\lk}{\left(}

\newcommand{\sli}{\sum\limits}
\newcommand{\pli}{\prod\limits}
\newcommand{\il}{\int\limits}

\renewcommand{\e}{\text{e}}

\newcommand{\nn}{\nonumber}

\newcommand{\vth}{\vartheta}
\newcommand{\lla}{\left\langle}
\newcommand{\rra}{\right\rangle}
\newcommand{\del}{\partial}
\newcommand{\rarr}{\to}

\def\dslash{\partial\kern-.5em\slash}
\def\kslash{k\kern-.5em\slash}
\def\pslash{p\kern-.5em\slash}
\def\Dslash{D\kern-.5em\slash}

\def\abs#1{\mathopen| #1 \mathclose|}	
\def\dcg{{\it diagonal Coulomb gauge}\xspace} 
\def\pcg{{\it pure Coulomb gauge}\xspace}